\documentclass[journal,draftclsnofoot,onecolumn,12pt]{IEEEtran}
\usepackage{amsthm,amssymb,graphicx,multirow,amsmath,color,amsfonts}
\usepackage[caption=false,font=footnotesize]{subfig}
\usepackage[update,prepend]{epstopdf}
\usepackage[noadjust]{cite}
\usepackage[latin1]{inputenc}
\usepackage{tikz}
\usepackage{bbm} 
\usepackage{pdfpages}
\usepackage{multirow}
\usepackage{tabulary}
\usepackage{comment}
\usepackage{algorithm}
\usepackage{url}
\usepackage{algpseudocode}
\def\BSTATE{\STATE\hskip-\ALG@thistlm}
\makeatother

\def\nbv{{\mathbf{v}}}

\def\nbx{{\mathbf{x}}}

\def\nb0{{\mathbf{0}}}
\def\nb1{{\mathbf{1}}}

\def\nbA{{\mathbf{A}}}

\def\ncalL{{\mathcal{L}}}

\def\ncalQ{{\mathcal{Q}}}

\def\nbbB{{\mathbb{B}}}

\def\nbbE{{\mathbb{E}}}

\def\nbbR{{\mathbb{R}}}

\newtheorem{ndef}{Definition}

\newtheorem{theorem}{Theorem}
\newtheorem{prop}{Proposition}
\newtheorem{cor}{Corollary}

\newtheorem{remark}{Remark}

\def\a{\overset{({\rm a})}{=}}
\def\b{\overset{({\rm b})}{=}}

\allowdisplaybreaks 

\begin{document}
\graphicspath{{./Figures/}}
\title{
Age of Information in Multi-source Updating Systems Powered by Energy Harvesting  
}
\author{
Mohamed A. Abd-Elmagid and Harpreet S. Dhillon
\thanks{M. A. Abd-Elmagid and H. S. Dhillon are with Wireless@VT, Department of ECE, Virginia Tech, Blacksburg, VA. Email: \{maelaziz,\ hdhillon\}@vt.edu. The support of the U.S. NSF (Grants CPS-1739642 and CNS-1814477) is gratefully acknowledged. 
}
\vspace{-5mm}
}

\maketitle

\begin{abstract}
This paper considers a multi-source real-time updating system in which an energy harvesting (EH)-powered transmitter node has multiple sources generating status updates about several physical processes. The status updates are then sent to a destination node where the freshness of each status update is measured in terms of Age of Information (AoI). The status updates of each source and harvested energy packets are assumed to arrive at the transmitter according to independent Poisson processes, and the service time of each status update is assumed to be exponentially distributed. Unlike most of the existing queueing-theoretic analyses of AoI that focus on characterizing its average when the transmitter has a reliable energy source and is hence not powered by EH (referred henceforth as a non-EH transmitter), our analysis is focused on understanding the distributional properties of AoI in multi-source systems through the characterization of its moment generating function (MGF). In particular, we use the stochastic hybrid systems (SHS) framework to derive closed-form expressions of the average/MGF of AoI under several queueing disciplines at the transmitter, including non-preemptive and source-agnostic/source-aware preemptive in service strategies. The generality of our results is demonstrated by recovering several existing results as special cases.
\end{abstract}
\section{Introduction} \label{sec:intro}
A typical model for real-time status update systems consists of a {\it transmitter node} that generates real-time status updates about some physical process(es) of interest and sends them through a communication network to a {\it destination node}. Such a model can be used to analyze the performance of a plethora of emerging Internet of Things (IoT)-enabled real-time applications including healthcare, factory automation, autonomous vehicles, and smart homes, to name a few \cite{abd2018role}. As a concrete example of healthcare applications, large-scale IoT deployments could be useful in containing pandemics through efficient monitoring and contact/infection tracing \cite{Roy20}. The performance of these applications highly depends upon the freshness of the information status at the destination node about its monitored physical process(es). Because of that, the main design objective of such real-time status update systems is to ensure timely delivery of status updates from the transmitter node to the destination node. To measure the freshness of information at the destination node, the authors of \cite{kaul2012real} introduced the concept of AoI which accounts for the generation time of each status update (which was ignored by conventional performance metrics, specifically throughput and delay). In particular, for a queuing-theoretic model in which status updates are generated at the transmitter node according to a Poisson process, AoI was defined in \cite{kaul2012real} as the time elapsed since the latest successfully received status update at the destination node was generated at the transmitter node.

As will be discussed next in detail, the queueing-theoretic analyses of AoI have mostly been focused on the characterization of its average in the case of having a non-EH transmitter. However, it is infeasible to ensure the availability of a reliable energy source at the transmitter node in many practical IoT scenarios. For instance, a transmitter node could represent an aggregator deployed at a hard-to-reach place in a large-scale IoT network, where it is impractical to replace or recharge the energy battery at the aggregator \cite{abd2018coverage}. To enable a sustainable operation of real-time status update systems in such scenarios, EH has been considered as a promising solution for powering the transmitter nodes (majority of which are low-power nodes). While there are a handful of prior works analyzing AoI for the system in which a transmitter node is powered by EH, their analyses have been limited to the evaluation of its average and that too to the special case where the transmitter has a single source that generates status updates about a single physical process. Motivated by this, we provide the first queueing-theoretic analysis of the distributional properties of AoI for a generic setup in which an EH-powered transmitter has multiple sources which generate status updates about multiple physical processes.
\subsection{Related Work} 
For systems in which a non-EH transmitter has a single source that generates status updates about some physical process, referred to as single-source systems, the authors of \cite{kaul2012real} first derived a closed-form expression of the average AoI under first-come-first-served (FCFS) queueing discipline. The average of AoI or peak AoI (an AoI-related metric introduced in \cite{costa2016age} to capture the peak values of AoI over time) is then characterized under several queueing disciplines in a series of subsequent prior works \cite{kaul2012status,soysal,costa2016age,chen2016age,kam2018age,kavitha2018controlling,zou2019waiting}. Further, a handful of recent works aimed to characterize the distribution (or some distributional properties) of AoI/peak AoI  \cite{Inoue19,kosta2020non,Champati19,Chiariotti_dist,ayan2020probability,olga20}. While AoI has been extensively analyzed in single-source systems, the prior work on the analysis of AoI in multi-source systems has been fairly limited \cite{yates2012real,Moltafet_multisource,pappas2015age,yates2017status,kosta2019age,najm2018status,huang2015optimizing,Xu_21,Akar21,Ozancan_2021}. Note that a multi-source system refers to the setup where a non-EH transmitter has multiple sources generating status updates about multiple physical processes. The average AoI was characterized for the M/M/1 FCFS queueing model in \cite{yates2012real}, the M/G/1 FCFS queueing model in \cite{Moltafet_multisource}, and the M/M/1 FCFS with preemption in waiting queueing model (where the transmitter has a buffer that only keeps the latest generated status update from each source) in \cite{pappas2015age}. The authors of \cite{yates2017status} and \cite{kosta2019age} analyzed the average AoI under scheduled and random multiaccess strategies for delivering the status updates generated from different sources at the transmitter. The average peak AoI was derived for the M/G/1 last-come-first-served (LCFS) queueing model with (without) preemption in service in \cite{najm2018status} (in \cite{huang2015optimizing}), and for the priority FCFS and LCFS queueing models (where the sources of information are prioritized at the transmitter) in \cite{Xu_21}. Further, the distributions of AoI and PAoI were numerically characterized for various discrete time queues in \cite{Akar21}, and for a probabilistically preemptive queueing model in \cite{Ozancan_2021} where a new arriving status update preempts the one in service with some probability.
Different from \cite{kaul2012status,soysal,costa2016age,chen2016age,kam2018age,kavitha2018controlling,zou2019waiting,Inoue19,kosta2020non,Champati19,Chiariotti_dist,ayan2020probability,olga20,yates2012real,huang2015optimizing,pappas2015age,yates2017status,najm2018status,kosta2019age,Moltafet_multisource,Xu_21,Akar21,Ozancan_2021}, our focus in this paper is on the analytical characterization of distributional properties of AoI in the case where the transmitter has multiple sources of information and is powered by EH.

The analyses of the above works were mainly based on identifying the properties of the AoI sample functions and applying geometric arguments, which often involve convoluted calculations of joint moments. This has motivated the authors of \cite{yates2018age} and \cite{yates2020age} to build on the SHS framework\footnote{A detailed description of the SHS will be provided in Section \ref{sec:problem_statement}.} in \cite{hespanha2006modelling}, and derive promising results allowing the use of the SHS approach for the queueing-theoretic analyses of AoI. Following \cite{yates2018age,yates2020age}, the SHS approach was then used to evaluate the average AoI for a variety of queueing disciplines in \cite{SHS_6,SHS_7,SHS_5}, and the MGF of AoI for a two-source system with status update management in \cite{SHS_8}. Compared to the analyses of \cite{SHS_6,SHS_7,SHS_5,SHS_8} that considered a non-EH transmitter, the analysis of AoI using the SHS approach becomes much more challenging when we consider an EH-powered transmitter. This is due to the fact that the joint evolution of the battery state at the transmitter and the system occupancy with respect to the status updates has to be incorporated in the process of decision-making (i.e., the decisions of discarding or serving the new arriving status updates at the transmitter). This, in turn, requires analyzing a two-dimensional continuous-time Markov chain (modeling the system discrete state that is represented by the number of energy packets in the battery and the number of status updates in the system) with new transitions associated with the events of harvested energy packet arrivals/departures, compared to the conventional one-dimensional Markov chain used in \cite{SHS_6,SHS_7,SHS_5,SHS_8} to track the number of status updates in a system with a non-EH transmitter.
\begin{table*}
\centering
{\caption{A Summary of the Queueing Theory-based Analyses of AoI in the Existing Literature.} 
\label{table:summary}
\scalebox{.85}
{ \begin{tabular}{ |p{5cm}|p{3cm}|p{3cm}|p{3cm}|p{3cm}|}
\hline
    & \multicolumn{2}{c|}{A non-EH transmitter}  & \multicolumn{2}{c|}{EH-powered transmitter}\\
    & Single-source & Multi-source & Single-source & Multi-source \\ \hline
Average of AoI/peak AoI& \cite{kaul2012real,kaul2012status,soysal,costa2016age,chen2016age,kam2018age,kavitha2018controlling,zou2019waiting} & \cite{yates2012real,huang2015optimizing,pappas2015age,yates2017status,najm2018status,kosta2019age,Moltafet_multisource,Xu_21,yates2018age,SHS_6,SHS_7,SHS_5} & \cite{Yates_EH,zheng2019closed,farazi2018average,farazi2018bverage} & This paper \\ \hline
Distribution/distributional properties of AoI/peak AoI& \cite{Inoue19,kosta2020non,Champati19,Chiariotti_dist,ayan2020probability,olga20} &  \cite{Akar21,Ozancan_2021,yates2020age,SHS_8} &\cite{abdelmagid_2021a} & This paper\\ \hline
\end{tabular}}} 
\end{table*} 

For the case where the transmitter is powered by EH, there are a handful of prior works \cite{Yates_EH,zheng2019closed,farazi2018average,farazi2018bverage,abdelmagid_2021a} analyzing AoI by applying geometric arguments \cite{Yates_EH,zheng2019closed}, and by using the SHS approach \cite{farazi2018average,farazi2018bverage,abdelmagid_2021a}. However, the analyses of \cite{Yates_EH,zheng2019closed,farazi2018average,farazi2018bverage} have been limited to the evaluation of the average AoI in single-source systems, and the analysis of \cite{abdelmagid_2021a} was focused on the characterization of the distributional properties of AoI in single-source systems. Different from these, this paper makes the first attempt at deriving the distributional properties of AoI for a variety of queueing disciplines in multi-source systems with an EH-powered transmitter. Table \ref{table:summary} further highlights the gap in the literature that we aim to fill in this paper. Before going into more details about our contributions, it is instructive to note that besides the above queueing theory-based analyses of AoI, there have also been efforts to optimize AoI or some other AoI-related metrics in different communication systems that deal with time critical information (see \cite{roy_survey} for a comprehensive survey). For instance, AoI has been studied in the context of EH systems \cite{Baran_EH,Jing_EH,Leng19,arafa2019age,AbdElmagid2019Globecom_a,hatami2020age,abd2019tcom,AbdElmagid_joint,Elvina21,khorsandmanesh2020average,nouri2020age}, age-optimal transmission scheduling policies \cite{sun2017update,Qing_he,han2020fairness}, remote estimation \cite{ornee2019sampling}, ultra-reliable low-latency vehicular networks \cite{abdel2018ultra}, unmanned aerial vehicle (UAV)-assisted communication systems \cite{abd2018average,AbdElmagid2019Globecom_b,ferdowsi2021neural}, large-scale analysis of IoT networks \cite{emara2019spatiotemporal,mankar2020stochastic_GC2,Praful_GC1}, cache updating systems \cite{tang2020age,ma2020age,bastopcu2020information}, and timely communication in federated learning \cite{yang2020age,buyukates2020timely}.
\subsection{Contributions}
This paper analyzes the AoI performance of a multi-source status update system in which an EH-powered transmitter is equipped with a battery of finite capacity to store the harvested energy packets. In particular, we characterize the AoI performance under the LCFS without (LCFS-WP) and with [source-agnostic (LCFS-PS)/source-aware (LCFS-SA)] preemption in service queueing disciplines. An arriving status update at the transmitter preempts the one being served (regardless of its generating source index) under the LCFS-PS queueing discipline, whereas the preemption in service under the LCFS-SA queueing discipline only occurs when the two status updates (the arriving one and the one being served) are generated from the same source. In our analysis, the harvested energy packets/status updates generated from each source are assumed to arrive at the transmitter according to a Poisson process, and the service time of each status update is assumed to be exponentially distributed. For this setup, our main contributions are listed next.

{\it A novel analysis for deriving the average/MGF of AoI associated with each source at the destination.} We use the SHS framework to first derive closed-form expressions of the average AoI of each source for each of the considered queueing disciplines. We then extend our analysis to understand the distributional properties of AoI through the characterization of its MGF under each queuing discipline. These results allow us to gain useful insights about the achievable AoI performance by each of the considered queueing disciplines. For instance, we analytically characterize the gaps between the achievable average AoI performances by the considered queueing disciplines as functions of the system parameters. Further, using the MGF of AoI expressions, we also characterize the relationship between the achievable second moments of AoI by the considered queueing disciplines.

{\it Asymptotic results demonstrating the generality of the derived expressions.} We demonstrate that as the aggregate generating rate of status updates from all the sources other than the source of interest approaches zero, the average AoI expressions derived in this paper reduce to their counterparts in \cite{farazi2018average} and \cite{abdelmagid_2021a} for single-source systems with an EH-powered transmitter, and the derived MGF of AoI expressions reduce to their counterparts in \cite{abdelmagid_2021a}. We further demonstrate that as the arrival rate of harvested energy packets at the transmitter node becomes large, the derived AoI results converge to their counterparts in \cite{costa2016age} and \cite{yates2018age} for single-source and multi-source systems with a non-EH transmitter, respectively. 

{\it System design insights.} Our numerical results provide several useful system design insights. For instance, they show that the achievable AoI performance by each queueing discipline improves with the increase in either the battery capacity or the arrival rate of harvested energy packets at the transmitter node. They also show that the superiority of the LCFS-PS queueing discipline over the LCFS-WP and LCFS-SA queueing disciplines in terms of the achievable AoI performance comes at the expense of having unfair achievable average AoI values among different sources. Further, they reveal that as the number of sources increases, the LCFS-SA queueing discipline becomes more effective (compared to the LCFS-PS) in achieving fairness between the achievable AoI performances by different sources. Finally, the results demonstrate the importance of incorporating the higher moments of AoI in the implementation/optimization of multi-source real-time status updates systems rather than just relying on its average.
\section{System Model}\label{sec:Model}
\subsection{Network Model}
\begin{figure}[t!]
\centering
\includegraphics[width=0.65\columnwidth]{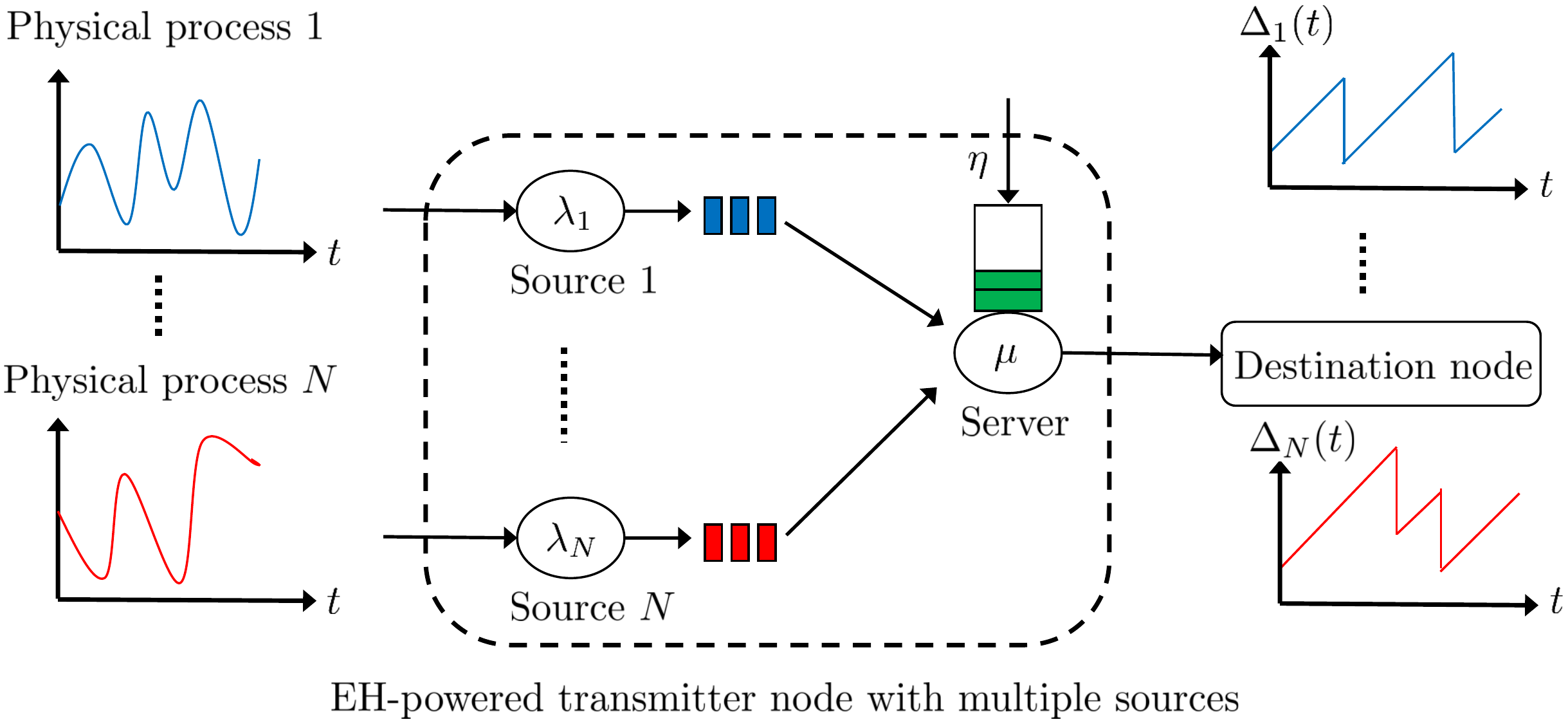}
\caption{An illustration of the system setup.}
\label{f:sys_setup}
\end{figure}

We consider a real-time status update system in which an EH-powered transmitter node observes $N$ physical processes, and sends its measurements to a destination node in the form of status update packets. As shown in Fig. \ref{f:sys_setup}, the transmitter node contains $N$ sources and a single server; each source generates status updates about one physical process, and the server delivers the status updates generated from all the sources to the destination. In particular, each status update packet generated by source $i$ carries some information about the value of the $i$-th physical process and a time stamp indicating the time at which that information was measured. 
This system setup can be mapped to many scenarios of practical interest, such as an IoT network in which an aggregator (represents the transmitter in our model) delivers measurements sensed/generated by the $N$ IoT devices (represent the sources) in its vicinity to a destination node. 

Source $i$ is assumed to generate status update packets at the transmitter node as a rate $\lambda_i$ Poisson process. Further, the transmitter harvests energy in the form of energy packets such that each energy packet contains the energy required for sending one status update packet from any of the sources to the destination node \cite{Yates_EH,zheng2019closed,farazi2018average,farazi2018bverage}. In particular, the harvested energy packets are assumed to arrive at the transmitter according to a Poisson process with rate $\eta$, and are stored in a battery queue of length $B$ packets at the server (for serving the update packets generated by the different sources). Given that the transmitter node has at least one energy packet in its battery queue, the time needed by its server to send a status update packet is assumed to be a rate $\mu$ exponential random variable \cite{kaul2012real,kaul2012status,costa2016age}. Let $\rho = \frac{\lambda}{\mu}$ and $\beta = \frac{\eta}{\mu}$ respectively denote the server utilization and energy utilization factors, where $\lambda = \sum_{i=1}^{N}{\lambda_i}$. Further, we have $\rho_i = \frac{\lambda_i}{\mu}$, $\lambda_{-i} = \sum_{j=1,\;j\neq i}^{N}{\lambda_j}$, and $\rho_{-i} = \frac{\lambda_{-i}}{\mu}$.

We quantify the freshness of information status about each physical process at the destination node (as a consequence of receiving status update packets from the transmitter node) using the concept of AoI. Formally, AoI is defined as follows \cite{kaul2012real}.
\begin{ndef}
Let $t_{i,k}$ and $t'_{i,k}$ denote the arrival and reception time instants of the $k$-th update packet of source $i$ at the transmitter and destination, respectively. Further, define $L_{i}(t)$ to be the index of the source $i$'s latest update packet received at the destination by time $t$, i.e., $L_{i}(t) = {\rm max}\{k|t'_{i,k} \leq t, \forall k\}$. Then, the AoI associated with the physical process observed by source $i$ at the destination node (referred henceforth as the AoI of source $i$) is defined as the following random process
\begin{align}
\Delta_i(t) = t - t_{L_i(t)}.
\end{align}
\end{ndef}
\subsection{Queueing Disciplines Considered in this Paper}\label{sub:disciplines}
For the above system setup, we analyze the AoI performance at the destination under three different queueing disciplines for managing update packet arrivals at the transmitter node. These queueing disciplines are described next.
\begin{itemize}
    \item {\it LCFS-WP queueing discipline}: Under this queueing discipline, a new arriving update packet at the transmitter (from any of the sources) enters service upon its arrival if the server is idle (i.e., there are no status update packets in the system) and the battery contains at least one energy packet; otherwise, the new arriving update packet is discarded. 
    \item {\it LCFS-PS queueing discipline}: When the server is idle, the management of a new arriving update packet under this queueing discipline is similar to the LCFS-WP one. However, when the server is busy, a new arriving update packet replaces the current packet being served and the old packet in service is discarded. 
    \item {\it LCFS-SA queueing discipline}: This queueing discipline is similar to the LCFS-PS one with the only difference that a new arriving update packet preempts the packet in service only if the two packets (the new arriving packet and the one in service) are generated from the same source.
\end{itemize}

Note that according to the LCFS-PS queueing discipline, status updates of a source $i$ with a small $\lambda_i$ are more likely to be preempted in service by status updates of a source $j$ with $\lambda_j \gg \lambda_i$. Since this issue is resolved under the LCFS-SA queueing discipline by only allowing preemption in service between the status updates generated from the same source, we expect that the LCFS-SA queueing discipline will be more effective (compared to the LCFS-PS) in achieving fairness between the achievable AoI performances by different sources (as will be demonstrated in Section \ref{sec:numerical}). As already conveyed, we consider that an energy packet contains the amount of energy required for sending one status update to the destination. Therefore, we assume that the length of the energy battery queue reduces by one whenever a status update is successfully transmitted to the destination. Further, with regards to the EH process, we consider that the transmitter can harvest energy only if its server is idle\footnote{The case where the transmitter can harvest energy anytime (i.e., even its server is busy) is left as a promising direction of future work.}. This case corresponds to the scenario where the transmitter is equipped with a single radio frequency (RF) chain and a single antenna, and thus can either transmit a status update or harvest energy at a certain time instant. 
\section{Problem Statement and Solution Approach}\label{sec:problem_statement}
Our goal is to analytically characterize the AoI performance of each source at the destination node as a function of: i) the rates of generating status update packets by the $N$ sources $\{\lambda_i\}$, ii) the rate of harvesting energy packets $\eta$, iii) the rate of serving status update packets $\mu$, and iv) the finite capacity of the energy battery queue $B$, at the transmitter node. Unlike most of the analyses of AoI in the existing literature which were focused on deriving its average, our analysis is focused on deriving distributional properties of AoI through the characterization of its MGF. 
To derive the MGF of AoI for the considered queueing disciplines at the transmitter node (presented in Subsection \ref{sub:disciplines}), we resort to the SHS framework in \cite{hespanha2006modelling}, which was first tailored for the analysis of AoI by \cite{yates2018age} and \cite{yates2020age}. In the following, we provide a very brief\footnote{Interested readers are advised to refer to \cite{yates2018age} and \cite{yates2020age} for a detailed discussion about the use of the SHS approach in the analysis of AoI.} introduction of the SHS framework, which will be useful in understanding our AoI MGF analysis in the subsequent sections. The SHS technique is used to analyze hybrid queueing systems that can be modeled by a combination of discrete and continuous state parameters. In particular, the SHS technique models the discrete state of the system $q(t) \in \ncalQ = \{1,\cdots,m\}$ by a continuous-time finite-state Markov chain, where $\ncalQ$ is the discrete state space. This continuous-time Markov chain governs the dynamics of the system discrete state that usually describes the occupancy of the system, e.g., $q(t)$ represents the numbers of status update and energy packets in the system for our problem. On the other hand, the evolution of the continuous state of the system is described by a continuous process ${\bf x}(t) = [x_0(t), \cdots,x_n(t)] \in \nbbR^{1 \times \left(n+1\right)}$, e.g., $x(t)$ models the evolution of the age-related processes in our system setting. 

A transition $l \in \ncalL$ from state $q_l$ to state $q'_l$ (in the Markov chain modeling $q(t)$) occurs due to the arrival of a status update/energy packet or the delivery of a status update to the destination (i.e., the departure of a status update from the system), where $\ncalL$ denotes the set of all transitions. Since the time elapsed between arrivals/departures is exponentially distributed, a transition $l$ takes place with a rate $\lambda^{(l)} \delta_{q_l,q(t)}$, where the Kronecker delta function $\delta_{q_l,q(t)}$ ensures that $l$
occurs only when the discrete state $q(t)$ is equal to $q_l$.
As a consequence of the occurrence of transition $l$, the discrete state of the system moves from state $q_l$  to state $q'_
l$, and the continuous state $\nbx$ is reset to $\nbx'$ according to a binary
reset map matrix $\nbA_l \in \nbbB^{(n+1)\times(n+1)}$ as $\nbx'=\nbx \nbA_l$. Further, $\overset{\cdot}{\nbx}(t) \triangleq \dfrac{\partial\nbx(t)}{\partial t} = {\bf 1}$ holds
as long as the state $q(t)$ is unchanged, where ${\bf 1}$ is the row vector $[1, \cdots,1] \in \nbbR^{1\times(n+1)}$. Different from ordinary continuous-time Markov chains, an inherent feature of SHSs is the possibility of having self-transitions in the Markov chain modeling the system discrete state. In particular, although a self-transition keeps $q(t)$ unchanged, it causes a change in the continuous process $x(t)$.

Now, we define some useful quantities for the characterization of the MGF of AoI at the destination node using the SHS technique. Denote by $\pi_q(t)$ the probability of being in state $q$ of the continuous-time Markov chain at time $t$. Further, let $\nbv_q(t) = [v_{q0}(t), \cdots,v_{qn}(t)] \in \nbbR^{1\times(n+1)}$ denote the correlation vector between $q(t)$ and $x(t)$, and $\nbv^s
_q(t) = [v^s_{q0}(t), \cdots, v^s_{qn}(t)] \in
\nbbR^{1\times(n+1)}$ denote the correlation vector between $q(t)$ and the exponential function $e^{s\nbx(t)}$, where $s \in \nbbR$. Thus, we can respectively express $\pi_q(t)$, $\nbv_q(t)$ and $\nbv^s_q(t)$ as
\begin{align}
\pi_q(t) = {\rm Pr}\left(q(t) = q\right) = \nbbE[\delta_{q,q(t)}],\; \forall q \in \ncalQ,
\end{align}
\begin{align}
\nbv_q(t) = [v_{q0}(t),\cdots,v_{qn}(t)] = \nbbE[\nbx(t)\delta_{q,q(t)}],\; \forall q \in \ncalQ,
\end{align}
\begin{align}
\nbv^s_q(t) = [v^s_{q0}(t),\cdots,v^s_{qn}(t)] = \nbbE[e^{s\nbx(t)}\delta_{q,q(t)}],\; \forall q \in \ncalQ.
\end{align}

According to the ergodicity assumption of the continuous-time Markov chain modeling $q(t)$ in the AoI analysis \cite{yates2018age,yates2020age}, the state probability vector $\pi(t) =
[\pi_0(t),\cdots,\pi_m(t)]$ converges uniquely to the stationary vector $\bar{\pi} = [\bar{\pi}_0,\cdots,\bar{\pi}_m]$ satisfying
\begin{align}\label{gen_steady}
\bar{\pi}_q\sum_{l\in\ncalL_q}{\lambda^{(l)}} = \sum_{l\in\ncalL'_q}{\lambda^{(l)}\bar{\pi}_{q_l}},\; q \in \ncalQ,\;\; \sum_{q\in\ncalQ}{\bar{\pi}_q} = 1,
\end{align}
where $\ncalL'_q = \{l\in\ncalL: q'_l = q\}$ and $\ncalL_q = \{l\in\ncalL: q_l = q\}$ denote the sets of incoming and outgoing transitions for state $q, \forall q \in \ncalQ$. 

Using the above notations, it has been shown in \cite[Theorem~1]{yates2020age} that under the
ergodicity assumption of the Markov chain modeling $q(t)$, if we can find a non-negative limit $\bar{\nbv}_q = [\bar{v}_{q0},\cdots,\bar{v}_{qn}],\;\forall q \in \ncalQ$, for the correlation vector $\nbv_q(t)$ satisfying
\begin{align}\label{gen_vavg}
\bar{\nbv}_q \sum_{l \in \ncalL_q}{\lambda^{(l)}} = \bar{\pi}_q {\bf 1} + \sum_{l\in\ncalL'_q}{\lambda^{(l)}} \bar{\nbv}_{q_l}\nbA_l,\; q \in \ncalQ,
\end{align} 
then:
\begin{itemize}
    \item The expectation of $x(t)$, $\nbbE[x(t)]$, converges to the following stationary vector:
    \begin{align}\label{gen_conv_Mom1}
\nbbE[x] = \sum_{q \in \ncalQ}{\bar{\nbv}_q}.
\end{align}
    \item There exists $s_0 > 0$ such that for all $s < s_0$, $\nbv_q^s(t)$ converges to $\bar{\nbv}^s_q$ that satisfies
\begin{align}\label{gen_vMGF}
\bar{\nbv}_q^s \sum_{l \in \ncalL_q}{\lambda^{(l)}} = s\bar{\nbv}_q^s + \sum_{l\in\ncalL'_q}{\lambda^{(l)}} [\bar{\nbv}_{q_l}^s\nbA_l + \bar{\pi}_{q_l} {\bf 1}\hat{\nbA}_l],\; q \in \ncalQ,
\end{align}
where $\hat{\nbA}_l \in \nbbB^{(n+1)\times(n+1)}$ is a binary matrix whose elements are constructed as
\begin{align}\label{A_hat}
\hat{\nbA}_l(k,j)=\begin{cases} 
1, \;\;& k=j, \;\text{and the j-th column of}\; \nbA_l\;\text{is a zero vector},\\
0, \;\;& \text{otherwise}.
\end{cases}
\end{align}

Further, the MGF of the state $\nbx(t)$, which can be obtained as $\nbbE[e^{s\nbx(t)}]$, converges to the following stationary vector:
\begin{align}\label{gen_conv_MGF}
\nbbE[e^{s\nbx}] = \sum_{q \in \ncalQ}{\bar{\nbv}_q^s}.
\end{align}
\end{itemize}

From (\ref{gen_conv_Mom1}) and (\ref{gen_conv_MGF}), when the first element of the continuous state $\nbx(t)$
represents the AoI at the destination node, the expectation and the MGF of AoI at the destination node respectively converge to:
\begin{align}\label{gen_Mom1}
\Delta_{1} = \sum_{q\in\ncalQ}{\bar{v}_{q0}},
\end{align}
\begin{align}\label{gen_MGF}
M(s) = \sum_{q\in\ncalQ}{\bar{v}^{s}_{q0}}.
\end{align}

\section{The Average AoI for the Queueing Disciplines Considered in this Paper}\label{sec:Avg}
It is clear from  \cite[Theorem~1]{yates2020age} (stated in Section \ref{sec:problem_statement}) that in order to use (\ref{gen_vMGF}) to derive the MGF of AoI at the destination, one needs to find a non-negative limit $\bar{\nbv}_q$ ($\forall q \in \ncalQ$) satisfying (\ref{gen_vavg}), which directly characterizes the average AoI as observed from (\ref{gen_conv_Mom1}). Thus, we first show in this section that this condition holds for the three queueing disciplines considered in this paper, which will immediately lead to the average AoI characterization for each queueing discipline. Afterwards, we extend our analysis in the next section to derive the MGF of AoI. 

Without loss of generality, we consider that source 1 is the source of interest in the AoI analysis in the sequel. The AoI performance of the other sources can then be obtained using the same expressions derived for source 1, as will be clear shortly. While analyzing the AoI of source 1, the status update packets associated with the other sources are generated according to a Poisson process with rate $\lambda_{-1} = \sum_{j=2}^{N}{\lambda_j}$. Using the notations of the SHS approach (presented in Section \ref{sec:problem_statement}), the continuous process $\nbx(t)$ in each queueing discipline is given by $\nbx(t)=[x_0(t),x_1(t)]$, where $x_0(t)$ represents the value of the source 1's AoI at the destination node at time instant $t$ (i.e., $\Delta_1(t)$), and $x_1(t)$ indicates the value that the source 1's AoI at the destination will become if the existing update packet in the system completes its service at time instant $t$ (i.e., the packet is delivered to the destination at $t$). Recall from Section \ref{sec:problem_statement} that as long as there is no change (due to the arrival/departure of an update/energy packet) in the discrete state $q(t)$, we have $\dfrac{\partial \nbx(t)}{\partial t} = 1$, i.e., the elements of the age vector $\nbx(t)$ increase linearly with time.  
\begin{figure}[t!]
\centering
\includegraphics[width=0.6\columnwidth]{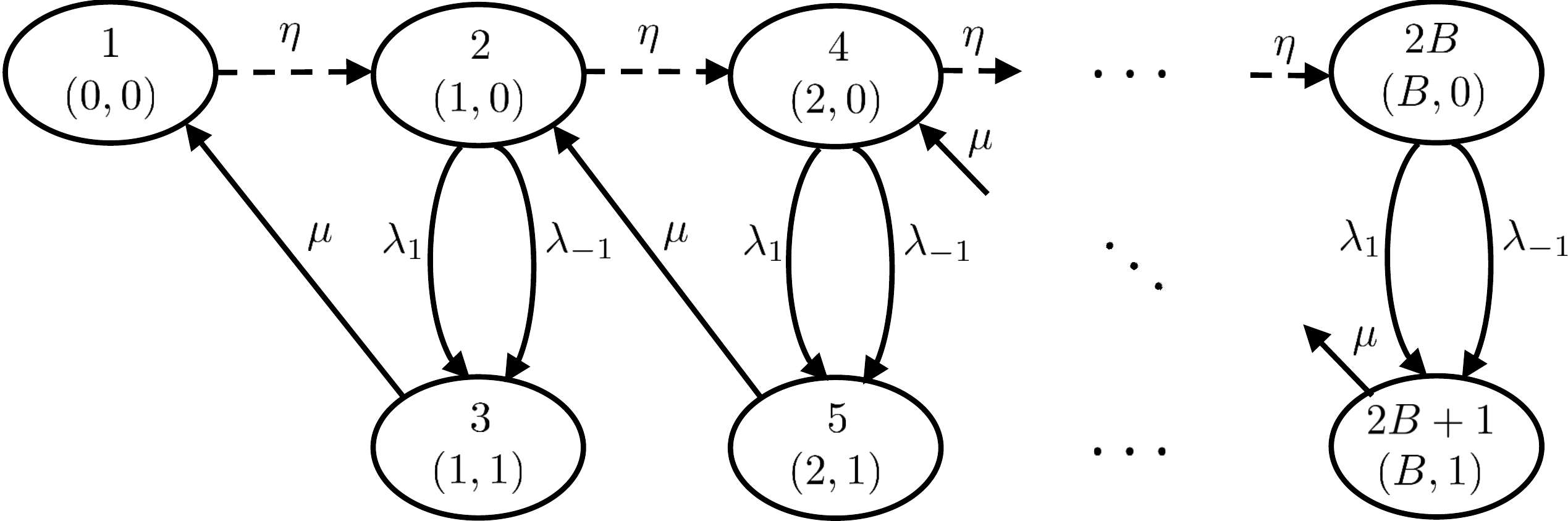}
\caption{The Markov chain modeling the discrete state in the LCFS-WP queueing discipline.}
\label{f:WP_MC}
\end{figure}
\subsection{LCFS-WP Queueing Discipline}\label{sub:Avg_WP}
The continuous-time Markov chain modeling the discrete state of the system $q(t) \in \ncalQ$ under the LCFS-WP queueing discipline is depicted in Fig. \ref{f:WP_MC}. Each state in $\ncalQ$ represents a potential combination of the number of update packets in the system and the number of energy packets in the battery queue at the server. For instance, a state $q=(e_q,u_q)$ indicates that the system has $u_q$ status update packets and the energy battery queue at the server contains $e_q$ energy packets. Note that since the system can have at most one status update packet at any time instant in the LCFS-WP queueing discipline and there is no need to track the source index from which the update packet in service was generated, we have $u_q \in \{0,1\}$. In particular, $u_q=0$ indicates that the system is empty and hence the server is idle, and $u_q=1$ indicates that the server is serving the existing update packet in the system. Since the battery queue at the server has a capacity of $B$ packets, we have $e_q \in \{0,1,\cdots,B\}$. We denote the set of states in the $i-$th row of the Markov chain by ${\rm r}_i$. Further, Table \ref{table:WP} presents the set of different transitions $\ncalL$ and their impact on the values of both $q(t)$ and $\nbx(t)$. Before proceeding into evaluating $\bar{\nbv}_q$, $\forall q \in \ncalQ$, satisfying (\ref{gen_vavg}), we first describe the set of transitions as follows:
\begin{itemize}
   \item $l = 4k -3$: This subset of transitions takes place between the states of the Markov chain in ${\rm r}_1$, corresponding to the time when the system is empty. In particular, a transition from this set of transitions occurs when a new energy packet is harvested by the transmitter. Clearly, since harvesting a new energy packet does not impact the value of $\Delta_1(t)$, we observe that the first element in the updated value of the age vector $\nbx \nbA_l$ (as a consequence of this transition) is $x_0$, i.e., this transition does not induce any change in the source 1's AoI at the destination. Further, since the server is idle in the states of ${\rm r}_1$, the second component of $\nbx(t)$ (quantifying the age of the source 1's packet in service, if any) becomes irrelevant for such set of states. Note that whenever a component of $\nbx(t)$ is/becomes irrelevant after the occurrence of some transition $l$, its value in the updated age vector $\nbx \nbA_l$ can be set arbitrarily (except for $l = 4k - 1$, as will be clear shortly). Following the convention \cite{yates2018age}, we set the value corresponding to such irrelevant components in the updated age value to 0, and thus we observe that the second component of $\nbx \nbA_{4k-3}$ is 0.
   \item $l = 4k - 2$: A transition from this subset of transitions occurs when there is a new arriving update packet of source 1 at the transmitter node. Since the age of this new arriving update packet at the transmitter is 0 and it does not have any impact on $\Delta_1(t)$, we note that the updated age vector $\nbx \nbA_{4k - 2}$ is set to be $[x_0,0]$.
    \item $l = 4k - 1$: A transition from this subset of transitions occurs when any of the sources other than source 1 generates a new update packet at the transmitter node. We note that the first component of $\nbx \nbA_{4k - 1}$ is $x_0$ since this transition does not have any impact on $\Delta_1(t)$. Further, to ensure that the value of $\Delta_1(t)$ does not change when this new arriving update packet is received by the destination, we set the second component of $\nbx \nbA_{4k - 1}$ to $x_0$, i.e., the value of the source 1's AoI at the arrival instant of this new update packet.
  \item $l = 4k$: This subset of transitions occurs when the update packet in service is delivered to the destination. When the update packet received at the destination belongs to source 1, the AoI of source 1 is reset to its age; otherwise, the AoI of source 1 does not change. Note that the latter case is achieved by setting the second component of $\nbx \nbA_{4k - 1}$ to $x_0$. In addition, since the system becomes empty after the occurrence of this transition, the second component of the age vector $\nbx(t)$ becomes irrelevant, and thus its corresponding value in the updated age vector $\nbx \nbA_{4k}$ is 0.
  \end{itemize}
\begin{table*}
\centering
{\caption{Transitions of the LCFS-WP queueing discipline in Fig. \ref{f:WP_MC} $(2 \leq k \leq B)$.} 
\label{table:WP}
\scalebox{.8}
{ \begin{tabular}{ |c |c|c|c|c|c|c|c|}
\hline
 $l$   & $q_l\rightarrow q'_l$  & $\lambda^{(l)}$ & $\nbx \nbA_l$ & $\nbA_l$ & $\hat{\nbA}_l$ & $\bar{\nbv}_{q_l} \nbA_l$ & $\bar{\pi}_{q_l} {\bf 1} \hat{\nbA}_l$\\ \hline
1& 1 $\rightarrow$ 2& $\eta$&$[x_0,0]$&
$\begin{bmatrix}
1 & 0\\ 0 & 0\\
\end{bmatrix}$
& $\begin{bmatrix}
0 & 0\\ 0 & 1\\
\end{bmatrix}$ &$[\bar{v}_{10},0]$ &$[0,\bar{\pi}_1]$ \\ \hline
2& 2 $\rightarrow$ 3&$\lambda_1$&$[x_0,0]$& $\begin{bmatrix}
1 & 0\\ 0 & 0\\
\end{bmatrix}$& $\begin{bmatrix}
0 & 0\\ 0 & 1\\
\end{bmatrix}$&$[\bar{v}_{20},0]$ &$[0,\bar{\pi}_2]$ \\ \hline
3& 2 $\rightarrow$ 3&$\lambda_{-1}$&$[x_0,x_0]$& $\begin{bmatrix}
1 & 1\\ 0 & 0\\
\end{bmatrix}$& $\begin{bmatrix}
0 & 0\\ 0 & 0\\
\end{bmatrix}$&$[\bar{v}_{20},\bar{v}_{20}]$ &$[0,0]$ \\ \hline
4& 3 $\rightarrow$ 1&$\mu$&$[x_1,0]$ & $\begin{bmatrix}
0 & 0\\ 1 & 0\\
\end{bmatrix}$&$\begin{bmatrix}
0 & 0\\ 0 & 1\\
\end{bmatrix}$ &$[\bar{v}_{31},0]$ &$[0,\bar{\pi}_3]$ \\ \hline
$4k-3$& $2k-2 \rightarrow 2k$& $\eta$&$[x_0,0]$& $\begin{bmatrix}
1 & 0\\ 0 & 0\\
\end{bmatrix}$& $\begin{bmatrix}
0 & 0\\ 0 & 1\\
\end{bmatrix}$&$[\bar{v}_{2k-2,0},0]$ &$[0,\bar{\pi}_{2k-2}]$ \\ \hline
$4k-2$& $2k \rightarrow 2k+1$&$\lambda_1$& $[x_0,0]$&$\begin{bmatrix}
1 & 0\\ 0 & 0\\
\end{bmatrix}$ &$\begin{bmatrix}
0 & 0\\ 0 & 1\\
\end{bmatrix}$ &$[\bar{v}_{2k,0},0]$ &$[0,\bar{\pi}_{2k}]$ \\ \hline
$4k-1$& $2k \rightarrow 2k+1$&$\lambda_{-1}$& $[x_0,x_0]$&$\begin{bmatrix}
1 & 1\\ 0 & 0\\
\end{bmatrix}$ &$\begin{bmatrix}
0 & 0\\ 0 & 0\\
\end{bmatrix}$ &$[\bar{v}_{2k,0},\bar{v}_{2k,0}]$ &$[0,0]$ \\ \hline
$4k$& $2k+1 \rightarrow 2k-2$&$\mu$&$[x_1,0]$ &$\begin{bmatrix}
0 & 0\\ 1 & 0\\
\end{bmatrix}$ &$\begin{bmatrix}
0 & 0\\ 0 & 1\\
\end{bmatrix}$ & $[\bar{v}_{2k+1,1},0]$&$[0,\bar{\pi}_{2k+1}]$ \\ \hline
\end{tabular}}} 
\end{table*} 

Now, in order to obtain $\bar{\nbv}_q$ satisfying (\ref{gen_vavg}), the steady state  probabilities $\{\bar{\pi}_q\}$ and the vector $\bar{\nbv}_{q_l} \nbA_l$ (associated with each transition $l$ in $\ncalL$) need to be computed. The calculations of  $\bar{\nbv}_{q_l} \nbA_l$, $l \in \ncalL$, are listed in Table \ref{table:WP}, and $\{\bar{\pi}_q\}$ are given by the following proposition.
\begin{prop}\label{prop1}
The steady state probabilities $\{\bar{\pi}_q\}$ can be expressed as
\begin{align}\label{prop1_1}
    \bar{\pi}_{2k}= \left(\frac{\beta}{\rho}\right)^{k} \bar{\pi}_1,
\end{align}
\begin{align}\label{prop1_2}
    \bar{\pi}_{2k+1}= \rho \left(\frac{\beta}{\rho}\right)^{k} \bar{\pi}_1,
\end{align}
where $1 \leq k \leq B$ and $\bar{\pi}_1$ is given by
\begin{align}\label{prop1_3}
    \bar{\pi}_{1}= \begin{cases}
    \dfrac{1}{1 + B (1 + \rho)}, & \;{\rm if}\; \rho = \beta,\\
    \dfrac{\rho^{B}\left(\beta - \rho\right)}{\rho^{B}\left(\beta - \rho\right) + \beta \left(1 + \rho\right) \left(\beta^{B} - \rho^{B}\right)}, & \;{\rm otherwise}.
    \end{cases}
\end{align}
\end{prop}
\begin{IEEEproof}
The expressions in (\ref{prop1_1})-(\ref{prop1_3}) follow from solving the set of equations in (\ref{gen_steady}). A detailed proof can be found in Appendix A of \cite{abdelmagid_2021a}.
\end{IEEEproof}

Having the steady state probabilities $\{\bar{\pi}_q\}$ in Proposition \ref{prop1} and the set of transitions $\ncalL$ in Table \ref{table:WP}, we are now ready to derive $\bar{\nbv}_q$ satisfying (\ref{gen_vavg}) as well as to characterize the average value of $\Delta_1(t)$ in the following theorem.
\begin{theorem}\label{theorem:Avg_WP}
Under the LCFS-WP queueing discipline, there exists a non-negative limit $\bar{\nbv}_q, \forall q \in 
\ncalQ$, satisfying (\ref{gen_vavg}) and the average AoI of source 1 is given by
\begin{align}\label{theorem:Avg_WP_1}
\overset{{\rm WP}}{\Delta}_{1,1} = \dfrac{1+\rho}{\mu \rho_1} + \dfrac{\sum_{q \in {\rm r}_2}{\bar{\pi}_q}}{\mu} + \dfrac{\bar{\pi}_1}{c_0 \mu \rho_{-1}} + \sum_{j=1}^{B}{\dfrac{\bar{\pi}_{2j}\left(\mu \rho_{-1}\right)^{j-1}}{\prod_{h=0}^{j}{c_{2h}}}} + \sum_{j=0}^{B-1}{\dfrac{\bar{\pi}_{2j+3}\left(\mu \rho_{-1}\right)^{j-1}}{\prod_{h=0}^{j}{c_{2h}}}},
\end{align}
where the set $\{c_0,c_2,\cdots,c_{2B}\}$ is defined as
\begin{align}\label{eq:c_2h}
c_{2h} = 
\begin{cases}
\lambda,\;\; & h = B,\\
\eta \left(1 - \dfrac{\lambda_{-1}}{c_{2h+2}}\right) + \lambda ,\;\; & 1 \leq h \leq B - 1,\\
\eta\left(\dfrac{1}{\lambda_{-1}} - \dfrac{1}{c_2}\right),\;\; & h = 0.
\end{cases}
\end{align}
\end{theorem}
\begin{IEEEproof}
See Appendix \ref{app:theorem:Avg_WP}.
\end{IEEEproof}

Note that the average AoI performance for source $i \in \{2,3,\cdots,N\}$ can be obtained directly using (\ref{theorem:Avg_WP_1}) by replacing $\lambda_1$ with $\lambda_{i}$ (which results in replacing $\{\lambda_{-1},\rho_1,\rho_{-1}\}$ with $\{\lambda_{-i},\rho_i,\rho_{-i}\}$ as well). This argument applies to all the results derived in this paper for source 1. 
\begin{cor}\label{cor:Avg_WP_single}
For the single source case where $\rho_{-1} = 0$ and $\rho = \rho_1$, $\overset{{\rm WP}}{\Delta}_{1,1}$ in (\ref{theorem:Avg_WP_1}) reduces to 
\begin{align}\label{cor:Avg_WP_single_1}
\overset{{\rm WP}}{\Delta}_{1,1} = 
\begin{cases}
\dfrac{2B\rho^2 + 2\left(1 + B\right)\rho + B + 2}{\mu\big[B\rho^2 + (1 + B)\rho\big]},\;\; & {\rm if}\; \rho = \beta,\\
\dfrac{\beta^{B+2}\left(2\rho^2+2\rho+1\right) - \rho^{B+2} \left(2\beta^2 + 2\beta + 1\right)}{\mu\big[\beta^{B+2}\left(\rho^2+\rho\right) - \rho^{B+2} \left(\beta^2+\beta\right)\big]},\;\; & {\rm otherwise}.
\end{cases}
\end{align}

Note that the expression of $\overset{{\rm WP}}\Delta_{1,1}$ in (\ref{cor:Avg_WP_single_1}) is identical to the average AoI expression derived in \cite[Theorem~3]{farazi2018average} under the LCFS-WP queueing discipline (for the case of having an EH-powered transmitter with a single source).
\end{cor}
\begin{IEEEproof}
We note from (\ref{eq:c_2h}) that when $\rho_{-1} = 0$, we have $c_{2h} = \eta + \lambda, 1 \leq h \leq B-1,$ and $c_0 = \infty$. Thus, $\overset{{\rm WP}}{\Delta}_{1,1}$ in (\ref{theorem:Avg_WP_1}) reduces to: $\overset{{\rm WP}}{\Delta}_{1,1} =  \dfrac{1+\rho}{\mu \rho_1} + \dfrac{\sum_{q \in {\rm r}_2}{\bar{\pi}_q}}{\mu} + \dfrac{\bar{\pi}_1 + \bar{\pi}_3}{\eta}$. The final expression in (\ref{cor:Avg_WP_single_1}) can be obtained by substituting $\{\bar{\pi}_q\}$ from Proposition \ref{prop1}, followed by some algebraic simplifications.
\end{IEEEproof}
\begin{cor}\label{cor:Avg_WP_noEH}
When $\beta \rightarrow \infty$, $\overset{{\rm WP}}{\Delta}_{1,1}$ in (\ref{theorem:Avg_WP_1}) reduces to 
\begin{align}\label{cor:Avg_WP_noEH_1}
\underset{\beta \rightarrow \infty}{\rm lim} \overset{{\rm WP}}{\Delta}_{1,1} = \frac{1 + \rho}{\mu \rho_1} + \frac{\rho}{\mu\left(1 + \rho\right)}.
\end{align}

Note that the expression in (\ref{cor:Avg_WP_noEH_1}) is identical to the average AoI expression in the case where a non-EH transmitter with multiple sources employs the LCFS-WP queueing discipline. Further, by setting $\rho_1$ in (\ref{cor:Avg_WP_noEH_1}) to $\rho$, we obtain $\underset{\beta \rightarrow \infty}{\rm lim} \overset{{\rm WP}}{\Delta}_{1,1} = \dfrac{2\rho^2+2\rho+1}{\mu\left(\rho^2+\rho\right)}$, which is the average AoI expression derived in \cite{costa2016age} for the M/M/1/1 case (where a non-EH transmitter with single source employing the LCFS-WP queueing discipline was considered).
\end{cor}
\begin{IEEEproof}
The result follows from noting that: $\underset{\beta \rightarrow \infty}{\rm lim} \dfrac{\bar{\pi}_1}{c_0 \mu \rho_{-1}} = \underset{\beta \rightarrow \infty}{\rm lim} \sum_{j=1}^{B}{\dfrac{\bar{\pi}_{2j}\left(\mu \rho_{-1}\right)^{j-1}}{\prod_{h=0}^{j}{c_{2h}}}} = \underset{\beta \rightarrow \infty}{\rm lim} \sum_{j=0}^{B-1}{\dfrac{\bar{\pi}_{2j+3}\left(\mu \rho_{-1}\right)^{j-1}}{\prod_{h=0}^{j}{c_{2h}}}} = 0$ and $\underset{\beta \rightarrow \infty}{\rm lim}\dfrac{\sum_{q \in {\rm r}_2}{\bar{\pi}_q}}{\mu} = \dfrac{\rho}{\mu\left(1 + \rho\right)}$.
\end{IEEEproof}
\begin{figure}[t!]
\centering
\includegraphics[width=0.6\columnwidth]{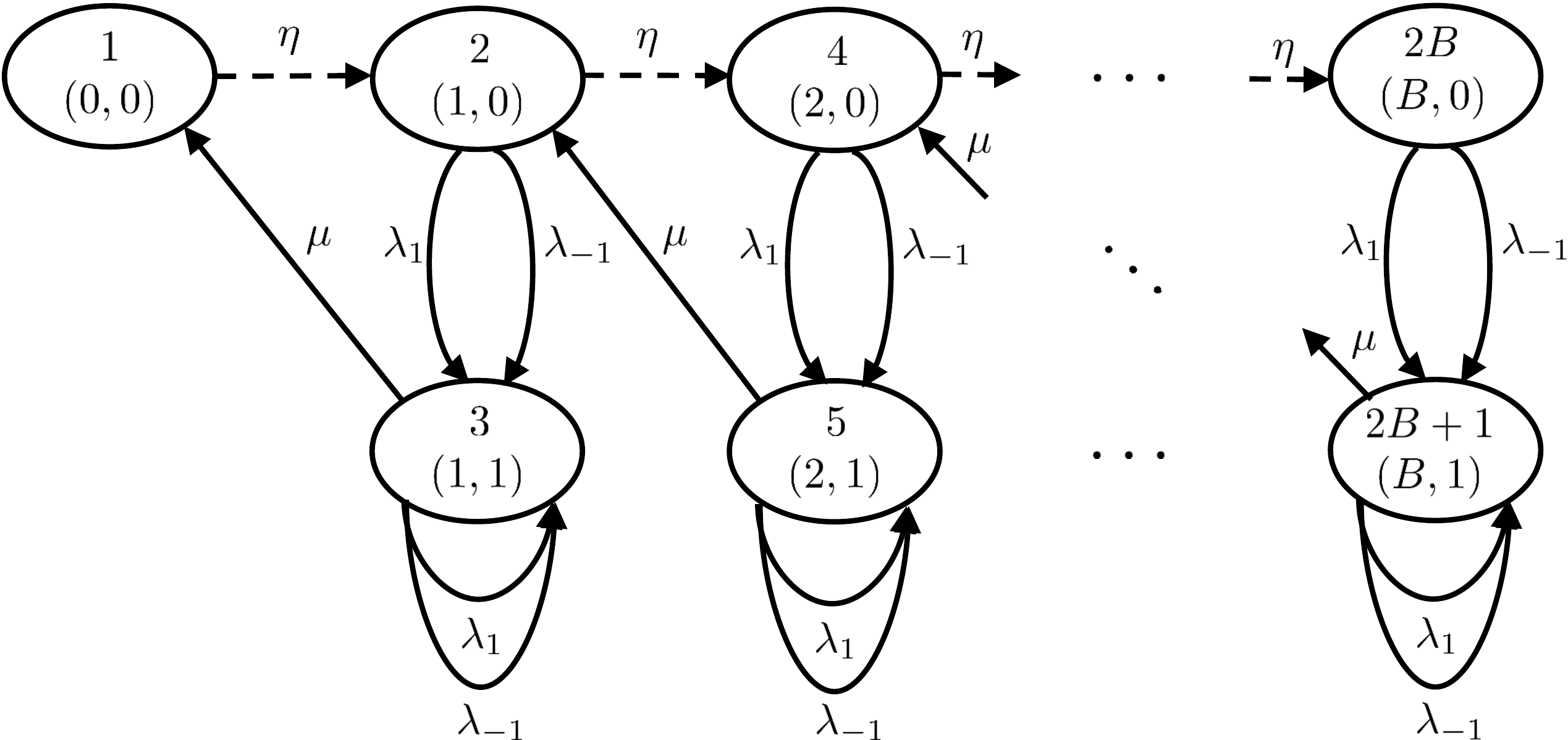}
\caption{The Markov chain modeling the discrete state in the LCFS-PS queueing discipline.}
\label{f:PS_MC}
\end{figure}
\subsection{LCFS-PS Queueing Discipline}\label{sub:Avg_PS}
Fig. \ref{f:PS_MC} depicts the Markov chain representing the discrete state of the system under the LCFS-PS queueing discipline, where the structure of $\ncalQ$ is similar to the one associated with the LCFS-WP queueing discipline. Further, the set of transitions in the LCFS-PS queueing discipline can be constructed using Tables \ref{table:WP} and \ref{table:PS}. The subset of transitions in Table \ref{table:PS} refers to the event of having a new arriving update packet at the transmitter node while its server is serving another update packet. According to the mechanism of the LCFS-PS queueing discipline, the status update that is currently being served will be discarded, and the new arrival will enter service upon its arrival. From (\ref{gen_steady}), we note that the self-transitions do not impact the values of the steady state probabilities $\{\bar{\pi}_q\}$, and hence $\{\bar{\pi}_q\}$ in this case can be obtained using Proposition \ref{prop1}. That said, the average value of $\Delta_1(t)$ is provided in the next theorem.
\begin{table*}
\centering
{\caption{Transitions of the LCFS-PS queueing discipline in Fig. \ref{f:PS_MC} $(2 \leq k \leq B).$} 
\label{table:PS}
\scalebox{.8}
{ \begin{tabular}{ |c |c|c|c|c|c|c|c|}
\hline
 $l$   & $q_l\rightarrow q'_l$  & $\lambda^{(l)}$ & $\nbx \nbA_l$ & $\nbA_l$ & $\hat{\nbA}_l$ & $\bar{\nbv}_{q_l} \nbA_l$ & $\bar{\pi}_{q_l} {\bf 1} \hat{\nbA}_l$\\ \hline
 $4B+2k-3$&$2k-1 \rightarrow 2k-1$ &$\lambda_1$ &$[x_0,0]$ &$\begin{bmatrix}1 &0 \\0 & 0\\ \end{bmatrix}$ & $\begin{bmatrix}0 &0 \\0 & 1\\ \end{bmatrix}$&$[\bar{v}_{2k-1,0},0]$ &$[0,\bar{\pi}_{2k-1}]$ \\ \hline
  $4B+2k-2$&$2k-1 \rightarrow 2k-1$ &$\lambda_{-1}$ &$[x_0,x_0]$ &$\begin{bmatrix}1 &1 \\0 & 0\\ \end{bmatrix}$ & $\begin{bmatrix}0 &0 \\0 & 0\\ \end{bmatrix}$&$[\bar{v}_{2k-1,0},\bar{v}_{2k-1,0}]$ &$[0,0]$ \\ \hline
\end{tabular}}} 
\end{table*} 
\begin{theorem}\label{theorem:Avg_PS}
Under the LCFS-PS queueing discipline, there exists a non-negative limit $\bar{\nbv}_q, \forall q \in 
\ncalQ$, satisfying (\ref{gen_vavg}) and the average AoI of source 1 is given by
\begin{align}\label{theorem:Avg_PS_1}
\overset{{\rm PS} }{\Delta}_{1,1} = \dfrac{1+\rho}{\mu \rho_1} + \dfrac{\bar{\pi}_1}{c_0 \mu \rho_{-1}} + \sum_{j=1}^{B}{\dfrac{\bar{\pi}_{2j}\left(\mu \rho_{-1}\right)^{j-1}}{\prod_{h=0}^{j}{c_{2h}}}} + \dfrac{1 + \rho_{-1}}{1 + \rho}\sum_{j=0}^{B-1}{\dfrac{\bar{\pi}_{2j+3}\left(\mu \rho_{-1}\right)^{j-1}}{\prod_{h=0}^{j}{c_{2h}}}},
\end{align}
where the set $\{c_0,c_2,\cdots,c_{2B}\}$ is defined as in (\ref{eq:c_2h}).
\end{theorem}
\begin{IEEEproof}
See Appendix \ref{app:theorem:Avg_PS}.
\end{IEEEproof}
\begin{cor}\label{rem:Avg_PS_single}
For the single source case where $\rho_{-1} = 0$ and $\rho = \rho_1$, $\overset{{\rm PS}}{\Delta}_{1,1}$ in (\ref{theorem:Avg_PS_1}) reduces to 
\begin{align}\label{rem:Avg_PS_single_1}
\overset{{\rm PS}}{\Delta}_{1,1} = 
\begin{cases}
\dfrac{B\rho^3 + \left(3B + 1\right)\rho^2 + \left(3B + 4\right) \rho+ B + 2}{\mu \rho \left(1 + \rho\right) \left(\rho B + B + 1\right)},\;\; & {\rm if}\; \rho = \beta,\\
\dfrac{\beta^{B+2}\left(1 + \rho\right)^{3} - \rho^{B+2} \big[\left(\beta^2 + \beta\right) \left(\rho + 2\right) + 1 + \rho \big]}{\mu \left(1 + \rho\right)\big[\beta^{B+2}\left(\rho^2+\rho\right) - \rho^{B+2} \left(\beta^2+\beta\right)\big]},\;\; & {\rm otherwise}.
\end{cases}
\end{align}

Note that the expression of $\overset{{\rm PS}}\Delta_{1,1}$ in (\ref{rem:Avg_PS_single_1}) is identical to the average AoI expression derived in \cite[Corollary~3]{abdelmagid_2021a} under the LCFS-PS queueing discipline (for the case of having an EH-powered transmitter with a single source).
\end{cor}
\begin{cor}\label{rem:Avg_PS_noEH}
When $\beta \rightarrow \infty$, $\overset{{\rm PS}}{\Delta}_{1,1}$ in (\ref{theorem:Avg_PS_1}) reduces to 
\begin{align}\label{rem:Avg_PS_noEH_1}
\underset{\beta \rightarrow \infty}{\rm lim} \overset{{\rm PS}}{\Delta}_{1,1} = \frac{1 + \rho}{\mu \rho_1}.
\end{align}

Note that the expression in (\ref{rem:Avg_PS_noEH_1}) is identical to the average AoI expression derived in \cite[Theorem~2(a)]{yates2018age} for the case where a non-EH transmitter with multiple sources employs the LCFS-PS queueing discipline. 
\end{cor}
\begin{remark}\label{rem:Avg_comp_WP,PS}
Note that from Theorems \ref{theorem:Avg_WP} and \ref{theorem:Avg_PS}, we have
\begin{align}\label{rem:Avg_comp_1}
\overset{{\rm WP}}{\Delta}_{1,1} -  \overset{{\rm PS}}{\Delta}_{1,1} =     \dfrac{\sum_{q \in {\rm r}_2}{\bar{\pi}_q}}{\mu} + \frac{\rho_1}{1 + \rho} \sum_{j=0}^{B-1}{\dfrac{\bar{\pi}_{2j+3}\left(\mu \rho_{-1}\right)^{j-1}}{\prod_{h=0}^{j}{c_{2h}}}}.
\end{align}

Since the set $\{c_{0},c_2,\cdots,c_{2B}\}$ contains positive real numbers, we observe from (\ref{rem:Avg_comp_1}) that $\overset{{\rm WP}}{\Delta}_{1,1} -  \overset{{\rm PS}}{\Delta}_{1,1} \geq 0$ for any choice of values of the system parameters. This, in turn, indicates the superiority of the LCFS-PS queueing discipline over LCFS-WP in terms of the achievable average AoI at the destination node.
\end{remark}
\subsection{LCFS-SA Queueing Discipline}\label{sub:Avg_SA}
\begin{figure}[t!]
\centering
\includegraphics[width=0.7\columnwidth]{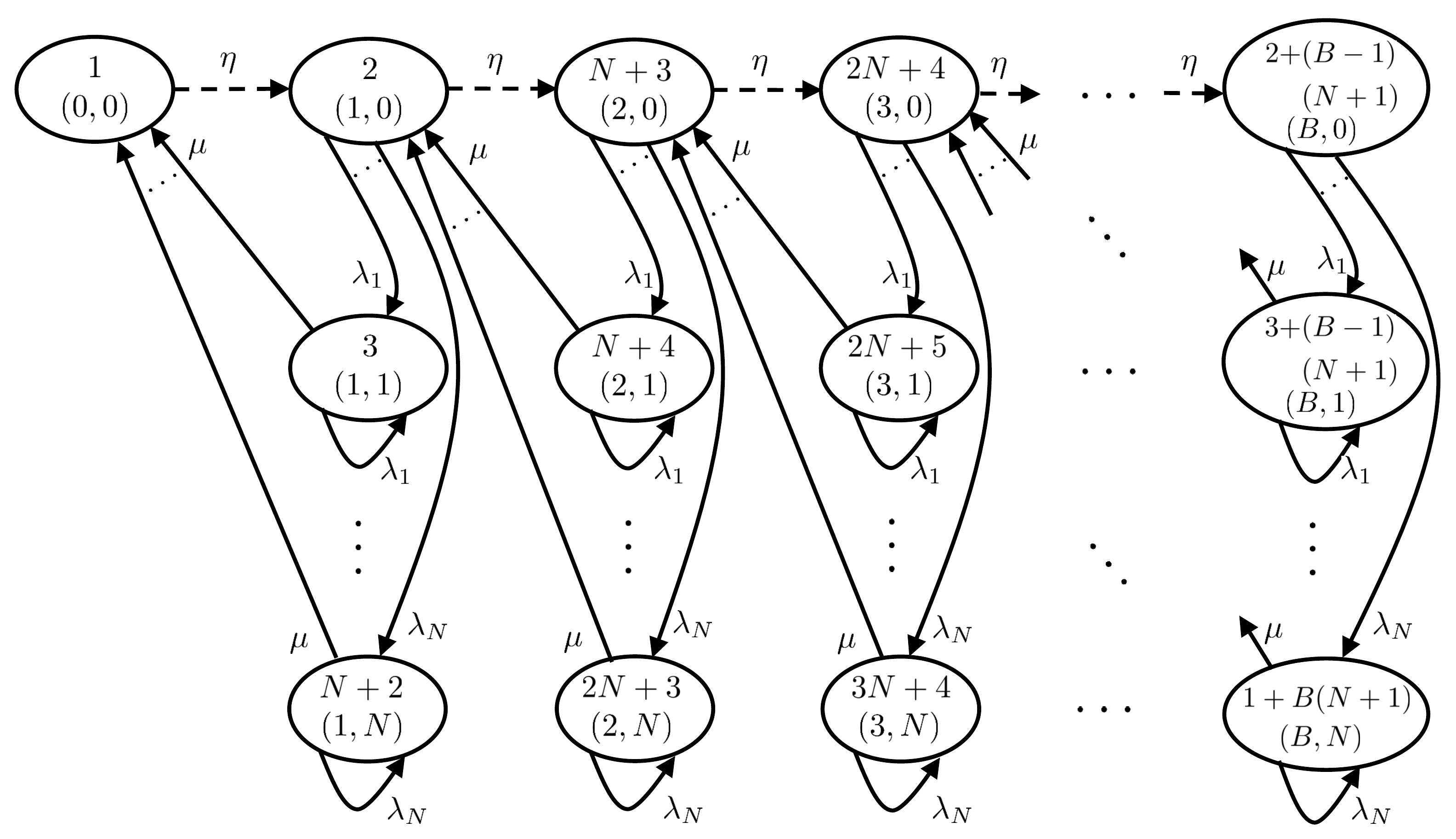}
\caption{The Markov chain modeling the discrete state in the LCFS-SA queueing discipline.}
\label{f:SA_MC}
\end{figure}
\begin{table*}
\centering
{\caption{Transitions of the LCFS-SA queueing discipline in Fig. \ref{f:SA_MC} ($2 \leq i \leq N$,\; $2 \leq k \leq B$).} 
\label{table:SA}
\scalebox{.72}
{ \begin{tabular}{ |c |c|c|c|c|c|c|c|}
\hline
 $l$   & $q_l\rightarrow q'_l$  & $\lambda^{(l)}$ & $\nbx \nbA_l$ & $\nbA_l$ & $\hat{\nbA}_l$ & $\bar{\nbv}_{q_l} \nbA_l$ & $\bar{\pi}_{q_l} {\bf 1} \hat{\nbA}_l$\\ \hline
1& 1 $\rightarrow$ 2& $\eta$&$[x_0,0]$&
$\begin{bmatrix}
1 & 0\\ 0 & 0\\
\end{bmatrix}$
& $\begin{bmatrix}
0 & 0\\ 0 & 1\\
\end{bmatrix}$ &$[\bar{v}_{10},0]$ &$[0,\bar{\pi}_1]$ \\ \hline
2& 2 $\rightarrow$ 3&$\lambda_1$&$[x_0,0]$& $\begin{bmatrix}
1 & 0\\ 0 & 0\\
\end{bmatrix}$& $\begin{bmatrix}
0 & 0\\ 0 & 1\\
\end{bmatrix}$&$[\bar{v}_{20},0]$ &$[0,\bar{\pi}_2]$   \\ \hline
$1+i$& 2 $\rightarrow$ $2+i$&$\lambda_i$&$[x_0,0]$& $\begin{bmatrix}
1 & 0\\ 0 & 0\\
\end{bmatrix}$& $\begin{bmatrix}
0 & 0\\ 0 & 1\\
\end{bmatrix}$&$[\bar{v}_{20},0]$ &$[0,\bar{\pi}_2]$\\ \hline
$2+N$& 3 $\rightarrow$ 3&$\lambda_1$&$[x_0,0]$& $\begin{bmatrix}
1 & 0\\ 0 & 0\\
\end{bmatrix}$& $\begin{bmatrix}
0 & 0\\ 0 & 1\\
\end{bmatrix}$&$[\bar{v}_{30},0]$ &$[0,\bar{\pi}_3]$   \\ \hline
$1+N+i$& $2+i$ $\rightarrow$ $2+i$&$\lambda_i$&$[x_0,0]$& $\begin{bmatrix}
1 & 0\\ 0 & 0\\
\end{bmatrix}$& $\begin{bmatrix}
0 & 0\\ 0 & 1\\
\end{bmatrix}$&$[\bar{v}_{2+i,0},0]$ &$[0,\bar{\pi}_{2+i}]$\\ \hline
$2\left(N+1\right)$& 3 $\rightarrow$ 1&$\mu$&$[x_1,0]$ & $\begin{bmatrix}
0 & 0\\ 1 & 0\\
\end{bmatrix}$&$\begin{bmatrix}
0 & 0\\ 0 & 1\\
\end{bmatrix}$ &$[\bar{v}_{31},0]$ &$[0,\bar{\pi}_3]$ \\ \hline
$1+2N+i$& $2+i$ $\rightarrow$ 1&$\mu$&$[x_0,0]$ & $\begin{bmatrix}
1 & 0\\ 0 & 0\\
\end{bmatrix}$&$\begin{bmatrix}
0 & 0\\ 0 & 1\\
\end{bmatrix}$ &$[\bar{v}_{2+i,0},0]$ &$[0,\bar{\pi}_{2+i}]$\\ \hline
$\left(3N+1\right)k - 3N$& $2 + (N + 1)(k - 2)$ $\rightarrow$ $2 + (N + 1)(k - 1)$& $\eta$&$[x_0,0]$&
$\begin{bmatrix}
1 & 0\\ 0 & 0\\
\end{bmatrix}$
& $\begin{bmatrix}
0 & 0\\ 0 & 1\\
\end{bmatrix}$ &$[\bar{v}_{2 + (N + 1)(k - 2),0},0]$ &$[0,\bar{\pi}_{2 + (N + 1)(k - 2)}]$ \\ \hline
$\left(3N+1\right)k - 3N + 1$& $2 + (N + 1)(k - 2)$ $\rightarrow$ $3 + (N + 1)(k - 2)$&$\lambda_1$&$[x_0,0]$& $\begin{bmatrix}
1 & 0\\ 0 & 0\\
\end{bmatrix}$& $\begin{bmatrix}
0 & 0\\ 0 & 1\\
\end{bmatrix}$&$[\bar{v}_{2 + (N + 1)(k - 2),0},0]$ &$[0,\bar{\pi}_{2 + (N + 1)(k - 2)}]$   \\ \hline
$\left(3N+1\right)k - 3N + i$& $2 + (N + 1)(k - 2)$ $\rightarrow$ $2 + i + (N + 1)(k - 2)$&$\lambda_i$&$[x_0,0]$& $\begin{bmatrix}
1 & 0\\ 0 & 0\\
\end{bmatrix}$& $\begin{bmatrix}
0 & 0\\ 0 & 1\\
\end{bmatrix}$&$[\bar{v}_{2 + (N + 1)(k - 2),0},0]$ &$[0,\bar{\pi}_{2 + (N + 1)(k - 2)}]$\\ \hline
$\left(3N+1\right)k - 2N + 1$& $3 + (N + 1)(k - 2)$ $\rightarrow$ $3 + (N + 1)(k - 2)$&$\lambda_1$&$[x_0,0]$& $\begin{bmatrix}
1 & 0\\ 0 & 0\\
\end{bmatrix}$& $\begin{bmatrix}
0 & 0\\ 0 & 1\\
\end{bmatrix}$&$[\bar{v}_{3 + (N + 1)(k - 2),0},0]$ &$[0,\bar{\pi}_{3 + (N + 1)(k - 2)}]$   \\ \hline
$\left(3N+1\right)k - 2N + i$& $2 + i + (N + 1)(k - 2)$ $\rightarrow$ $2 + i + (N + 1)(k - 2)$&$\lambda_i$&$[x_0,0]$& $\begin{bmatrix}
1 & 0\\ 0 & 0\\
\end{bmatrix}$& $\begin{bmatrix}
0 & 0\\ 0 & 1\\
\end{bmatrix}$&$[\bar{v}_{2 + i + (N + 1)(k - 2),0},0]$ &$[0,\bar{\pi}_{2 + i + (N + 1)(k - 2)}]$\\ \hline
$(3N+1)k - N + 1$& $3 + (N + 1)(k - 1)$ $\rightarrow$ $2 + (N + 1)(k - 2)$ &$\mu$&$[x_1,0]$ & $\begin{bmatrix}
0 & 0\\ 1 & 0\\
\end{bmatrix}$&$\begin{bmatrix}
0 & 0\\ 0 & 1\\
\end{bmatrix}$ &$[\bar{v}_{3 + (N + 1)(k - 1),1},0]$ &$[0,\bar{\pi}_{3 + (N + 1)(k - 1)}]$ \\ \hline
$(3N+1)k - N + i$& $2 + i + (N + 1)(k - 1)$ $\rightarrow$ $2 + (N + 1)(k - 2)$&$\mu$&$[x_0,0]$ & $\begin{bmatrix}
1 & 0\\ 0 & 0\\
\end{bmatrix}$&$\begin{bmatrix}
0 & 0\\ 0 & 1\\
\end{bmatrix}$ &$[\bar{v}_{2 + i + (N + 1)(k - 1),0},0]$ &$[0,\bar{\pi}_{2 + i + (N + 1)(k - 1)}]$\\ \hline
\end{tabular}}} 
\end{table*} 
Under the LCFS-SA queueing discipline, the continuous-time Markov chain modeling the discrete state of the system $q(t) \in \ncalQ$ is depicted in Fig. \ref{f:SA_MC}. Recall that according to the mechanism of the LCFS-SA queueing discipline, a new arriving update packet preempts the packet in service only if the two packets are generated from the same source. Thus, the discrete state of the system needs to not only account for the number of update packets in the system (as it was the case for the LCFS-WP and the LCFS-PS queueing disciplines) but also track the index of the source which generated the current packet in service. Because of that, we observe from Fig. \ref{f:SA_MC} that for a state $q=(e_q,u_q)$, we have $u_{q} \in \{0,1,\cdots,N\}$. In particular, $u_q=0$ indicates that the system is empty and hence the server is idle, and $u_q=i$ indicates that there is an update packet in service and the index of its generating source is $i$. Further, due to the finite capacity of the battery queue at the server, we have $e_q \in \{0,1,\cdots,B\}$. Table \ref{table:SA} presents the set of transitions $\ncalL$ and their impact on the values of both $q(t)$ and $\nbx(t)$. While the description of most transitions in Table \ref{table:SA} is similar to the description of their counterparts in Tables \ref{table:WP} and \ref{table:PS}, there are some key differences. The first difference is that the second component of the updated age vector $\nbx \nbA_l$ in Table \ref{table:SA} is set to 0 (rather than $x_0$ as in Tables \ref{table:WP} and \ref{table:PS}) when a new update packet is generated from source $i \in \{2,3,\cdots,N\}$ at the time when the system is empty (i.e., transition $l = (3N + 1)k - 3N + i$) or another packet generated from source $i$ is being served (i.e., transition $l = (3N + 1)k - 2N + i$). This is because under the LCFS-SA queueing discipline, we know the index of the source that generated the packet in service, and hence we can safely set the irrelevant second component of $\nbx \nbA_l$ to 0 when such transitions $l$ occur. The second difference, which has a similar interpretation to the first one, is that the first component of the updated age vector $\nbx \nbA_l$ in Table \ref{table:SA} is set to $x_0$ (rather than $x_1$ as in Table \ref{table:WP}) when an update packet generated from source $i \in \{2,3,\cdots,N\}$ is delivered to the destination (i.e., transition $l = (3N + 1)k - N + i$). We start the analysis by characterizing the steady state probabilities $\{\bar{\pi}_q\}$ in the following proposition.
\begin{prop}\label{prop2}
The steady state probabilities $\{\bar{\pi}_q\}$ can be expressed as
\begin{align}\label{prop2_1}
    \bar{\pi}_{2+\left(k-1\right)\left(N+1\right)}= \left(\frac{\beta}{\rho}\right)^{k} \bar{\pi}_1,
\end{align}
\begin{align}\label{prop2_2}
    \bar{\pi}_{2+i+\left(k-1\right)\left(N+1\right)}= \rho_i \left(\frac{\beta}{\rho}\right)^{k} \bar{\pi}_1,
\end{align}
where $1 \leq k \leq B$, $1 \leq i \leq N$, and $\bar{\pi}_1$ is given by (\ref{prop1_3}).
\end{prop}
\begin{IEEEproof}
The expressions in (\ref{prop2_1}) and (\ref{prop2_2}) follow directly from the solution of (\ref{gen_steady}).
\end{IEEEproof}
 Now, using Table \ref{table:SA} and Proposition \ref{prop2}, the average AoI is obtained in the following theorem.
\begin{theorem}\label{theorem:Avg_SA}
Under the LCFS-SA queueing discipline, there exists a non-negative limit $\bar{\nbv}_q, \forall q \in 
\ncalQ$, satisfying (\ref{gen_vavg}) and the average AoI of source 1 is given by
\begin{align}\label{theorem:Avg_SA_1}
\overset{{\rm SA} }{\Delta}_{1,1} = \dfrac{1+\rho}{\mu \rho_1\left(1+\rho_1\right)} +  \dfrac{\left(1+\rho\right)\sum_{q \in \ncalQ / {\rm r}_2}{\bar{\pi}_q}}{\mu \left(1+\rho_1\right)} + \dfrac{\sum_{q \in \ncalQ / {\rm r}_1}{\bar{\pi}_q}}{\mu} + \bar{v}_{10}, 
\end{align}
where $\bar{v}_{10}$ is given by  
\begin{align}\label{theorem:Avg_SA_2}
\bar{v}_{10} = \dfrac{\bar{\pi}_1}{\bar{c}_{-1} \mu  \rho_{-1}} + \sum_{j=1}^{B}{\dfrac{\bar{\pi}_{2+\left(j - 1\right)\left(N + 1\right)}\left(\mu \rho_{-1}\right)^{j-1}}{\prod_{h=0}^{j}{\bar{c}_{h-1}}}} + \sum_{j=0}^{B-1}{\dfrac{\frac{\bar{\pi}_{3+j\left(N+1\right)}}{1+\rho_1} + \sum_{m=4+j\left(N+1\right)}^{1+\left(j+1\right)\left(N+1\right)}{\bar{\pi}_m}}{\prod_{h=0}^{j}{\bar{c}_{h-1}}}\left(\mu \rho_{-1}\right)^{j-1}},
\end{align}
where the set $\{\bar{c}_{-1},\bar{c}_0,\cdots,\bar{c}_{B-1}\}$ is defined as
\begin{align}\label{eq:cbar}
\bar{c}_{h} = 
\begin{cases}
\lambda,\;\; & h = B - 1,\\
\eta \left(1 - \dfrac{\lambda_{-1}}{\bar{c}_{h+1}}\right) + \lambda ,\;\; & 0 \leq h \leq B - 2,\\
\eta\left(\dfrac{1}{\lambda_{-1}} - \dfrac{1}{\bar{c}_0}\right),\;\; & h = -1.
\end{cases}
\end{align}
\end{theorem}
\begin{IEEEproof}
See Appendix \ref{app:theorem:Avg_SA}.
\end{IEEEproof}
\begin{cor}\label{rem:Avg_SA_asym}
Note that for the single source case where $\rho_{-1} = 0$ and $\rho = \rho_1$, the LCFS-PS and LCFS-SA queueing disciplines are similar. Because of that, $\overset{{\rm SA}}{\Delta}_{1,1}$ in (\ref{theorem:Avg_SA_1}) reduces to (\ref{rem:Avg_PS_single_1}) when $\rho=\rho_1$. Further, when $\beta \rightarrow \infty$, $\overset{{\rm SA}}{\Delta}_{1,1}$ in (\ref{theorem:Avg_SA_1}) reduces to
\begin{align}\label{rem:Avg_SA_asym_1}
\underset{\beta \rightarrow \infty}{\rm lim} \overset{{\rm SA}}{\Delta}_{1,1} = \frac{1 + \rho}{\mu \rho_1} + \frac{\rho_{-1}}{\mu\left(1+\rho\right)\left(1+\rho_1\right)},
\end{align}
which indicates that $\underset{\beta \rightarrow \infty}{\rm lim} \overset{{\rm PS}}{\Delta}_{1,1} \leq \underset{\beta \rightarrow \infty}{\rm lim} \overset{{\rm SA}}{\Delta}_{1,1} \leq \underset{\beta \rightarrow \infty}{\rm lim} \overset{{\rm WP}}{\Delta}_{1,1}$.
\end{cor}
\begin{remark}\label{rem:Avg_comp_SA,WP,PS}
Note that from Theorems \ref{theorem:Avg_WP}, \ref{theorem:Avg_PS} and \ref{theorem:Avg_SA}, we have
\begin{align}\label{rem:Avg_comp_WP_SA}
\overset{{\rm WP}}{\Delta}_{1,1} -  \overset{{\rm SA}}{\Delta}_{1,1} = \dfrac{\rho_1\left(1+\rho\right)\sum_{k=1}^{B}{(\frac{\beta}{\rho})^k}}{\mu\left(1+
\rho_1\right)\left[1 + \left(1+\rho\right)\sum_{k=1}^B{\left(\frac{\beta}{\rho}\right)^k}\right]}  + \dfrac{\bar{\pi}_1 \rho_1^2}{1+\rho_1}\sum_{j=0}^{B-1}{\dfrac{\left(\frac{\beta}{\rho}\right)^{j+1} \left(\mu \rho_{-1}\right)^{j-1}}{\prod_{h=0}^{j}{\bar{c}_{h-1}}}},
\end{align}
\begin{align}\label{rem:Avg_comp__SA_PS}
\overset{{\rm SA}}{\Delta}_{1,1} -  \overset{{\rm PS}}{\Delta}_{1,1} = \dfrac{\rho_{-1}\sum_{k=1}^{B}{(\frac{\beta}{\rho})^k}}{\mu\left(1+
\rho_1\right)\left[1 + \left(1+\rho\right)\sum_{k=1}^B{\left(\frac{\beta}{\rho}\right)^k}\right]}  + \dfrac{\bar{\pi}_1 \rho_1 \rho_{-1}}{\left(1+\rho_1\right)\left(1+\rho\right)}\sum_{j=0}^{B-1}{\dfrac{\left(\frac{\beta}{\rho}\right)^{j+1} \left(\mu \rho_{-1}\right)^{j-1}}{\prod_{h=0}^{j}{\bar{c}_{h-1}}}}.
\end{align}

Since the set $\{\bar{c}_{-1},\bar{c}_0,\cdots,\bar{c}_{B-1}\}$ contains positive real numbers, we observe from (\ref{rem:Avg_comp_WP_SA}) and (\ref{rem:Avg_comp__SA_PS}) that $\overset{{\rm PS}}{\Delta}_{1,1} \leq \overset{{\rm SA}}{\Delta}_{1,1} \leq \overset{{\rm WP}}{\Delta}_{1,1} $ for any choice of values of the system parameters. 
\end{remark}
\section{The MGF of AoI for the Queueing Disciplines Considered in this Paper}\label{sec:MGF}
This section is dedicated to the analysis of the MGF of AoI under each of the queueing disciplines considered in this paper.
\subsection{LCFS-WP Queueing Discipline}\label{sub:MGF_WP}
According to Theorem \ref{theorem:Avg_WP}, there exists a non-negative limit $\bar{\nbv}_q, \forall q \in \ncalQ$, satisfying (\ref{gen_vavg}), under the LCFS-WP queueing discipline. Thus, the MGF of AoI can be evaluated using (\ref{gen_vMGF}) as in the following theorem, where the calculations required to solve the set of equations (i.e., $\bar{\nbv}^s_{q_l} \nbA_l$ and $\bar{\pi}_{q_l}{\bf 1}\hat{\nbA}_l$, $l \in \ncalL$) in (\ref{gen_vMGF}) are listed in Table \ref{table:WP}.
\begin{theorem}\label{theorem:MGF_WP}
The MGF of AoI of source 1 for the LCFS-WP queueing discipline is given by
\begin{align}\label{theorem:MGF_WP_1}
\overset{{\rm WP}}{M}_1(\bar{s}) = \dfrac{\rho_1\left(1 + \rho - \bar{s}\right)\sum_{q \in {\rm r}_1 / \{1\}}{\bar{\pi}_q} + \bar{v}^s_{10} \rho_{1} \left(1 - \bar{s}\right)}{\left(1 - \bar{s}\right)\Big[\left(1 - \bar{s}\right) \left(\rho - \bar{s}\right) - \rho_{-1}\Big]},
\end{align}
where $\bar{s} = \frac{s}{\mu}$ and $\bar{v}^s_{10}$ is given by 
\begin{align}\label{theorem:MGF_WP_v10}
\bar{v}^s_{10} = \dfrac{\rho_1}{\rho_{-1}} \sum_{j=0}^{B-1}{\dfrac{\bar{\pi}_{2j+2}}{\prod_{h=0}^{j}{c^s_{2h}}}\left(\dfrac{\mu \rho_{-1}}{1 - \bar{s}}\right)^j},
\end{align}
where the set $\{c^s_0,c^s_2,\cdots,c^s_{2B}\}$ is defined as
\begin{align}\label{eq:cs_2h}
c^s_{2h} = 
\begin{cases}
\lambda - s,\;\; & h = B,\\
\eta + \lambda - s - \dfrac{\mu \eta \lambda_{-1}}{c^s_{2h+2}\left(\mu - s\right)} ,\;\; & 1 \leq h \leq B - 1,\\
\dfrac{\left(\mu - s\right)\left(\eta - s\right)}{\mu \lambda_{-1}} - \dfrac{\eta}{c^s_2},\;\; & h = 0.
\end{cases}
\end{align}
\end{theorem}
\begin{IEEEproof}
See Appendix \ref{app:theorem:MGF_WP}.
\end{IEEEproof}
\begin{cor}\label{cor:WP_MGF_single}
When $\rho_{-1} = 0$ (i.e., $\rho_1 = \rho$), $\overset{{\rm WP}}{M}_1(\bar{s})$ in (\ref{theorem:MGF_WP_1}) reduces to 
the following MGF of AoI derived in \cite[Theorem~1]{abdelmagid_2021a} for the case where an EH-powered transmitter with a single source employs the LCFS-WP queueing discipline
\begin{align}\label{cor:WP_MGF_single_1}
\overset{{\rm WP}}{M}_1(\bar{s}) = \dfrac{\rho \bar{\pi}_1 \Big[\bar{s}^{2} \theta - \bar{s} \theta \left(1 + \rho + \beta\right) + \beta \left(1+ \theta + \theta \rho\right)\Big]}{\left(1 - \bar{s}\right)^{2} \left(\rho - \bar{s}\right) \left(\beta - \bar{s}\right)},
\end{align}
where $\theta$ can be expressed as
\begin{align}\label{cor:WP_MGF_single_2}
\theta = 
\begin{cases}
B,\;\; &{\rm if}\; \rho = \beta,\\
\dfrac{\beta\left(\beta^{B} - \rho^{B}\right)}{\rho^{B}\left(\beta - \rho\right)},\;\; &{\rm otherwise}.
\end{cases}
\end{align}
\end{cor}
\begin{IEEEproof}
When $\rho_{-1} = 0$, we first note from (\ref{eq:cs_2h}) that we have: $c_{2h}^s = \eta + \lambda - s, 1 \leq h \leq B-1,$ and $c^s_0 = \infty$. As a result, $\bar{v}_{10}^s$ reduces to $\dfrac{\rho_1 \bar{\pi}_2}{\left(1 - \bar{s}\right)\left(\beta - \bar{s}\right)}$. The final expression in (\ref{cor:WP_MGF_single_1}) can be obtained by defining $\sum_{q \in {\rm r}_1 / \{1\}}{\bar{\pi}_q} = \theta \bar{\pi}_1$ and substituting $\bar{\pi}_2$ from Proposition \ref{prop1} as $\frac{\beta}{\rho} \bar{\pi}_1$.
\end{IEEEproof}
\begin{cor}\label{cor:WP_moments}
When $B = 2$, the first and second moments of AoI  of source 1 under the LCFS-WP queueing discipline can be respectively expressed as
%
\begin{align}\label{mom1_WP_B2}
\overset{{\rm WP}}{\Delta}_{1,1} = 
\begin{cases}
\dfrac{\rho_1^2 + 4 \rho_1 \rho^3 + 2 \rho_{-1} \rho + \rho^2 \left(4 \rho^4 + 12 \rho + 7\right)}{2\mu\rho_1\rho^2\left(3 + 2 \rho\right)},\;\; & {\rm if}\; \rho = \beta,\\
\dfrac{\sum_{n=0}^5{\beta^n \gamma_n}}{\mu\rho_1 \beta\left(\beta + \rho\right)^2\big[\rho^2 + \beta\left(1 + \rho\right)\left(\beta + \rho\right)\big]},\;\; & {\rm otherwise}.
\end{cases}
\end{align}
\begin{align*}
&\gamma_5 = \rho_1\rho + \left(1 + \rho\right)^2, \;\gamma_4 = 3\rho_1\rho^2 + 3 \rho \left(1 + \rho\right)^2, \; \gamma_3 = \rho_1^2 + \rho_1 \rho \left(3 \rho^2 - 2\right) + \rho^2 \left(1 + \rho\right)\left(5 + 3\rho \right),\\&\gamma_2= \rho_1^2\rho + \rho_1 \rho^2 \left(\rho^2 - 2\right) + \rho^3\left(1 + \rho\right)\left(5 + \rho\right), \;
\gamma_1 = \rho^4\left(3 + 2 \rho\right),\; \gamma_0 = \rho^5,
\end{align*}
\begin{align}\label{mom2_WP_B2}
\overset{{\rm WP}}{\Delta}_{1,2} = 
\begin{cases}
\dfrac{2\rho_1^3\left(1 + \rho\right) + \rho_1^2\rho\left(8\rho^3+2\rho+3\right) + 8\rho_1\rho^5 + 2\rho_{-1}\rho^2\left(6 + 13\rho\right) + \zeta_0}{2\mu^2\rho_1^2\rho^3\left(3 + 2 \rho\right)},\;\; & {\rm if}\; \rho = \beta,\\
\dfrac{2\sum_{n=0}^7{\beta^n \psi_n}}{\mu^2\rho_1^2\beta^2\left(\beta + \rho\right)^3\big[\rho^2 + \beta \left(1 + \rho\right) \left(\beta + \rho\right)\big]},\;\; & {\rm otherwise}.
\end{cases}
\end{align}
\begin{align*}
&\zeta_0 = \rho^3 \left(8\rho^3 + 36 \rho^2 + 28 \rho + 15\right), \;\psi_7 = \rho_1^2\rho + \rho_1\left(\rho^2 - 1\right) + \left(1 + \rho\right)^3, \;\psi_6 = 4\rho_1^2\rho^2 + 4\rho_1\rho\left(\rho^2 - 1\right) \\& + 4\rho\left(1 + \rho\right)^3, \; \psi_5 = \rho_1^3 + \rho_1^2\left(6\rho^3 + \rho + 2 \right) + 2\rho_1\rho\left(3\rho^3 - 6\rho - 2\right) + 3\rho^2\left(1 + \rho\right)^2\left(3 + 2\rho\right),\\&
\psi_4= 2\rho_1^3\left(1 + \rho\right) + \rho_1^2\rho\left(4\rho^3 + 2\rho + 1\right) + 2\rho_1\rho^2\left(2\rho^3 -9\rho - 4\right) + \rho^3\left(1 + \rho\right)^2\left(13 + 4\rho\right),\\& \psi_3 = \rho_1^3\rho\left(2 + \rho\right) + \rho_1^2 \rho^2\left(\rho^3 + \rho + 1\right) + \rho_1\rho^3\left(\rho^3 - 12\rho - 8\right) + \rho^4\left(\rho^3 + 12\rho^2 + 24 \rho + 13\right), \\&\psi_2 = 2\rho_1^2\rho^3 - 4 \rho_1\rho^4\left(1 + \rho\right) + 3\rho^5\left(1 + \rho\right)\left(3 + \rho\right),\;
\psi_1=\rho^6\left(4 + 3 \rho\right) - \rho_1 \rho^6,\; \psi_0 = \rho^7.
\end{align*}
\end{cor}
\begin{IEEEproof}
The expressions in (\ref{mom1_WP_B2}) and (\ref{mom2_WP_B2}) follow from the fact that the expression of $\overset{\rm WP}{M}_1(\bar{s})$ (derived in Theorem \ref{theorem:MGF_WP}) can be used to compute the $k$-th moment of AoI of source 1 (denoted by $\overset{{\rm WP}}{\Delta}_{1,k}$) as follows
\begin{align}\label{MGF_derav}
\overset{{\rm WP}}{\Delta}_{1,k} = \frac{1}{\mu^{k}}\times\dfrac{{\rm d}^k\big[\overset{\rm WP}{M}_1(\bar{s})\big]}{{\rm d}\bar{s}^k} \Big|_{\bar{s}=0},
\end{align}
where $\frac{{\rm d}^k}{{\rm d}\bar{s}^k}$ denotes the $k$-th derivative with respect to $\bar{s}$.
\end{IEEEproof}
\subsection{LCFS-PS Queueing Discipline}\label{sub:MGF_PS}
Based on Theorem \ref{theorem:Avg_PS}, the MGF of AoI under the LCFS-PS queueing discipline is derived in the following theorem by solving the set of equations in (\ref{gen_vMGF}) using the calculations in Tables \ref{table:WP} and \ref{table:PS}.
\begin{theorem}\label{theorem:MGF_PS}
The MGF of AoI of source 1 for the LCFS-PS queueing discipline is given by
\begin{align}\label{theorem:MGF_PS_1}
\overset{{\rm PS}}{M}_1(\bar{s}) = \dfrac{\rho_1\left(1 - \bar{\pi}_1 + \bar{v}^s_{10}\right)}{\left(1 - \bar{s}\right) \left(\rho - \bar{s}\right) - \rho_{-1}},
\end{align}
where $\bar{v}^s_{10}$ is given by 
\begin{align}\label{theorem:MGF_PS_v10}
\bar{v}^s_{10} = \dfrac{\mu \rho_1}{1 + \rho - \bar{s}} \sum_{j=0}^{B-1}{\dfrac{\bar{\pi}_{2j+2} + \bar{\pi}_{2j+3}}{\prod_{h=0}^{j}{c^s_{2h}}}\left(\dfrac{\mu \rho_{-1}}{1 - \bar{s}}\right)^{j-1}},
\end{align}
where the set $\{c^s_0,c^s_2,\cdots,c^s_{2B}\}$ is defined as in (\ref{eq:cs_2h}).
\end{theorem}
\begin{IEEEproof}
See Appendix \ref{app:MGF_PS}.
\end{IEEEproof}
\begin{cor}\label{rem:PS_MGF_single}
When $\rho_{-1} = 0$ (i.e., $\rho_1 = \rho$), $\overset{{\rm PS}}{M}_1(\bar{s})$ in (\ref{theorem:MGF_PS_1}) reduces to 
the following MGF of AoI derived in \cite[Theorem~3]{abdelmagid_2021a} for the case where an EH-powered transmitter with a single source employs the LCFS-PS queueing discipline
\begin{align}\label{rem:PS_MGF_single_1}
\overset{{\rm PS}}{M}_1(\bar{s}) = \frac{\rho \left(1 + \rho\right)\bar{\pi}_1 \Big[\bar{s}^{2} \theta - \bar{s} \theta \left(1 + \rho + \beta\right) + \beta \left(1+ \theta + \theta \rho\right)\Big]}{\left(1 - \bar{s}\right) \left(\rho - \bar{s}\right) \left(1 + \rho - \bar{s}\right)\left(\beta - \bar{s}\right)},
\end{align}
where $\theta$ is given by (\ref{cor:WP_MGF_single_2}).
\end{cor}
\begin{cor}\label{cor:PS_moments}
When $B = 2$, the first and second moments of AoI  of source 1 under the LCFS-PS queueing discipline can be respectively expressed as
%
\begin{align}\label{mom1_PS_B2}
\overset{{\rm PS}}{\Delta}_{1,1} = 
\begin{cases}
\dfrac{\rho_1^2 \left(1 + \rho\right)+ 2 \rho_{-1} \rho \left(\rho^2 + \rho + 1\right) + \rho^2 \left(4\rho^3 + 14\rho^2 + 19\rho + 7\right)}{2\mu\rho_1\rho^2\left(1 + \rho\right)\left(3 + 2 \rho\right)},\;\; & {\rm if}\; \rho = \beta,\\
\dfrac{\sum_{n=0}^5{\beta^n \gamma'_n}}{\mu\rho_1 \beta\left(1 + \rho\right)\left(\beta + \rho\right)^2\big[\rho^2 + \beta\left(1 + \rho\right)\left(\beta + \rho\right)\big]},\;\; & {\rm otherwise}.
\end{cases}
\end{align}
\begin{align*}
&\gamma'_5 = \left(1 + \rho\right)^3, \;\gamma'_4 = 3\rho\left(1 + \rho\right)^3, \; \gamma'_3 = \rho_1^2 \left(1 + \rho\right) - \rho_1 \rho \left(\rho^2 + 2\rho + 2\right) + \rho^2 \left(1 + \rho\right)^2\left(5 + 3\rho \right),\\&\gamma'_2= \rho_1^2\rho \left(1 + \rho\right) - 2 \rho_1 \rho^2 \left(\rho^2 + \rho + 1\right) + \rho^3\left(1 + \rho\right)^2\left(5 + \rho\right), \;
\gamma'_1 = \rho^4\left(1 + \rho\right)\left(3 + 2 \rho\right) - \rho_1 \rho^5,\\&\gamma'_0 = \rho^5\left(1 + \rho\right),
\end{align*}
\begin{align}\label{mom2_PS_B2}
\overset{{\rm PS}}{\Delta}_{1,2} = 
\begin{cases}
\dfrac{2\rho_1^3\left(1 + \rho\right)\left(1 + 2\rho\right) + \rho_1^2\rho\left(2\rho^3+11\rho^2+8\rho+3\right) + \rho_{-1} \zeta'_1 + \zeta'_0}{2\mu^2\rho_1^2\rho^3\left(1 + \rho \right)^2\left(3 + 2 \rho\right)},\;\; & {\rm if}\; \rho = \beta,\\
\dfrac{2\sum_{n=0}^7{\beta^n \psi'_n}}{\mu^2\rho_1^2\beta^2\left(\beta + \rho\right)^3\left(1 + \rho\right)^2\big[\rho^2 + \beta \left(1 + \rho\right) \left(\beta + \rho\right)\big]},\;\; & {\rm otherwise}.
\end{cases}
\end{align}
\begin{align*}
&\zeta'_1 = 2\rho^2\left(1 + \rho\right)\left(8\rho^3+22\rho^2+19\rho+6\right),\; \zeta'_0 = \rho^3 \left(1 + \rho\right)\left(3 + 2\rho\right)\left(4\rho^3 + 8\rho^2 + 11\rho + 5\right), \\&\psi'_7 = \left(1 + \rho\right)^5 - \rho_1\left(1 + \rho\right)^3, \;\psi'_6 =4 \rho \left(1 + \rho\right)^5 - 4\rho_1\rho \left(1 + \rho\right)^3, \; \psi'_5 = \rho_1^3\left(1 + \rho\right) + \rho_1^2  \\& \left(2\rho^3 + 6\rho^2 + 5\rho + 2\right) - 4 \rho_1 \rho \left(1 + \rho\right)\left(2\rho^3 + 5\rho^2 + 4\rho + 1\right) + 3\rho^2 \left(1 + \rho\right)^4\left(3 + 2\rho\right),\\&
\psi'_4= 2\rho_1^3\left(1 + \rho\right)\left(1 + 2\rho\right) + \rho_1^2\rho\left(3\rho^3 + 9\rho^2 + 4\rho + 1\right) - 2\rho_1\rho^2\left(1 + \rho\right)\left(5\rho^3 + 14\rho^2 +13\rho + 4\right) +  \\& \rho^3\left(1 + \rho\right)^2\left(4\rho^3 + 21\rho^2 + 30\rho + 13\right),\; \psi'_3 = \rho_1^3\rho\left(1 + \rho\right)\left(2 + 3\rho\right) + \rho_1^2 \rho^2\left(5\rho^2 + 3 \rho + 1\right) - \rho_1\rho^3\\&\left(1 + \rho\right)\left(7 \rho^3 + 20 \rho^2 + 20\rho + 8\right) + \rho^4\left(1 + \rho\right)^2\left(\rho^3 + 12\rho^2 + 24 \rho + 13\right), \; \psi'_2 = \rho_1^2\rho^3 \big(-\rho^3 + 2 \rho^2 + \\&4\rho + 2\big)- 2 \rho_1\rho^4\left(1 + \rho\right) \left(\rho^3 + 4\rho^2 + 4\rho + 2\right) + 3\rho^5\left(1 + \rho\right)^3\left(3 + \rho\right),\;
\psi'_1=\rho^6\left(1 + \rho\right)^2\left(4 + 3 \rho\right) - \\&\rho_1 \rho^6\left(1 + \rho\right)\left(1 + 2\rho\right),\; \psi'_0 = \rho^7\left(1 + \rho\right)^2.
\end{align*}
\end{cor}
\subsection{LCFS-SA Queueing Discipline}
Based on Theorem \ref{theorem:Avg_SA}, the MGF of AoI under the LCFS-SA queueing discipline is derived in the following theorem by solving the set of equations in (\ref{gen_vMGF}) using the calculations in Table \ref{table:SA}.
\begin{theorem}\label{theorem:MGF_SA}
The MGF of AoI of source 1 for the LCFS-SA queueing discipline is given by
\begin{align}\label{theorem:MGF_SA_1}
\overset{{\rm SA}}{M}_1(\bar{s}) = \dfrac{\rho_1\Big[\left(1+\rho-\bar{s}\right)\sum_{q \in {\rm r}_1 \cup \;{\rm r}_2 /\{1\}}{\bar{\pi}_q} + \left(1 + \rho_1 - \bar{s}\right)\bar{v}^s_{10}\Big]}{\left(1 + \rho_1 - \bar{s}\right)\Big[\left(1 - \bar{s}\right) \left(\rho - \bar{s}\right) - \rho_{-1}\Big]},
\end{align}
where $\bar{v}^s_{10}$ is given by 
\begin{align}\label{theorem:MGF_SA_v10}
\bar{v}^s_{10} = \dfrac{\mu \rho_1}{1 + \rho_1 - \bar{s}} \sum_{j=0}^{B-1}{\dfrac{\bar{\pi}_{2+j\left(N+1\right)} + \bar{\pi}_{3+j\left(N+1\right)}}{\prod_{h=0}^{j}{\bar{c}^s_{h-1}}}\left(\dfrac{\mu \rho_{-1}}{1 - \bar{s}}\right)^{j-1}},
\end{align}
where the set $\{\bar{c}^s_{-1},\bar{c}^s_0,\cdots,\bar{c}^s_{B-1}\}$ is defined as 
\begin{align}\label{eq:csbar}
\bar{c}^s_{h} = 
\begin{cases}
\lambda - s,\;\; & h = B - 1,\\
\eta + \lambda - s - \dfrac{\mu \eta \lambda_{-1}}{\bar{c}^s_{h+1}\left(\mu - s\right)} ,\;\; & 0 \leq h \leq B - 2,\\
\dfrac{\left(\mu - s\right)\left(\eta - s\right)}{\mu \lambda_{-1}} - \dfrac{\eta}{\bar{c}^s_0},\;\; & h = -1.
\end{cases}
\end{align}
\end{theorem}
\begin{IEEEproof}
See Appendix \ref{app:theorem:MGF_SA}.
\end{IEEEproof}
\begin{cor}\label{rem:SA_MGF_single}
When $\rho_{-1} = 0$ (i.e., $\rho_1 = \rho$), $\overset{{\rm SA}}{M}_1(\bar{s})$ in (\ref{theorem:MGF_SA_1}) reduces to 
the MGF of AoI derived in \cite[Theorem~3]{abdelmagid_2021a} (given by (\ref{rem:PS_MGF_single_1})) for the case where an EH-powered transmitter with a single source employs the LCFS-PS queueing discipline.
\end{cor}
\begin{cor}\label{cor:SA_moments}
When $B =2$, the first and second moments of AoI  of source 1 under the LCFS-SA queueing discipline can be respectively expressed as
\begin{align}\label{mom1_SA_B2}
\overset{{\rm SA}}{\Delta}_{1,1} = 
\begin{cases}
\dfrac{\rho_1^3 + \rho_1^2 + \rho_1\rho^2\left(4\rho^2+10\rho+7\right) + \rho^2 \left(4\rho^2+12\rho+5\right) + 2 \rho_{-1}\rho\left[\rho_1\left(3\rho+1\right) + 2\right]}{2\mu\rho_1\rho^2\left(1 + \rho_1\right)\left(3 + 2 \rho\right)},& {\rm if}\; \rho = \beta,\\
\dfrac{\sum_{n=0}^5{\beta^n \bar{\gamma}_n}}{\mu\rho_1 \beta\left(1 + \rho_1\right)\left(\beta + \rho\right)^2\big[\rho^2 + \beta\left(1 + \rho\right)\left(\beta + \rho\right)\big]},\;\; & {\rm otherwise}.
\end{cases}
\end{align}
\begin{align*}
&\bar{\gamma}_5 = - \rho_1^2 + \rho_1 \left(\rho^2 + 3\rho + 1\right) +\left(1 +
\rho\right)^2, \;\bar{\gamma}_4 = -3\rho_1^2\rho + 3\rho_1\rho\left(\rho^2 + 3 \rho + 1\right) + 3 \rho \left(1 + \rho\right)^2, \\ &\bar{\gamma}_3 = \rho_1^3 + \rho_1^2\left(-4\rho^2 - 2\rho + 1\right) + \rho_1\rho\left(3\rho^3 + 11\rho^2 + 5\rho -2\right) + \rho^2 \left(1 + \rho\right)\left(5 + 3\rho\right),\; \bar{\gamma}_2= \rho_1^3\rho -\\ &\rho_1^2\rho\left(1 + \rho\right)\left(3\rho - 1\right) + \rho_1 \rho^2 \left(\rho^3 + 7 \rho^2 + 5\rho - 2\right) + \rho^3\left(1 + \rho\right)\left(5 + \rho\right), \;
\bar{\gamma}_1 = -\rho_1^2\rho^4 +  \rho^4\left(2\rho+3\right)\\&\left(1 + \rho_1\right), \; \bar{\gamma_0} = \rho^5\left(1 + \rho_1\right),
\end{align*}
\begin{align}\label{mom2_SA_B2}
\overset{{\rm SA}}{\Delta}_{1,2} = 
\begin{cases}
\dfrac{\sum_{n=0}^{5}{\rho_1^n \bar{\zeta}_n} + \rho_{-1}\rho_1^2 \bar{\zeta}_6 + \rho_{-1}^2\rho_1 \bar{\zeta}_7 + \rho_{-1}\rho_1 \bar{\zeta}_8 + \rho_{-1} \bar{\zeta}_9 }{2\mu^2\rho_1^2\rho^3\left(1 + \rho_1\right)^2\left(3 + 2 \rho\right)},\;\; & {\rm if}\; \rho = \beta,\\
\dfrac{2\sum_{n=0}^7{\beta^n \bar{\psi}_{n}}}{\mu^2\rho_1^2\beta^2\left(\beta + \rho\right)^3\left(1 + \rho_1\right)^2\big[\rho^2 + \beta \left(1 + \rho\right) \left(\beta + \rho\right)\big]},\;\; & {\rm otherwise}.
\end{cases}
\end{align}
\begin{align*}
&\bar{\zeta}_9 = 12\rho^2,\; \bar{\zeta}_8 = \rho^2\left(43\rho + 22\right),\; \bar{\zeta}_7 = \rho^2\left(18\rho + 4\right),\; \bar{\zeta}_6 = \rho^2\left(24\rho^2 + 74\rho + 8\right),\; \bar{\zeta}_5 = 2,\; \bar{\zeta}_4 = 6\rho^2 \\&+ 5 \rho + 4,\; \bar{\zeta}_3 = 2\left(4\rho + 1\right),\; \bar{\zeta}_2 = \rho \left(8\rho^5 + 20 \rho^4 + 3\right),\; \bar{\zeta}_1 = \rho^3 \left(16 \rho^3 + 62 \rho^2 + 61 \rho + 6\right),\; \bar{\zeta}_0 = \rho^3\\&\left(8\rho^3 + 36 \rho^2 + 54 \rho + 15\right),\; \bar{\psi}_{7} = -\rho_1^3\left(3 + 2\rho\right) + \rho_1^2 \left(\rho^3 + 4\rho^2 + 2\rho - 2\right) + \rho_1\left(1 + \rho\right)\left(2\rho^2 + 5\rho + 1\right) \\&+ \left(1 + \rho\right)^3,\; \bar{\psi}_{6} = -4\rho_1^3\rho\left(3 + 2\rho\right) + 4\rho_1^2\rho\big(\rho^3 + 4\rho^2 + 2\rho - 2\big) + 4\rho_1\rho\left(1 + \rho\right)\left(2\rho^2 + 5\rho + 1\right) + 4\rho \\&\left(1 + \rho\right)^3,\; \bar{\psi}_{5} = 3\rho_1^4\left(1 + \rho\right) -\rho_1^3\left(14\rho^3 + 27\rho^2 - 5\right) + \rho_1^2\left(6\rho^5 + 27 \rho^4 + 16\rho^3 - 23\rho^2 - 7\rho + 2\right) + \\&2\rho_1\rho\big(6\rho^4 + 24\rho^3 + 24 \rho^2 + 3\rho - 2 \big) + 3\rho^2 \left(3 + 2\rho\right) \left(1 + \rho\right)^2,\; \bar{\psi}_{4} = 2\rho_1^5 + \rho_1^4\left(6\rho^2 + 3\rho + 4\right) -\rho_1^3\big(14\rho^4 \\&+ 35\rho^3 -4\rho -2\big) + \rho_1^2\rho \left(4\rho^5 + 25\rho^4 + 20\rho^3 -33\rho^2 -14\rho + 1\right) + 2\rho_1\rho^2\big(4\rho^4 + 23\rho^3 + 30\rho^2 + 4\rho \\&- 4\big) + \rho^3\left(13 + 4\rho\right)\left(1 + \rho\right)^2,\; \bar{\psi}_{3} = 2\rho_1^5\rho + \rho_1^4\rho\left(3\rho^2 + 2\rho + 4\right) -\rho_1^3\rho\big(8\rho^4 + 24\rho^3 + 4\rho^2 -3\rho -2\big) +\\& \rho_1^2\rho^2\left(\rho^5 + 13 \rho^4 + 17\rho^3 - 19 \rho^2 - 15\rho + 1\right) + \rho_1 \rho^3 \big(2\rho^4 + 25\rho^3 + 48\rho^2 + 14\rho - 8\big) + \rho^4\left(1 + \rho\right)\big(\rho^2 + \\&11\rho + 13\big),\; \bar{\psi}_2 = 2\rho_1^4 \rho^3 - \rho_1^3\rho^3\left(2\rho^3 + 9\rho^2 + 4\rho - 4\right) + \rho_1^2\rho^3\left(3\rho^4 + 10\rho^3 -3\rho^2 -8\rho +2\right) + 2\rho_1\\&\rho^4\left(3\rho^3 + 12\rho^2 + 7\rho - 2\right) + 3\rho^5\left(1 + \rho\right)\left(3 + \rho\right),\; \bar{\psi}_1 = -2\rho_1^3\rho^6 + \rho_1^2\rho^6\left(1 + 3\rho\right) + \rho_1\rho^6\left(7 + 6\rho\right) + \\&\rho^6\left(4 + 3\rho\right),\; \bar{\psi}_0 = \rho^7\left(1 + \rho_1\right)^2.
\end{align*}
\end{cor}
\begin{remark}
Similar to Remark \ref{rem:Avg_comp_SA,WP,PS}, one can deduce from Corollaries \ref{cor:WP_moments}, \ref{cor:PS_moments} and \ref{cor:SA_moments} that $\overset{\rm PS}{\Delta}_{1,2} \leq \overset {\rm SA}{\Delta}_{1,2} \leq \overset {\rm WP}{\Delta}_{1,2}$ for any choice of values of the system parameters. Further, when $\rho_{-1} = 0$ (i.e., $N = 1$), we have $\overset {\rm WP}{\Delta}_{1,2} - \overset{\rm PS}{\Delta}_{1,2} = \overset {\rm WP}{\Delta}_{1,2} - \overset{\rm SA}{\Delta}_{1,2}$.
\end{remark}
\begin{figure*}[h!]
\centerline{
\subfloat[]{\includegraphics[width=0.37\textwidth]{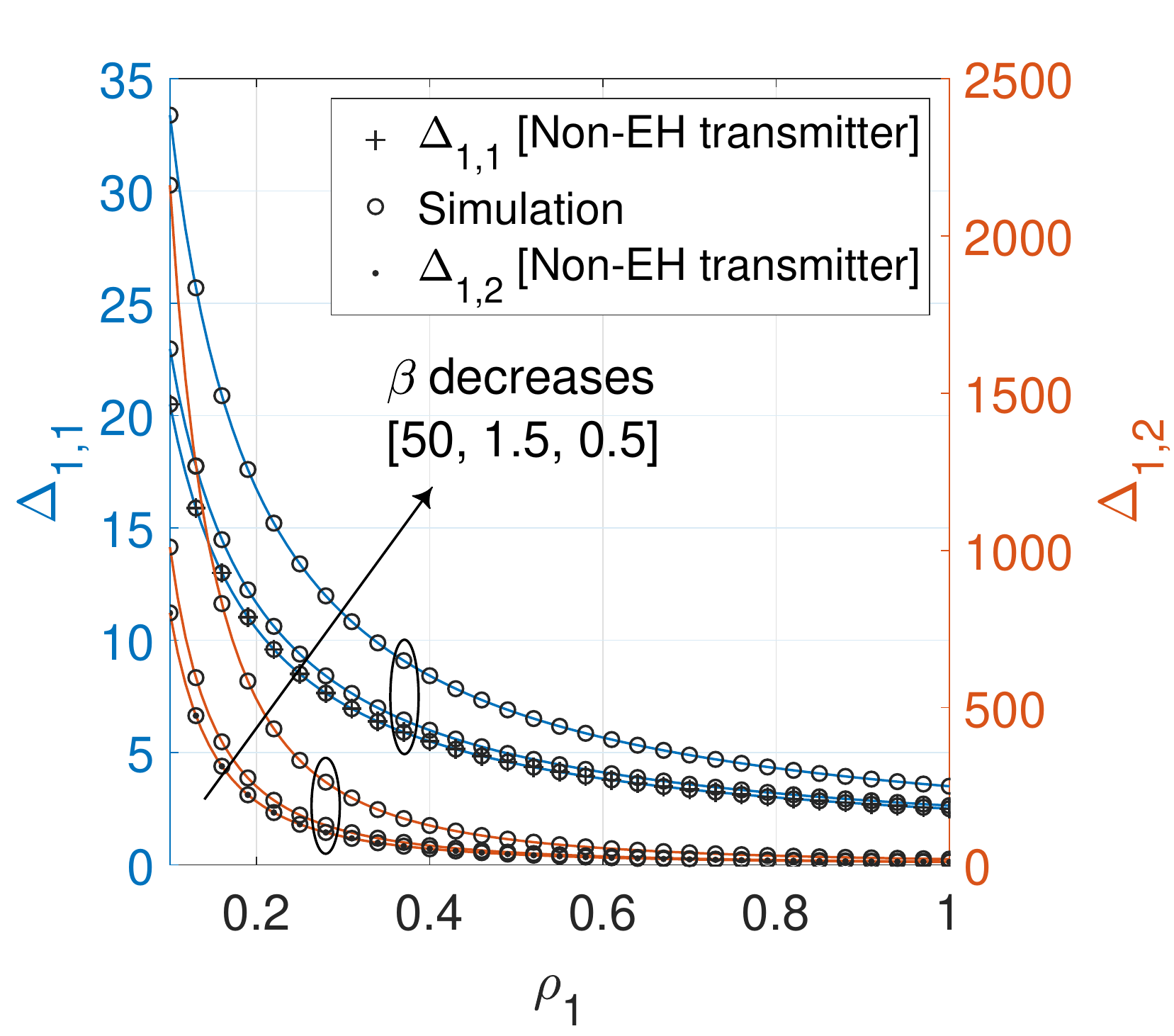}%
\label{f:WP_B2_ver}} \hfil
\subfloat[]{\includegraphics[width=0.37\textwidth]{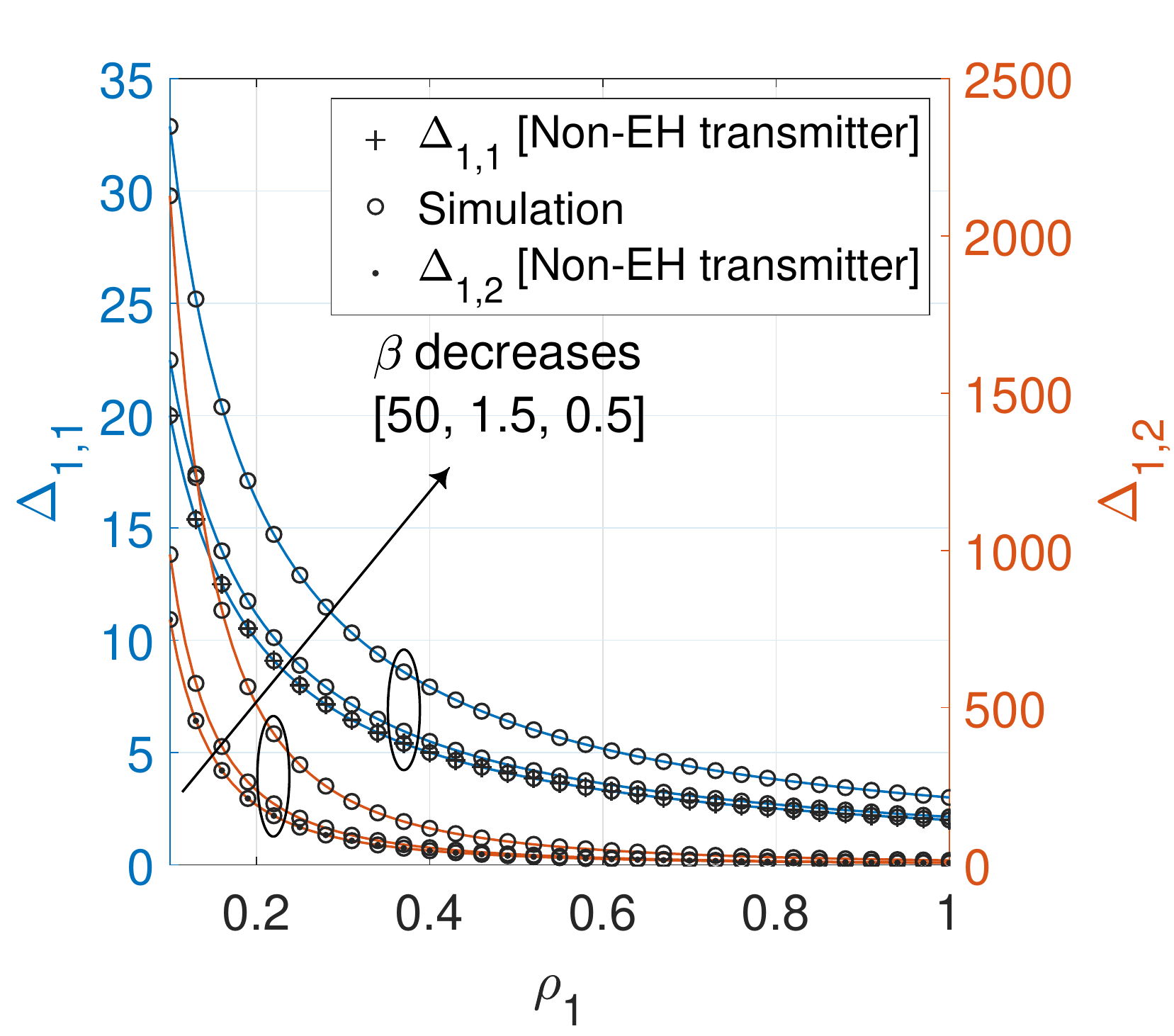}%
\label{f:PS_B2_ver}} \hfil
\subfloat[]{\includegraphics[width=0.375\textwidth]{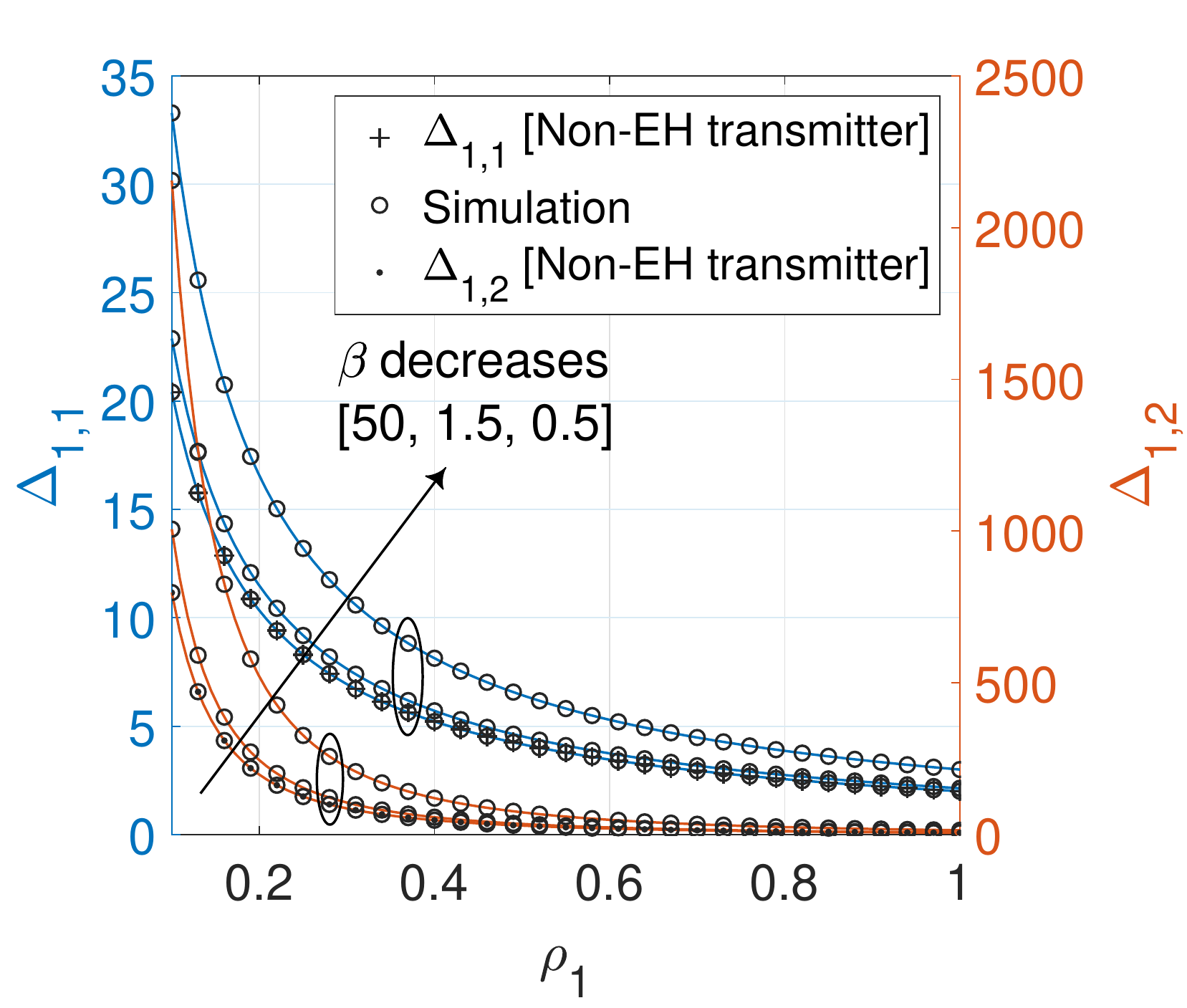}%
\label{f:SA_B2_ver}}} \vfil
\centerline{
\subfloat[]{\includegraphics[width=0.35\textwidth]{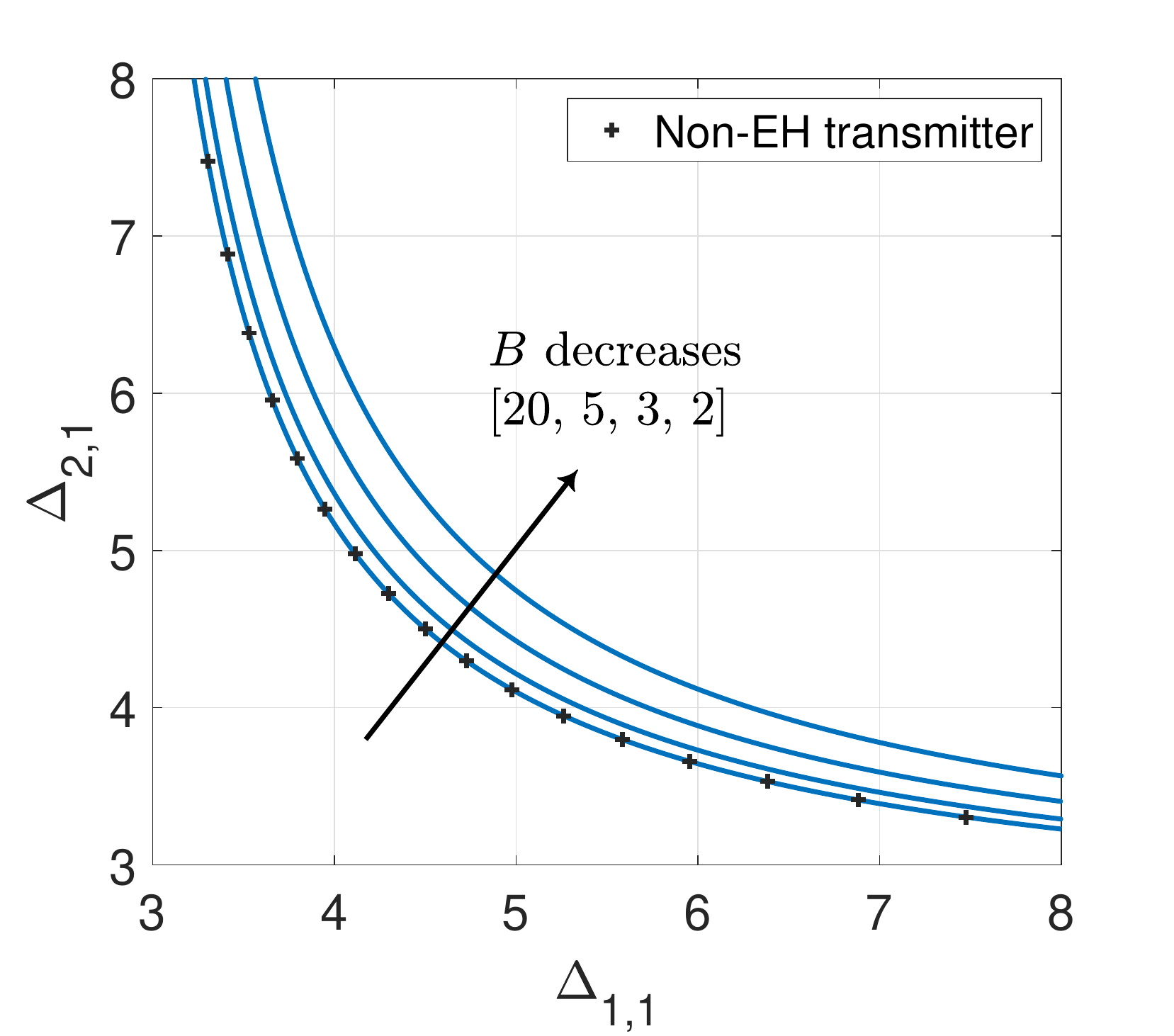}%
\label{f:WP_Bimpact}} \hfil
\subfloat[]{\includegraphics[width=0.35\textwidth]{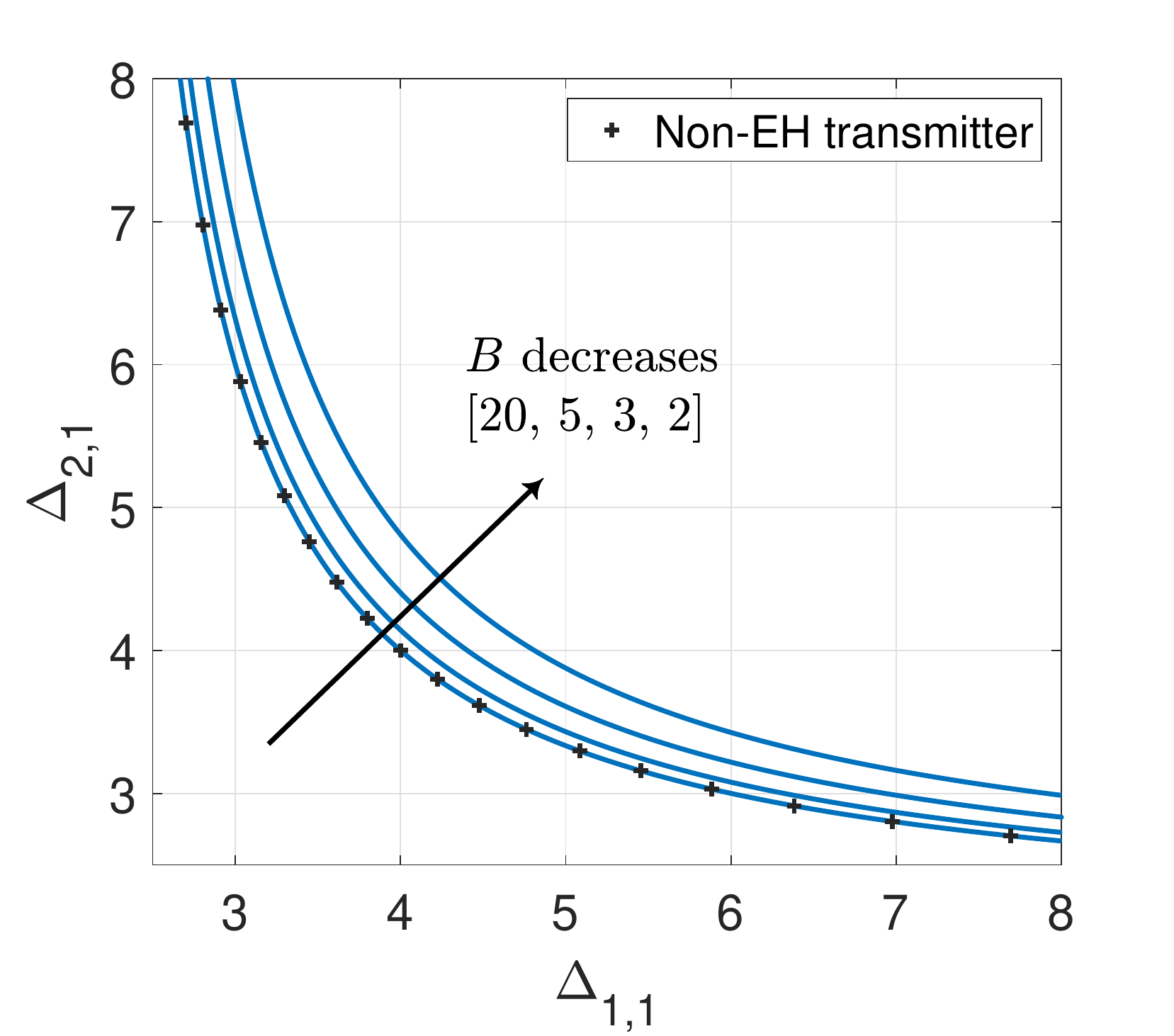}%
\label{f:PS_Bimpact}} \hfil
\subfloat[]{\includegraphics[width=0.35\textwidth]{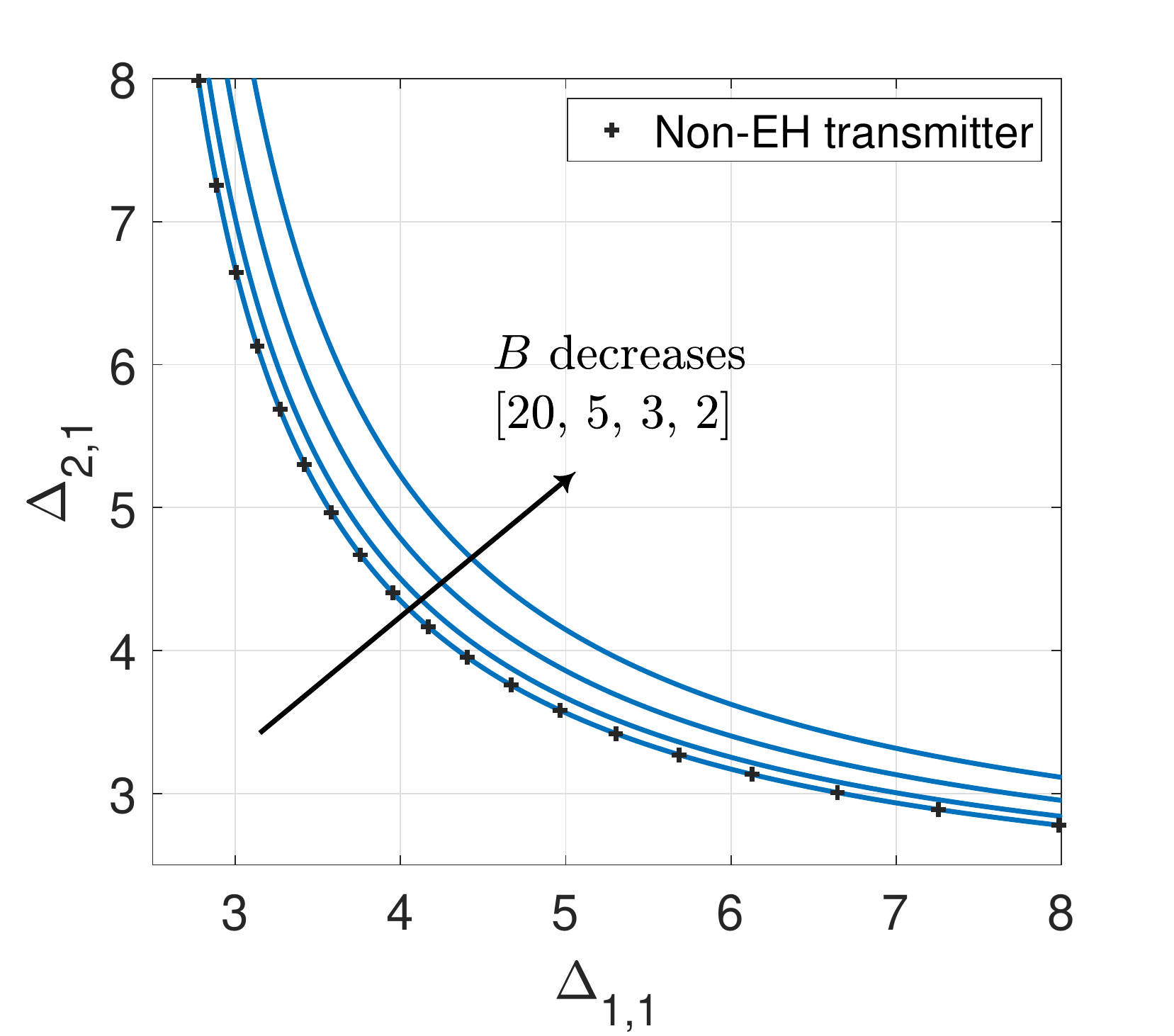}%
\label{f:SA_Bimpact}}}
\caption{Verification of the analytical results derived in Corollaries \ref{cor:WP_moments}, \ref{cor:PS_moments} and \ref{cor:SA_moments}: (a) the LCFS-WP queueing discipline, (b) the LCFS-PS queueing discipline, and (c) the LCFS-SA queueing discipline. In (a)-(c), $N$ can be chosen arbitrary and we use $\rho = 1$. Impact of battery capacity $B$: (d) the LCFS-WP queueing discipline, (e) the LCFS-PS queueing discipline, and (f) the LCFS-SA queueing discipline. In (d)-(f), we use $N = 2$, $\beta = 1.5$, and $\rho = 1$. }\label{f:verification}
\end{figure*} 
\section{Numerical Results}\label{sec:numerical}
In this section, we study the impact of the system design parameters on the achievable AoI performance under each of the three queueing disciplines considered in this paper. We use $\mu = 1$ in all the figures. In Fig \ref{f:verification}, we first verify the accuracy of the analytical expressions of the first and second moments of AoI for all the queueing disciplines (obtained in Corollaries \ref{cor:WP_moments}, \ref{cor:PS_moments} and \ref{cor:SA_moments} using the MGFs derived in Theorems \ref{theorem:MGF_WP}, \ref{theorem:MGF_PS} and \ref{theorem:MGF_SA}) in (a)-(c) by comparing them to their simulated counterparts (obtained numerically using \cite[Theorem~1]{yates2020age}). We then study the impact of battery capacity $B$ on the achievable pairs of average AoI $(\Delta_{1,1},\Delta_{2,1})$ when $N = 2$ and $\rho$ is fixed. We observe from Fig. \ref{f:verification} that the AoI performance of each queueing discipline improves with increasing $B$ or $\beta$ until it converges to its counterpart with a non-EH transmitter (as stated in Corollaries \ref{cor:Avg_WP_noEH}, \ref{rem:Avg_PS_noEH} and \ref{rem:Avg_SA_asym}). This happens since increasing $B$ or $\beta$ decreases the likelihood that the battery queue is empty upon the arrival of a new status update at the transmitter when the server is idle, and hence increases the likelihood of delivering new arriving updates to the destination.

\begin{figure*}[t!]
\centerline{
\subfloat[]{\includegraphics[width=0.37\columnwidth]{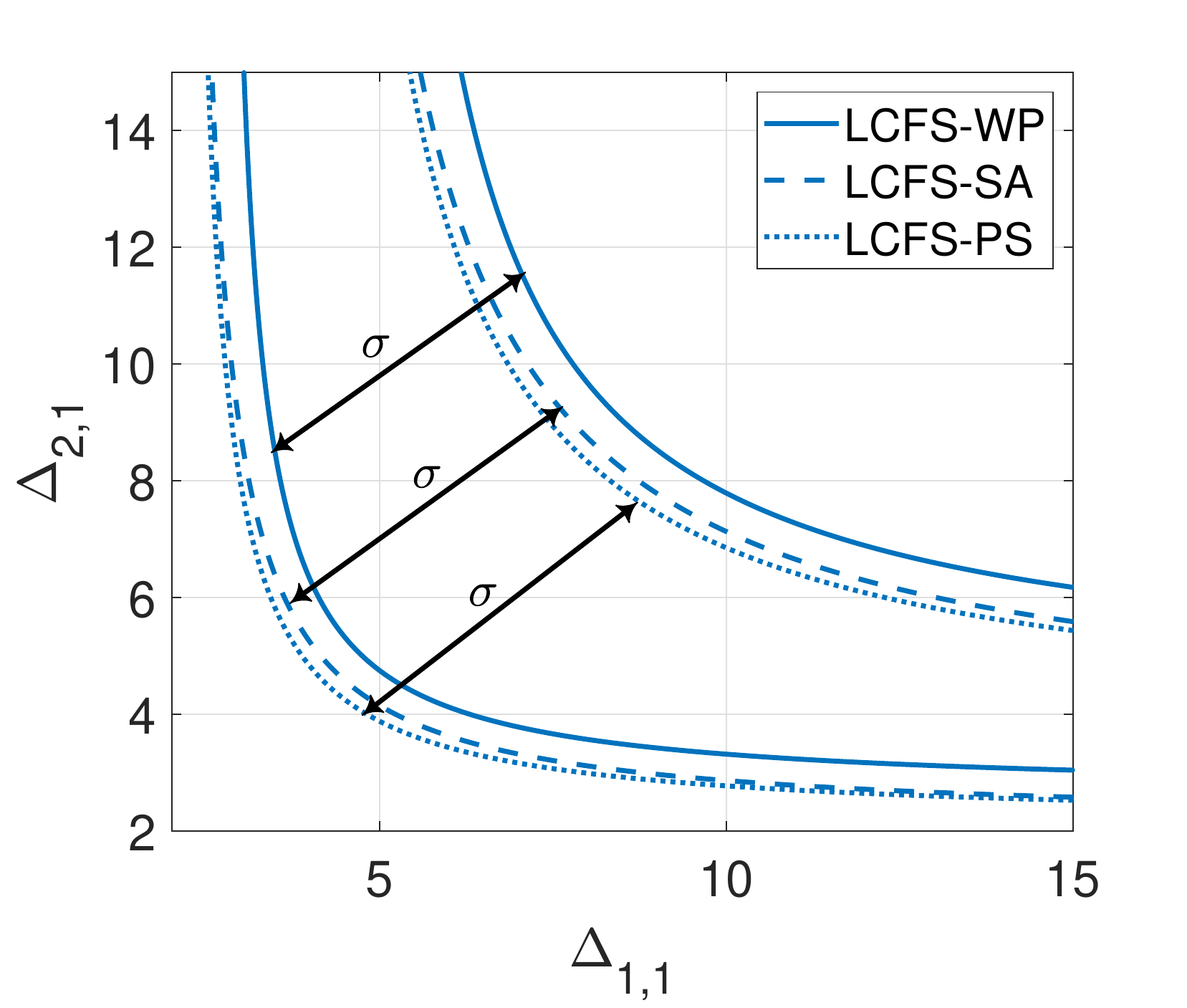}%
\label{f:comp_region_rho1}} \hfil
\subfloat[]{\includegraphics[width=0.37\columnwidth]{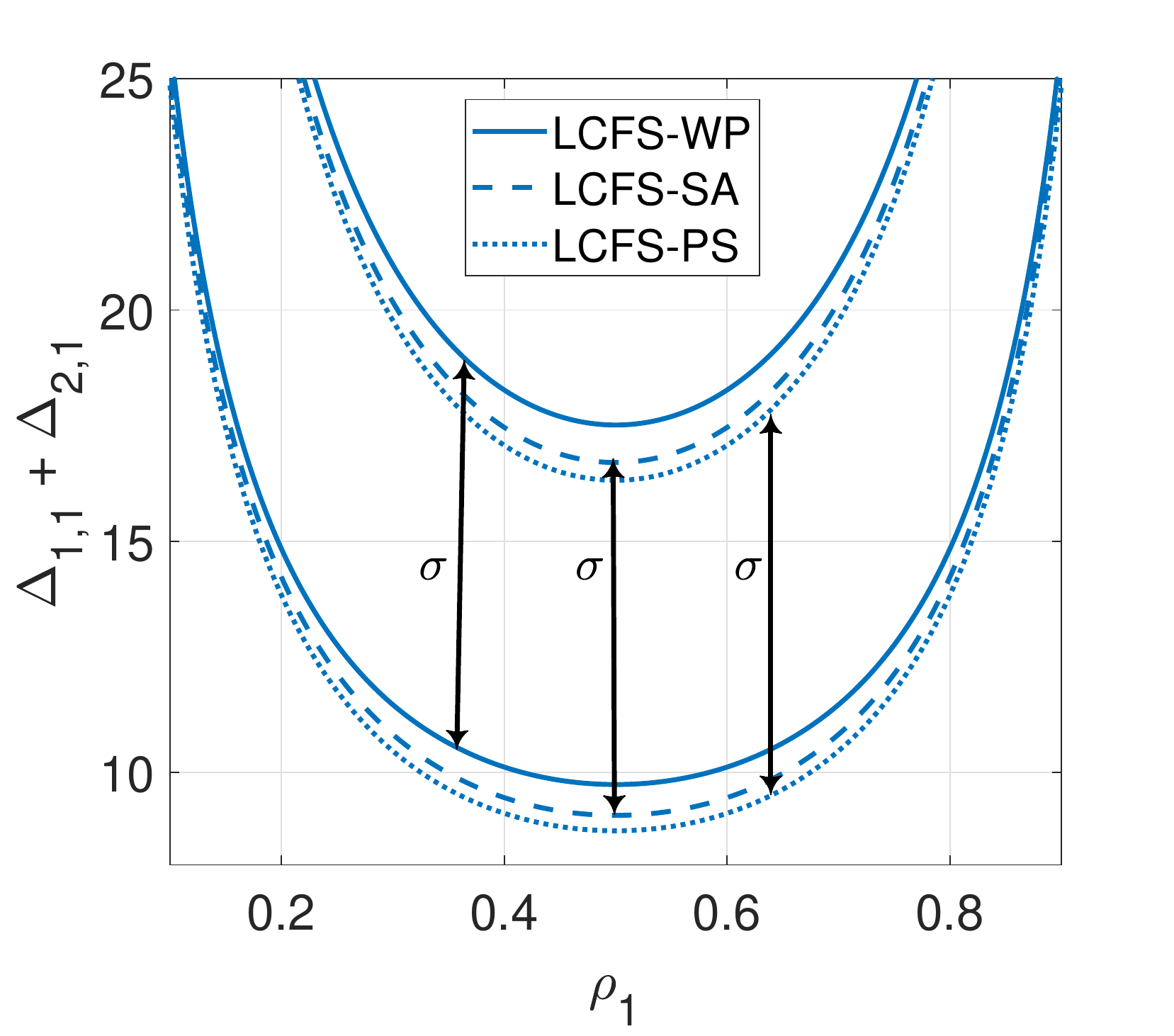}%
\label{f:comp_sumAoI_rho1}} \hfil
\subfloat[]{\includegraphics[width=0.37\columnwidth]{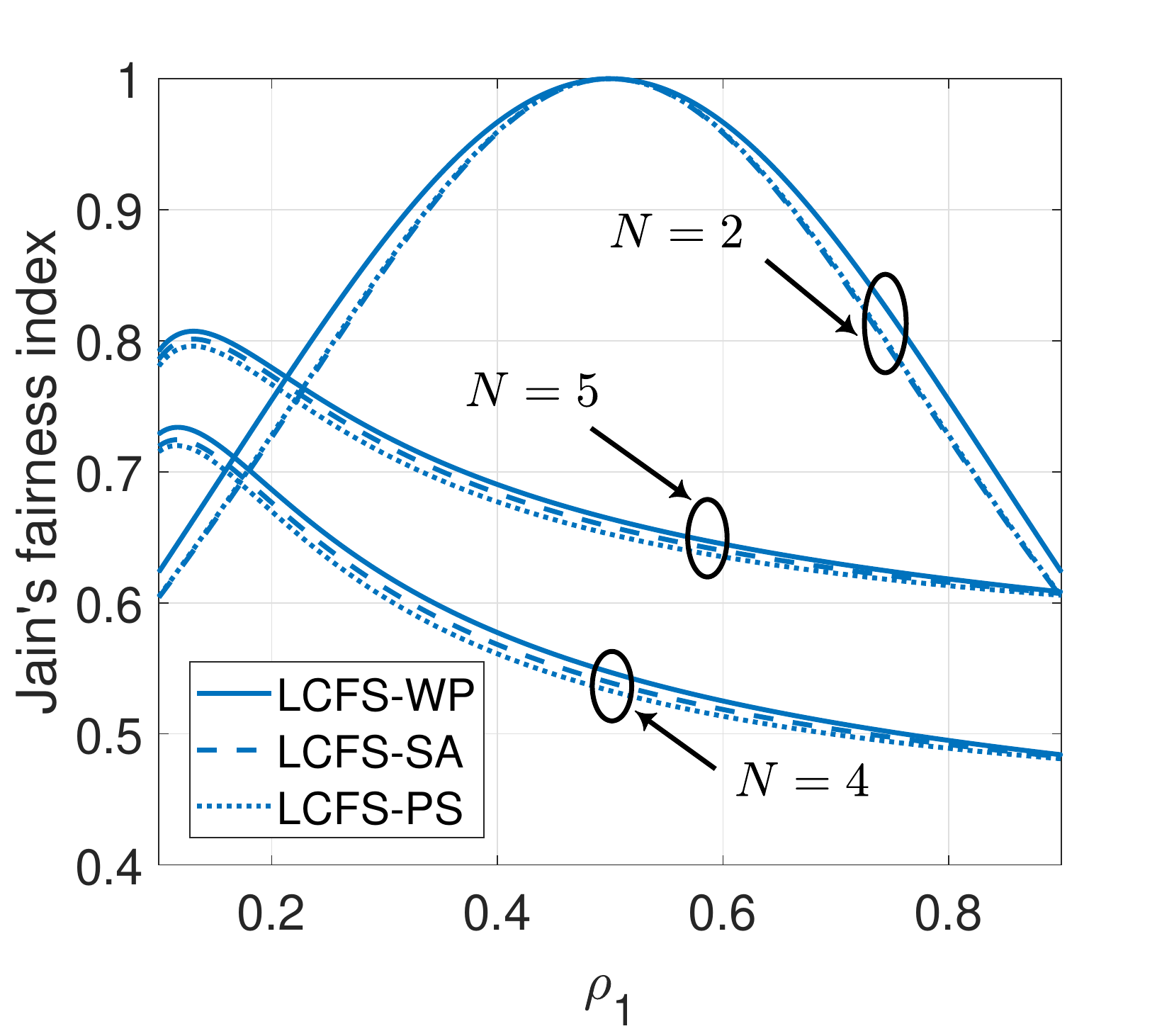}%
\label{f:comp_JFI_rho1}}} \vfil
\centerline{
\subfloat[]{\includegraphics[width=0.37\columnwidth]{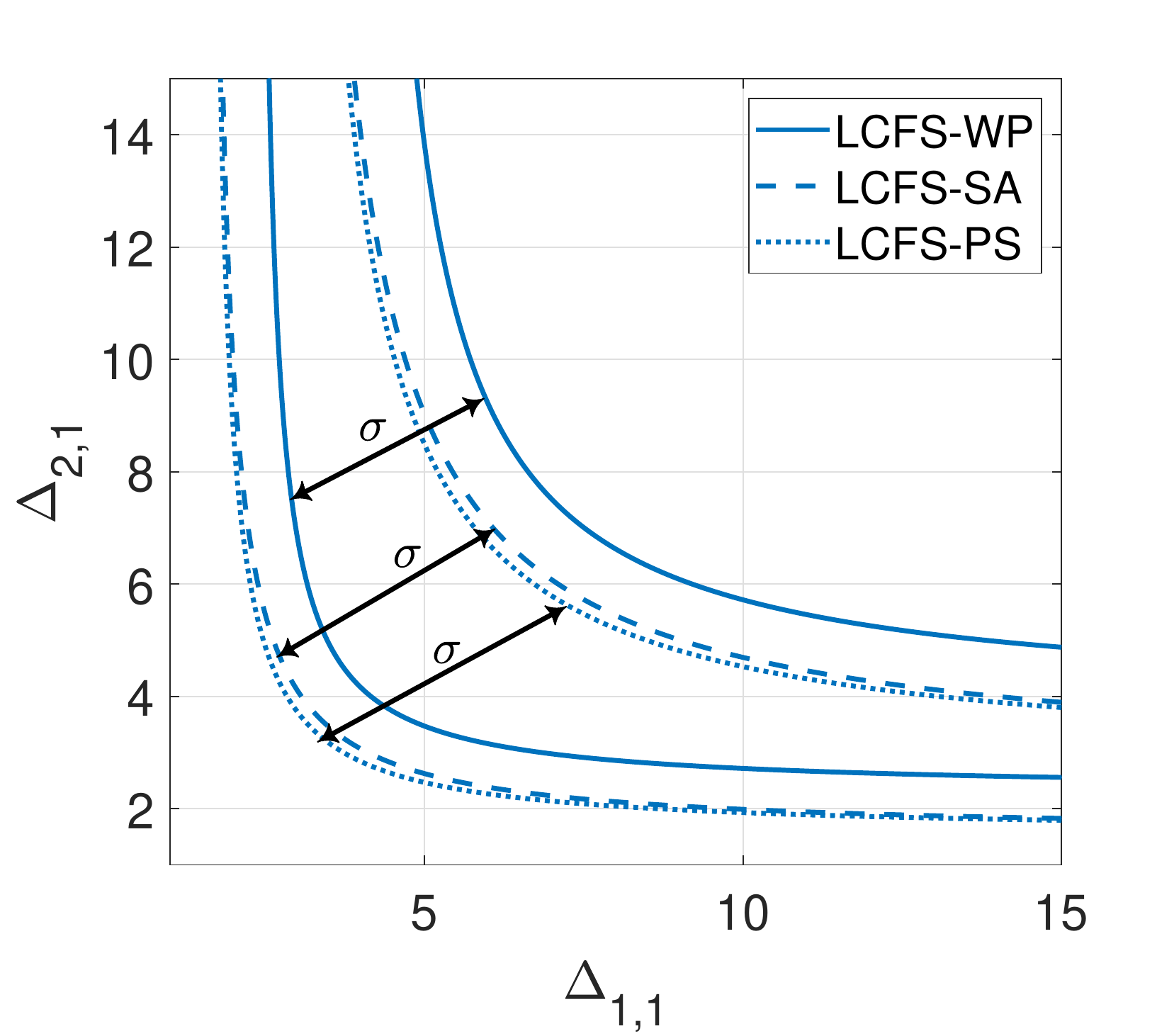}%
\label{f:comp_region_rho3}} \hfil
\subfloat[]{\includegraphics[width=0.37\columnwidth]{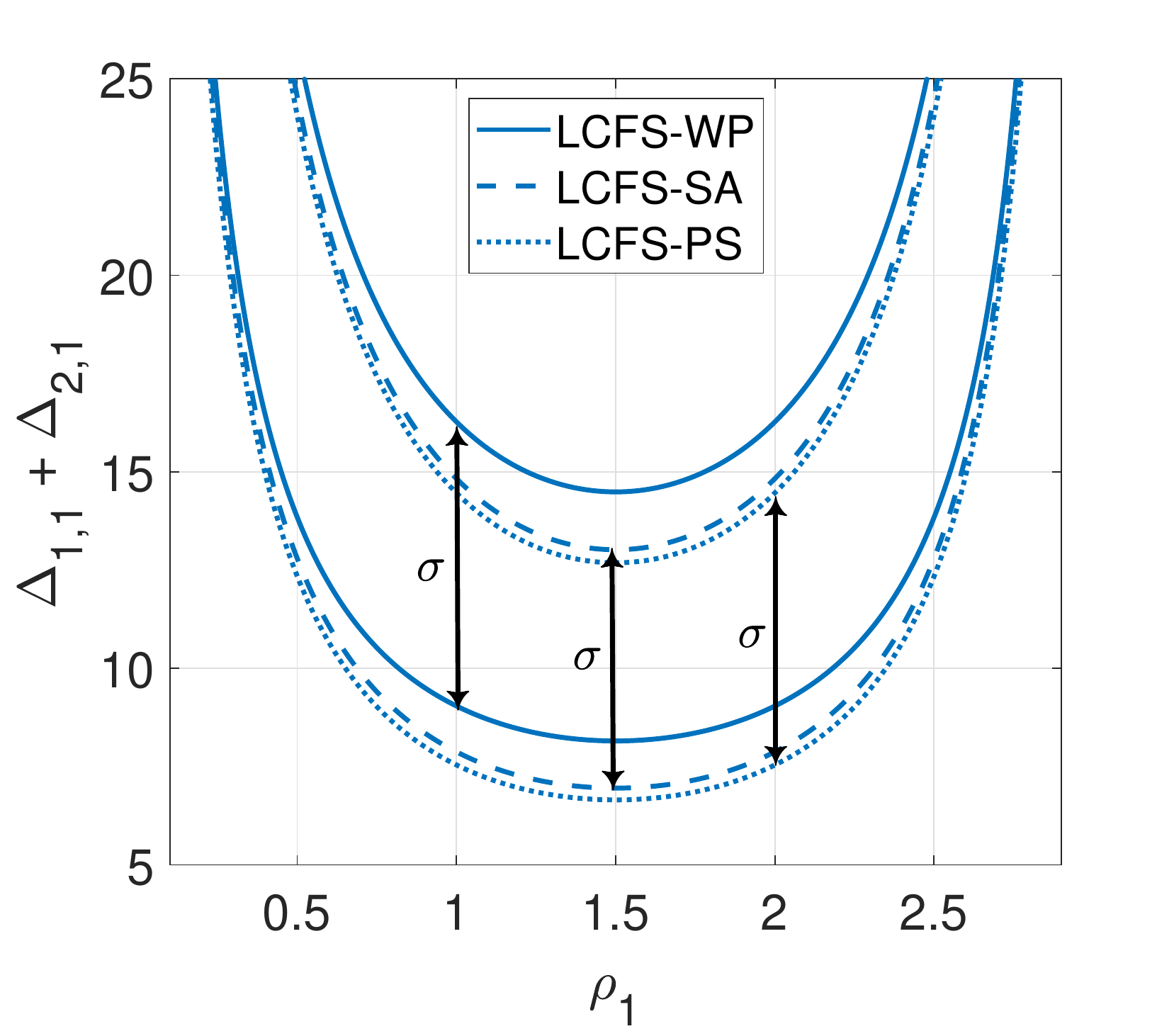}%
\label{f:comp_sumAoI_rho3}} \hfil
\subfloat[]{\includegraphics[width=0.37\columnwidth]{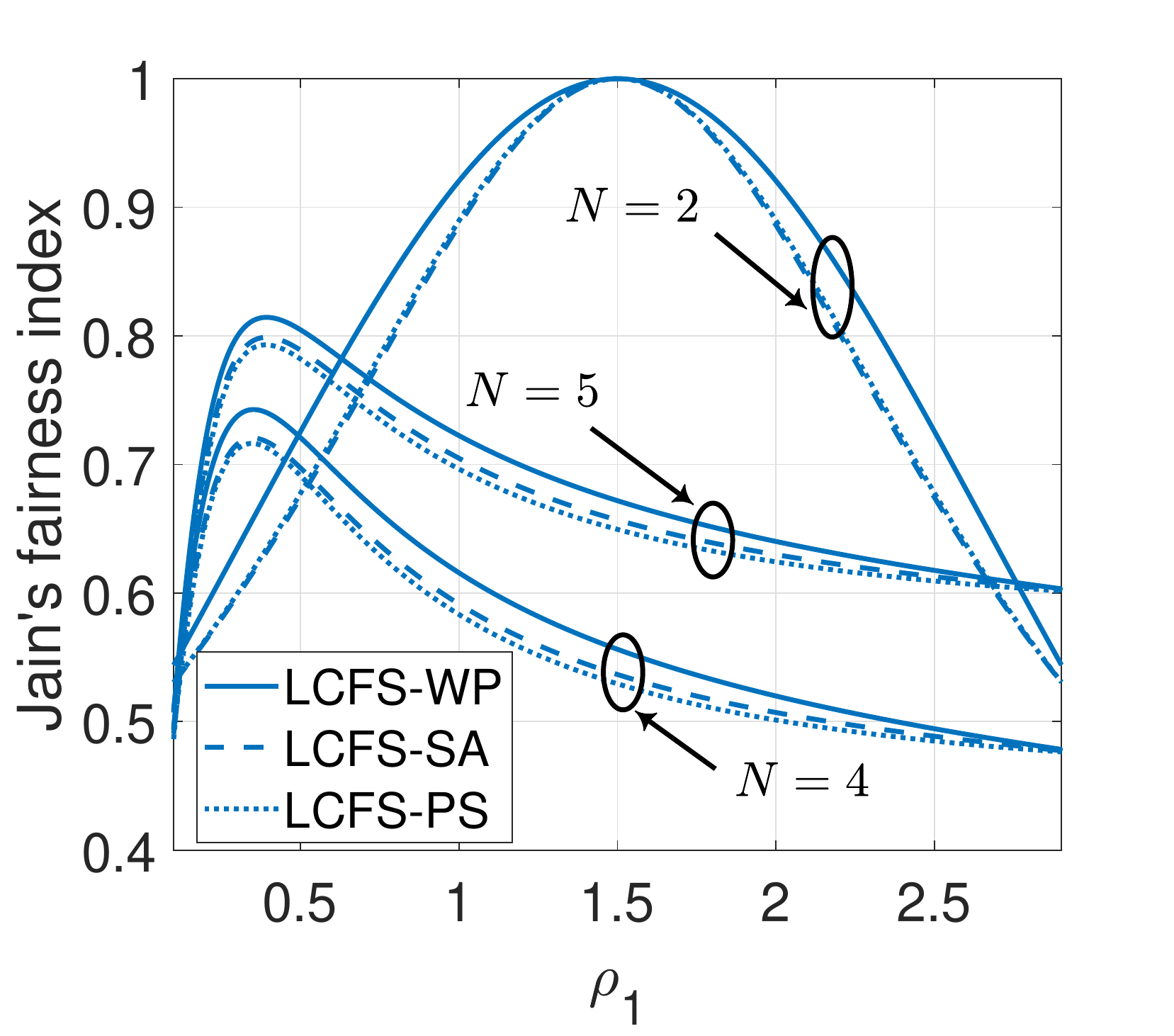}%
\label{f:comp_JFI_rho3}}
} \caption{Comparison between the achievable AoI performance of the queueing disciplines considered in this paper. We use $\beta = 1.5$, $B = 2$, and $\rho = 1$ $[\rho = 3]$ in (a), (b) and (c) [(d), (e) and (f)]. Further, we use $N = 2$ in (a), (b), (d) and (e). In (c) and (f) for $N \in \{4,5\}$, we set $\rho_2 = 0.1(\rho - \rho_1)$ and $\rho_i = \frac{0.9}{N - 2}(\rho - \rho_1)$, $3 \leq i \leq N$.} \label{f:comp}
\end{figure*} 
In Fig. \ref{f:comp}, we compare the three queueing disciplines studied in this paper when $N = 2$ in terms of: i) the achievable average AoI pairs $(\Delta_{1,1}, \Delta_{2,1})$ in Figs. \ref{f:comp_region_rho1} and \ref{f:comp_region_rho3}, ii) the average sum-AoI $\Delta_{1,1} + \Delta_{2,1}$ in Figs. \ref{f:comp_sumAoI_rho1} and \ref{f:comp_sumAoI_rho3}, and iii) the Jain's fairness index in Figs. \ref{f:comp_JFI_rho1} and \ref{f:comp_JFI_rho3}, which is defined as ${\rm JFI} = \dfrac{\left(\sum_{i=1}^{N}{\Delta_{i,1}}\right)^2}{N\sum_{i=1}^{N}{\Delta^2_{i,1}}}$ \cite{Jain}. Note that the ${\rm JFI} \in \left[N^{-1},1\right]$ is a measure of the fairness between the achievable average AoI values by different sources such that ${\rm JFI} = 1$ when the average AoI values of different sources are equal (the best scenario with respect to fairness). First, we observe from Figs. \ref{f:comp_region_rho1}, \ref{f:comp_sumAoI_rho1},  \ref{f:comp_region_rho3} and \ref{f:comp_sumAoI_rho3} the superiority of the LCFS-PS queueing discipline over the LCFS-WP and the LCFS-SA queueing disciplines in terms of the achievable average AoI performance (which supports our arguments in Remark \ref{rem:Avg_comp_SA,WP,PS}). However as indicated from Figs. \ref{f:comp_JFI_rho1} and \ref{f:comp_JFI_rho3}, such a superiority of the LCFS-PS queueing discipline comes at the expense of having unfair achievable average AoI values among different sources. Second, we observe from Figs. \ref{f:comp_JFI_rho1} and \ref{f:comp_JFI_rho3} that as $N$ increases, the superiority of the LCFS-SA queueing discipline over LCFS-PS in terms of the achievable fairness performance becomes more significant. As expected, this happens since the LCFS-SA queueing discipline does not allow source-agnostic preemption in service. Finally, as was the case in \cite{abdelmagid_2021a} for single-source systems with an EH-powered transmitter node, we observe from Figs. \ref{f:comp_region_rho1}, \ref{f:comp_sumAoI_rho1},  \ref{f:comp_region_rho3} and \ref{f:comp_sumAoI_rho3} that the standard deviation of AoI $\sigma$ associated with each queueing discipline in multi-source systems is relatively large with respect to the average value. This indicates that the implementation of multi-source status update systems based on just the average value of AoI does not ensure reliability, and it is crucial to incorporate the higher moments of AoI in the design of such systems. This insight demonstrates the significance of the analytical distributional properties of AoI derived in this paper.
\section{Conclusion}\label{sec:con}
This paper analytically characterized the AoI performance of multi-source EH updating systems, where an EH-powered transmitter sends status updates about several observed physical processes to a destination. In particular, we used the SHS approach to analyze AoI under non-preemptive and source-agnostic (LCFS-PS)/source-aware (LCFS-SA) preemptive in service queueing disciplines. We started our analysis by characterizing the average AoI for each considered queueing discipline in closed-form. We then extended the analysis to study the distributional properties of AoI through the characterization of its MGF. Our analytical results allowed us to obtain several useful insights regarding the achievable AoI performance under the considered queueing disciplines. For instance, the gaps between the achievable average AoI performances by the considered queueing disciplines were characterized in closed-form as functions of the system parameters. Further, our asymptotic results demonstrated the generality of the expressions derived in this paper by recovering several existing results for single source-systems with an EH-powered transmitter (when the aggregate generating rate of status updates from all the sources other than the source of interest approaches zero), and for multi-source systems with a non-EH transmitter (when the arrival rate of harvested energy packets becomes large).

Several key system design insights were also drawn from our numerical results. For instance, our results revealed a fundamental trade-off between obtaining a minimum average sum-AoI and having fair achievable average AoI values among different sources. Further, they showed the effectiveness of the LCFS-SA queueing discipline (compared to the LCFS-PS) in achieving fairness between different sources in term of the achievable AoI performance (especially when the number of sources is large). Finally, the results demonstrated that it is necessary to incorporate the higher moments of AoI in the implementation/optimization of multi-source real-time status updates systems rather than just relying on its average value.
\appendix
\subsection{Proof of Theorem~\ref{theorem:Avg_WP}} \label{app:theorem:Avg_WP}
We first show the existence of a non-negative limit $\bar{\nbv}_q, \forall q \in \ncalQ$, satisfying (\ref{gen_vavg}), and then we obtain the average AoI of source 1 using $\overset{\rm WP}{\Delta}_{1,1} = \sum_{q \in \ncalQ}{\bar{v}_{q0}}$. The set of equations in (\ref{gen_vavg}) can be expressed as
\begin{align}\label{theorem:Avg_WP_proof_1}
q_1: \;\; \eta [\bar{v}_{10},\bar{v}_{11}] = \mu[\bar{v}_{31},0] + [\bar{\pi}_1,\bar{\pi}_1],
\end{align}
\begin{align}\label{theorem:Avg_WP_proof_2}
q_2: \;\; \left(\eta + \lambda\right)[\bar{v}_{20},\bar{v}_{21}] = \mu[\bar{v}_{51},0] + \eta[\bar{v}_{10},0] +[\bar{\pi}_2,\bar{\pi}_2],
\end{align}
\begin{align}\label{theorem:Avg_WP_proof_3}
q_{2k}, 2 \leq k \leq B-1: \;\; \left(\eta + \lambda\right)[\bar{v}_{2k,0},\bar{v}_{2k,1}] = \mu[\bar{v}_{2k+3,1},0] + \eta[\bar{v}_{2k-2,0},0] + [\bar{\pi}_{2k},\bar{\pi}_{2k}],
\end{align}
\begin{align}\label{theorem:Avg_WP_proof_4}
q_{2B}: \;\; \lambda [\bar{v}_{2B,0},\bar{v}_{2B,1}] = \eta[\bar{v}_{2B-2,0},0] + [\bar{\pi}_{2B},\bar{\pi}_{2B}],
\end{align}
\begin{align}\label{theorem:Avg_WP_proof_5}
q_{2k+1}, 1 \leq k \leq B: \;\; \mu[\bar{v}_{2k+1,0},\bar{v}_{2k+1,1}] = [\lambda\bar{v}_{2k,0},\lambda_{-1}\bar{v}_{2k,0}] + [\bar{\pi}_{2k+1},\bar{\pi}_{2k+1}].
\end{align}

 From (\ref{theorem:Avg_WP_proof_4}), $\bar{v}_{2B,0}$ can be expressed as
\begin{align}\label{theorem:Avg_WP_proof_6}
\bar{v}_{2B,0} = \dfrac{\eta \bar{v}_{2B-2,0}}{c_{2B}} + \dfrac{\bar{\pi}_{2B}}{c_{2B}},    
\end{align}
where $c_{2B} = \lambda$. Substituting $k = B-1$ in (\ref{theorem:Avg_WP_proof_3}), $\bar{v}_{2B-2,0}$ can be expressed as
\begin{align}\label{theorem:Avg_WP_proof_7}
\bar{v}_{2B-2,0} = \dfrac{\eta \bar{v}_{2B-4,0}}{c_{2B-2}} + \dfrac{\bar{\pi}_{2B-2}}{c_{2B-2}} + \dfrac{\lambda_{-1} \bar{\pi}_{2B}}{c_{2B-2}c_{2B}} + \dfrac{\bar{\pi}_{2B+1}}{c_{2B-2}},   
\end{align}
where $\bar{v}_{2B+1,1}$ and $\bar{v}_{2B,0}$ were respectively substituted from (\ref{theorem:Avg_WP_proof_5}) and (\ref{theorem:Avg_WP_proof_6}), and $c_{2B-2} = \eta\left(1 - \frac{\lambda_{-1}}{c_{2B}}\right) + \lambda$. Repeated application of (\ref{theorem:Avg_WP_proof_3}) gives
\begin{align}\label{theorem:Avg_WP_proof_8}
\bar{v}_{2k,0} = \dfrac{\eta \bar{v}_{2k-2}}{c_{2k}} + \sum_{j=1}^{B+1-k}{\dfrac{\bar{\pi}_{2(k+j-1) }\lambda_{-1}^{j-1}}{\prod_{h=1}^{j}{c_{2(k+h-1)}}}} + \sum_{j=1}^{B-k}{\dfrac{\bar{\pi}_{2(k+j-1)+3 }\lambda_{-1}^{j-1}}{\prod_{h=1}^{j}{c_{2(k+h-1)}}}},\; 2 \leq k \leq B,
\end{align}
\begin{align}\label{theorem:Avg_WP_proof_9}
\bar{v}_{20} = \dfrac{\eta \bar{v}_{10}}{c_{2}} + \sum_{j=1}^{B}{\dfrac{\bar{\pi}_{2j }\lambda_{-1}^{j-1}}{\prod_{h=1}^{j}{c_{2h}}}} + \sum_{j=1}^{B-1}{\dfrac{\bar{\pi}_{2j+3 }\lambda_{-1}^{j-1}}{\prod_{h=1}^{j}{c_{2h}}}}.
\end{align}

After substituting $\bar{v}_{31}$ from (\ref{theorem:Avg_WP_proof_5}) into (\ref{theorem:Avg_WP_proof_1}) and then solving (\ref{theorem:Avg_WP_proof_1}) and (\ref{theorem:Avg_WP_proof_9}), $\bar{v}_{10}$ is given by
\begin{align}\label{theorem:Avg_WP_proof_10}
 \bar{v}_{10} = \dfrac{\bar{\pi}_1}{c_0 \lambda_{-1}} + \sum_{j=1}^{B}{\dfrac{\bar{\pi}_{2j }\lambda_{-1}^{j-1}}{\prod_{h=0}^{j}{c_{2h}}}} + \sum_{j=0}^{B-1}{\dfrac{\bar{\pi}_{2j+3 }\lambda_{-1}^{j-1}}{\prod_{h=0}^{j}{c_{2h}}}},
\end{align}
where $c_0 = \eta \left(\frac{1}{\lambda_{-1}} - \frac{1}{c_2}\right)$. Note that the set $\{c_0,c_2,\cdots,c_{2B}\}$ contains positive real numbers, and hence we observe from (\ref{theorem:Avg_WP_proof_10}) that $\bar{v}_{10} \geq 0$. Thus, from (\ref{theorem:Avg_WP_proof_8}) and (\ref{theorem:Avg_WP_proof_9}), we deduce that $\bar{v}_{q0} \geq 0, q \in {\rm r}_1$. As a result, we observe from (\ref{theorem:Avg_WP_proof_5}) that $\bar{v}_{q0} \geq 0$ and $\bar{v}_{q1} \geq 0, q \in {\rm r}_2$. Finally, from (\ref{theorem:Avg_WP_proof_1})-(\ref{theorem:Avg_WP_proof_4}), one can easily see that $\bar{v}_{q1} \geq 0, q \in {\rm r}_1$, and hence there exists a non-negative limit $\bar{\nbv}_q, \forall q \in \ncalQ,$ satisfying (\ref{theorem:Avg_WP_proof_1})-(\ref{theorem:Avg_WP_proof_5}). 

Now, we proceed with evaluating the average AoI of source 1. Summing (\ref{theorem:Avg_WP_proof_1})-(\ref{theorem:Avg_WP_proof_4}) gives
\begin{align}\label{theorem:Avg_WP_proof_11}
\lambda \sum_{q \in \;{\rm r}_1/\{1\}} {\bar{v}_{q0}}= \mu \sum_{q \in\;{\rm r}_2}{\bar{v}_{q1}} + \sum_{q \in \;{\rm r}_1}{\bar{\pi}_q}.
\end{align}

Further, by summing the set of equations in (\ref{theorem:Avg_WP_proof_5}), we have
\begin{align}\label{theorem:Avg_WP_proof_12}
\mu \sum_{q \in \;{\rm r}_2} {\bar{v}_{q0}}= \lambda \sum_{q \in \;{\rm r}_1 / \{1\}}{\bar{v}_{q0}} + \sum_{q \in\; {\rm r}_2}{\bar{\pi}_q},
\end{align}
\begin{align}\label{theorem:Avg_WP_proof_13}
\mu \sum_{q \in \;{\rm r}_2} {\bar{v}_{q1}}= \lambda_{-1} \sum_{q \in \;{\rm r}_1 / \{1\}}{\bar{v}_{q0}} + \sum_{q \in\; {\rm r}_2}{\bar{\pi}_q}.
\end{align}

Thus, the average AoI of source 1 can be evaluated as 
\begin{align}\label{theorem:Avg_WP_proof_14}
 \overset{\rm WP}{\Delta}_{1,1} = \sum_{q \in\; {\rm r}_1 \cup \;{\rm r}_2}{\bar{v}_{q0}} \a \left(1 + \rho\right) \sum_{q \in \;{\rm r}_1 / \{1\}}{\bar{v}_{q0}} + \dfrac{\sum_{q \in\; {\rm r}_2}{\bar{\pi}_q}}{\mu} + \bar{v}_{10} \b \dfrac{1 + \rho}{\mu \rho_1}  + \dfrac{\sum_{q \in\; {\rm r}_2}{\bar{\pi}_q}}{\mu} + \bar{v}_{10},
\end{align}
where step (a) follows from substituting $\sum_{q \in\; {\rm r}_2}{\bar{v}_{q0}}$ from (\ref{theorem:Avg_WP_proof_12}) into (\ref{theorem:Avg_WP_proof_14}), and step (b) follows from substituting $\sum_{q \in\; {\rm r}_1}{\bar{v}_{q0}}$ using (\ref{theorem:Avg_WP_proof_11}) and (\ref{theorem:Avg_WP_proof_13}) into (\ref{theorem:Avg_WP_proof_14}). The final expression of $\overset{\rm WP}{\Delta}_{1,1}$ in (\ref{theorem:Avg_WP_1}) can directly be obtained by substituting $\bar{v}_{10}$ from (\ref{theorem:Avg_WP_proof_10}) into (\ref{theorem:Avg_WP_proof_14}).
\hfill 
\IEEEQED
\subsection{Proof of Theorem~\ref{theorem:Avg_PS}} \label{app:theorem:Avg_PS}
By inspecting Fig. \ref{f:PS_MC}, we observe that the set of equations in (\ref{gen_vavg}) corresponding to the states in ${\rm r}_1$ are still given by (\ref{theorem:Avg_WP_proof_1})-(\ref{theorem:Avg_WP_proof_4}). Regarding the states in ${\rm r}_2$, we have 
\begin{align}\label{theorem:Avg_PS_proof_1}
q_{2k+1}:\;\;
\left(\lambda + \mu \right)[\bar{v}_{2k+1,0},\bar{v}_{2k+1,1}] = [\lambda\bar{v}_{2k,0},\lambda_{-1}\bar{v}_{2k,0}] +  [\lambda\bar{v}_{2k+1,0},\lambda_{-1}\bar{v}_{2k+1,0}] + [\bar{\pi}_{2k+1},\bar{\pi}_{2k+1}],
\end{align}
where $1 \leq k \leq B$. Similar to the procedure in (\ref{theorem:Avg_WP_proof_6})-(\ref{theorem:Avg_WP_proof_10}) in Appendix \ref{app:theorem:Avg_WP}, repeated application of (\ref{theorem:Avg_WP_proof_3}) gives
\begin{align}\label{theorem:Avg_PS_proof_2}
\bar{v}_{2k,0} = \dfrac{\eta \bar{v}_{2k-2}}{c_{2k}} + \sum_{j=1}^{B+1-k}{\dfrac{\bar{\pi}_{2(k+j-1) }\lambda_{-1}^{j-1}}{\prod_{h=1}^{j}{c_{2(k+h-1)}}}} + \dfrac{1+\rho_{-1}}{1+\rho}\sum_{j=1}^{B-k}{\dfrac{\bar{\pi}_{2(k+j-1)+3 }\lambda_{-1}^{j-1}}{\prod_{h=1}^{j}{c_{2(k+h-1)}}}},\; 2 \leq k \leq B,
\end{align}
\begin{align}\label{theorem:Avg_PS_proof_3}
\bar{v}_{20} = \dfrac{\eta \bar{v}_{10}}{c_{2}} + \sum_{j=1}^{B}{\dfrac{\bar{\pi}_{2j }\lambda_{-1}^{j-1}}{\prod_{h=1}^{j}{c_{2h}}}} + \dfrac{1+\rho_{-1}}{1+\rho} \sum_{j=1}^{B-1}{\dfrac{\bar{\pi}_{2j+3 }\lambda_{-1}^{j-1}}{\prod_{h=1}^{j}{c_{2h}}}}.
\end{align}

Thus, from (\ref{theorem:Avg_WP_proof_1}), (\ref{theorem:Avg_PS_proof_1}) and 
(\ref{theorem:Avg_PS_proof_3}), $\bar{v}_{10}$ can be expressed as
\begin{align}\label{theorem:Avg_PS_proof_4}
 \bar{v}_{10} = \dfrac{\bar{\pi}_1}{c_0 \lambda_{-1}} + \sum_{j=1}^{B}{\dfrac{\bar{\pi}_{2j }\lambda_{-1}^{j-1}}{\prod_{h=0}^{j}{c_{2h}}}} + \dfrac{1 + \rho_{-1}}{1 + \rho}\sum_{j=0}^{B-1}{\dfrac{\bar{\pi}_{2j+3 }\lambda_{-1}^{j-1}}{\prod_{h=0}^{j}{c_{2h}}}}.
\end{align}

Recalling that the set $\{c_0,c_2,\cdots,c_{2B}\}$ contains positive real numbers, we deduce from (\ref{theorem:Avg_WP_proof_1})-(\ref{theorem:Avg_WP_proof_4}) and (\ref{theorem:Avg_PS_proof_1})-(\ref{theorem:Avg_PS_proof_4}) that there exists a non-negative limit $\bar{\nbv}_q, \forall q \in \ncalQ,$ satisfying (\ref{gen_vavg}). Further, the average AoI of source 1 can be evaluated as follows. We first note that $\sum_{q \in\; {\rm r}_1/ \{1\}}{\bar{v}_{q0}}$ and $\sum_{q \in\; {\rm r}_2}{\bar{v}_{q0}}$ can be expressed as in (\ref{theorem:Avg_WP_proof_11}) and (\ref{theorem:Avg_WP_proof_12}), respectively. In addition, summing the set of equations in (\ref{theorem:Avg_PS_proof_1}) gives
\begin{align}\label{theorem:Avg_PS_proof_5}
\left(\mu + \lambda\right) \sum_{q \in \;{\rm r}_2} {\bar{v}_{q1}}= \lambda_{-1} \sum_{q \in \;{\rm r}_1 / \{1\}}{\bar{v}_{q0}} + \lambda_{-1} \sum_{q \in \;{\rm r}_2}{\bar{v}_{q0}} +  \sum_{q \in\; {\rm r}_2}{\bar{\pi}_q}.
\end{align}

By solving (\ref{theorem:Avg_WP_proof_11}), (\ref{theorem:Avg_WP_proof_12}) and (\ref{theorem:Avg_PS_proof_5}), we get \begin{align}\label{theorem:Avg_PS_proof_6}
\sum_{q\in\;{\rm r}_1 / \{1\}}{\bar{v}_{q0}} = \dfrac{1 + \rho_{-1}}{\mu \rho_1\left(1 + \rho\right)} + \dfrac{\sum_{q\in\;{\rm r}_1}{\bar{\pi}_q}}{\mu\left(1 + \rho\right)},\; \sum_{q \in \; {\rm r}_2}{\bar{v}_{q0}} = \dfrac{\rho\left(1 + \rho_{-1}\right)}{\mu \rho_1\left(1 + \rho\right)} +\dfrac{\rho  \sum_{q\in\;{\rm r}_1}{\bar{\pi}_q}}{\mu\left(1 + \rho\right)} + \dfrac{\sum_{q\in\;{\rm r}_2}{\bar{\pi}_q}}{\mu}.
\end{align}

From (\ref{theorem:Avg_PS_proof_6}), the average AoI of source 1 can be evaluated as
\begin{align}\label{theorem:Avg_PS_proof_7}
\overset{\rm PS}{\Delta}_{1,1} = \sum_{q \in\; {\rm r}_1 \cup \;{\rm r}_2}{\bar{v}_{q0}} = \dfrac{1 + \rho}{\mu \rho_1} + \bar{v}_{10}.
\end{align}

The expression of $\overset{\rm PS}{\Delta}_{1,1}$ in (\ref{theorem:Avg_PS_1}) can be obtained by substituting $\bar{v}_{10}$ from (\ref{theorem:Avg_PS_proof_4}) into (\ref{theorem:Avg_PS_proof_7}).
\hfill 
\IEEEQED

\subsection{Proof of Theorem~\ref{theorem:Avg_SA}} \label{app:theorem:Avg_SA}
We first note from Table \ref{table:SA} that the second component of the vector $\bar{\nbv}_{q_l} \nbA_l, \forall l\in\ncalL'_q$ and $q \in \ncalQ,$ is 0. Thus, we observe from (\ref{gen_vavg}) that $\bar{v}_{q1} \geq 0, \forall q \in \ncalQ$. The existence of a non-negative limit $\bar{\nbv}_q, \forall q \in \ncalQ$, satisfying (\ref{gen_vavg}) is then tied with having $\bar{v}_{q0} \geq 0, \forall q \in \ncalQ$. The set of equations in (\ref{gen_vavg}) corresponding to the states in ${\rm r}_1$ can be expressed as
\begin{align}\label{theorem:Avg_SA_proof_1}
q_1:\;\; \eta\bar{v}_{10} = \mu\bar{v}_{31} + \mu \sum_{j=4}^{N+2}{\bar{v}_{j0}} + \bar{\pi}_1,
\end{align}
\begin{align}\label{theorem:Avg_SA_proof_2}
q_2:\;\; \left(\eta+\lambda\right)\bar{v}_{20} = \eta\bar{v}_{10} + \mu\bar{v}_{N+4,1} + \mu \sum_{j=N+5}^{2N+3}{\bar{v}_{j0}} + \bar{\pi}_2,
\end{align}
\begin{align}\label{theorem:Avg_SA_proof_3}
\nonumber q_{2+k(N+1)}, 1 \leq k \leq B-2:\;\; \left(\eta+\lambda\right)\bar{v}_{2+k(N+1),0} &= \eta\bar{v}_{2+(k-1)(N+1),0} + \mu\bar{v}_{3+(k+1)(N+1),1} \\
&+ \mu \sum_{j=4+(k+1)(N+1)}^{N+2+(k+1)(N+1)}{\bar{v}_{j0}} + \bar{\pi}_{2+k(N+1)},
\end{align}
\begin{align}\label{theorem:Avg_SA_proof_4}
 q_{2+(B-1)(N+1)}:\;\; \lambda\bar{v}_{2+(B-1)(N+1),0} = \eta\bar{v}_{2+(B-2)(N+1),0} + \bar{\pi}_{2+(B-1)(N+1)}.
\end{align}

Further, the set of equations in (\ref{gen_vavg}) corresponding to the states in ${\rm r}_{i+1}$, $1 \leq i \leq N$, can be expressed as
\begin{align}\label{theorem:Avg_SA_proof_5}
q_{2+i+k(N+1)}, 0 \leq k \leq B-1:\;\; \mu \bar{v}_{2+i+k(N+1),0} = \lambda_i \bar{v}_{2+k(N+1),0} + \bar{\pi}_{2+i+k(N+1)}.   
\end{align}

By noting that $\bar{v}_{3+k(N+1),1} = \dfrac{\bar{\pi}_{3+k(N+1)}}{\mu + \lambda_1}, 0 \leq k \leq B-1$, (\ref{theorem:Avg_SA_proof_3}) can be rewritten as 
\begin{align}\label{theorem:Avg_SA_proof_6}
q_{2+k(N+1)}:\;\; \left(\eta+\lambda\right)\bar{v}_{2+k(N+1),0} \nonumber&= \eta\bar{v}_{2+(k-1)(N+1),0} + \lambda_{-1}\bar{v}_{2+(k+1)(N+1),0} + \bar{\pi}_{2+k(N+1)} \\&+ \dfrac{\mu\bar{\pi}_{3+(k+1)(N+1)}}{\mu+\lambda_{1}} + \sum_{j=4+(k+1)(N+1)}^{N+2+(k+1)(N+1)}{\bar{\pi}_{j}},
\end{align}
where $1 \leq k \leq B-2$, and $\sum_{j=4+(k+1)(N+1)}^{N+2+(k+1)(N+1)}{\bar{v}_{j0}}$ in (\ref{theorem:Avg_SA_proof_3}) was substituted from (\ref{theorem:Avg_SA_proof_5}). Now, repeated application of (\ref{theorem:Avg_SA_proof_6}) gives
\begin{align}\label{theorem:Avg_SA_proof_7}
\bar{v}_{2+k(N+1),0} \nonumber&= \frac{\eta \bar{v}_{2+(k-1)(N+1),0}}{\bar{c}_k} +  \sum_{j=1}^{B-k}{\dfrac{\bar{\pi}_{2+\left(k + j - 1\right)\left(N + 1\right)}\left(\mu \rho_{-1}\right)^{j-1}}{\prod_{h=1}^{j}{\bar{c}_{k + h-1}}}} \\&+ \sum_{j=1}^{B-k-1}{\dfrac{\frac{\bar{\pi}_{3+\left(k+j\right)\left(N+1\right)}}{1+\rho_1} + \sum_{m=4+\left(k+j\right)\left(N+1\right)}^{1+\left(k +j+1\right)\left(N+1\right)}{\bar{\pi}_m}}{\prod_{h=1}^{j}{\bar{c}_{k + h-1}}}\left(\mu \rho_{-1}\right)^{j-1}},\; 1 \leq k \leq B-1,
\end{align}
\begin{align}\label{theorem:Avg_SA_proof_8}
\bar{v}_{2,0} = \frac{\eta \bar{v}_{1,0}}{\bar{c}_0} +  \sum_{j=1}^{B}{\dfrac{\bar{\pi}_{2+\left(j - 1\right)\left(N + 1\right)}\left(\mu \rho_{-1}\right)^{j-1}}{\prod_{h=1}^{j}{\bar{c}_{h-1}}}} + \sum_{j=1}^{B-1}{\dfrac{\frac{\bar{\pi}_{3+j\left(N+1\right)}}{1+\rho_1} + \sum_{m=4+j\left(N+1\right)}^{1+\left(j+1\right)\left(N+1\right)}{\bar{\pi}_m}}{\prod_{h=1}^{j}{\bar{c}_{ h-1}}}\left(\mu \rho_{-1}\right)^{j-1}},
\end{align}
where the set $\{\bar{c}_h\}$ is defined in (\ref{eq:cbar}). The expression of $\bar{v}_{10}$ in (\ref{theorem:Avg_SA_2}) can be obtained by solving (\ref{theorem:Avg_SA_proof_1}) and (\ref{theorem:Avg_SA_proof_8}) while noting that $\bar{v}_{31} = \frac{\bar{\pi}_3}{\mu + \lambda_1}$ and $\mu \sum_{j=4}^{N+4}{\bar{v}_{j0}} = \lambda_{-1} \bar{v}_{20} + \sum_{j=4}^{N+2}{\bar{\pi}_j}$. Since the set $\{\bar{c}_{-1},\bar{c}_0,\cdots,\bar{c}_{B-1}\}$ contains positive real numbers, we have $\bar{v}_{10} \geq 0$. Therefore, from (\ref{theorem:Avg_SA_proof_5}), (\ref{theorem:Avg_SA_proof_7}) and (\ref{theorem:Avg_SA_proof_8}), we observe that $\bar{v}_{q0} \geq 0, \forall q \in \ncalQ,$ and hence there exits a non-negative limit $\bar{\nbv}_q, \forall q \in \ncalQ,$ satisfying (\ref{gen_vavg}). In the following, we evaluate the average AoI of source 1. By summing the equations in (\ref{theorem:Avg_SA_proof_5}), we have
\begin{align}\label{theorem:Avg_SA_proof_9}
\mu \sum_{q \in\;{\rm r}_{i+1}}{\bar{v}_{q0}} = \lambda_i \sum_{q\in\;{\rm r}_1 / \{1\}}{\bar{v}_{q0}} + \sum_{q \in\;{\rm r}_{i+1}}{\bar{\pi}_q},
\end{align}
where $1 \leq i \leq N$. Further, summing the equations in  (\ref{theorem:Avg_SA_proof_1})-(\ref{theorem:Avg_SA_proof_4}) gives
\begin{align}\label{theorem:Avg_SA_proof_10}
\lambda \sum_{q \in\;{\rm r}_1 / \{1\}}{\bar{v}_{q0}} = \mu \sum_{q \in\; {\rm r}_2}{\bar{v}_{q1}} + \mu \sum_{q \in\;\ncalQ/({\rm r}_1 \cup\; {\rm r}_2)}{\bar{v}_{q0}} + \sum_{q \in\;{\rm r}_1}{\bar{\pi}_q},    
\end{align}
where $\sum_{q \in\; {\rm r}_2}{\bar{v}_{q1}} = \dfrac{\sum_{q \in\;{\rm r}_2}{\bar{\pi}_q}}{\mu + \lambda_1}$. From (\ref{theorem:Avg_SA_proof_9}) and (\ref{theorem:Avg_SA_proof_10}), we get
\begin{align}\label{theorem:Avg_SA_proof_11}
\lambda_1 \sum_{q \in\;{\rm r}_1 / \{1\}}{\bar{v}_{q0}} = \dfrac{\sum_{q \in\;{\rm r}_2}{\bar{\pi}_q}}{1+\rho_1} + \sum_{q \in\;\ncalQ/{\rm r}_2}{\bar{\pi}_q}.
\end{align}

Hence, the average AoI of source 1 can be obtained as
\begin{align}\label{theorem:Avg_SA_proof_12}
\overset{{\rm SA}}{\Delta}_{1,1} = \sum_{q \in\; \ncalQ}{\bar{v}_{q0}} \a \dfrac{1+\rho}{\mu \rho_1\left(1+\rho_1\right)} +  \dfrac{\left(1+\rho\right)\sum_{q \in \ncalQ/ {\rm r}_2}{\bar{\pi}_q}}{\mu \left(1+\rho_1\right)} + \dfrac{\sum_{q \in \ncalQ / {\rm r}_1}{\bar{\pi}_q}}{\mu} + \bar{v}_{10},    
\end{align}
where step (a) follows from (\ref{theorem:Avg_SA_proof_9}) and (\ref{theorem:Avg_SA_proof_11}). This completes the proof.
\hfill 
\IEEEQED
\subsection{Proof of Theorem~\ref{theorem:MGF_WP}} \label{app:theorem:MGF_WP}
Using Table \ref{table:WP}, the set of equations in (\ref{gen_vMGF}) can be expressed as
\begin{align}\label{theorem_MGF_WP_proof_1}
q_1: \;\; \left(\eta - s\right)[\bar{v}^{s}_{10},\bar{v}^{s}_{11}] = \mu[\bar{v}^{s}_{31},\bar{\pi}_3],
\end{align}
\begin{align}\label{theorem_MGF_WP_proof_2}
q_2: \;\; \left(\eta + \lambda - s\right)[\bar{v}^{s}_{20},\bar{v}^{s}_{21}] = \mu[\bar{v}^{s}_{51},\bar{\pi}_5] + \eta[\bar{v}^{s}_{10},\bar{\pi}_1],
\end{align}
\begin{align}\label{theorem_MGF_WP_proof_3}
q_{2k}, 2 \leq k \leq B-1: \;\; \left(\eta + \lambda - s\right)[\bar{v}^{s}_{2k,0},\bar{v}^{s}_{2k,1}] = \mu[\bar{v}^{s}_{2k+3,1},\bar{\pi}_{2k+3}] + \eta[\bar{v}^{s}_{2k-2,0},\bar{\pi}_{2k-2}],
\end{align}
\begin{align}\label{theorem_MGF_WP_proof_4}
q_{2B}: \;\; \left(\lambda - s\right)[\bar{v}^{s}_{2B,0},\bar{v}^{s}_{2B,1}] = \eta[\bar{v}^{s}_{2B-2,0},\bar{\pi}_{2B-2}],
\end{align}
\begin{align}\label{theorem_MGF_WP_proof_5}
q_{2k+1}, 1 \leq k \leq B: \;\; \left(\mu - s\right)[\bar{v}^{s}_{2k+1,0},\bar{v}^{s}_{2k+1,1}] = \lambda_1[\bar{v}^{s}_{2k,0},\bar{\pi}_{2k}] + \lambda_2 [\bar{v}^{s}_{2k,0},\bar{v}^{s}_{2k,0}].
\end{align}

Summing the set of equations in (\ref{theorem_MGF_WP_proof_1})-(\ref{theorem_MGF_WP_proof_4}) gives
\begin{align}\label{WP_v0_r1}
\left(\lambda - s\right)\sum_{q \in \;{\rm r}_1} {\bar{v}^s_{q0}}= \mu \sum_{q \in\;{\rm r}_2}{\bar{v}^s_{q1}} + \lambda \bar{v}^s_{10}. 
\end{align}

Further, by summing the set of equations in (\ref{theorem_MGF_WP_proof_5}), we get
\begin{align}\label{WP_v0_r2}
\left(\mu - s\right)\sum_{q \in \;{\rm r}_2} {\bar{v}^s_{q0}}= \lambda \sum_{q \in \;{\rm r}_1}{\bar{v}^s_{q0}} - \lambda  \bar{v}^s_{10}.
\end{align}
\begin{align}\label{WP_v1_r2}
\left(\mu - s\right)\sum_{q \in \;{\rm r}_2} {\bar{v}^s_{q1}}= \lambda_1 \sum_{q \in \;{\rm r}_1/\{1\}}{\bar{\pi}_{q}} + \lambda_2 \sum_{q \in \;{\rm r}_1/\{1\}}{\bar{v}^s_{q0}}.
\end{align}

From (\ref{gen_MGF}), the MGF of AoI of source 1 at the destination can be evaluated as
\begin{align}\label{MGF_WP_v0}
 \overset{{\rm WP}}{M}_1(\bar{s}) = \sum_{q \in\; {\rm r}_1 \cup \;{\rm r}_2}{\bar{v}^s_{q0}} \nonumber&\a \frac{\left(\lambda + \mu - s\right)\sum_{q \in \;{\rm r}_1}{\bar{v}^s_{q0}} - \lambda \bar{v}^s_{10}}{\mu - s},
\\&\b \frac{\rho_1\left(1 + \rho - \bar{s}\right)\sum_{q \in {\rm r}_1 / \{1\}}{\bar{\pi}_q} + \bar{v}^s_{10} \rho_{1} \left(1 - \bar{s}\right)}{\left(1 - \bar{s}\right)\Big[\left(1 - \bar{s}\right) \left(\rho - \bar{s}\right) - \rho_{-1}\Big]},
\end{align}
where step (a) follows from substituting (\ref{WP_v0_r2}) into (\ref{MGF_WP_v0}), and step (b) follows from obtaining $\sum_{q \in \;{\rm r}_1}{\bar{v}^s_{q0}}$ from (\ref{WP_v0_r1})-(\ref{WP_v1_r2}) as $\frac{\rho_1\sum_{q \in {\rm r}_1 / \{1\}}{\bar{\pi}_q} + \bar{v}^s_{10}  \left(\rho_{1} - \rho \bar{s}\right)}{\left(1 - \bar{s}\right) \left(\rho - \bar{s}\right) - \rho_{-1}}$ and substituting it into (\ref{MGF_WP_v0}). Finally, $\bar{v}^s_{10}$ in (\ref{theorem:MGF_WP_v10}) can be obtained by following similar steps as in (\ref{theorem:Avg_WP_proof_6})-(\ref{theorem:Avg_WP_proof_10}). 
\hfill 
\IEEEQED
\subsection{Proof of Theorem~\ref{theorem:MGF_PS}} \label{app:MGF_PS}
Similar to Appendix \ref{app:theorem:Avg_PS}, We first note that the set of equations in (\ref{gen_vMGF}) corresponding to the states in ${\rm r}_1$ are given by (\ref{theorem_MGF_WP_proof_1})-(\ref{theorem_MGF_WP_proof_4}), and hence $\sum_{q \in\;{\rm r}_1}{\bar{v}^s_{q0}}$ can be expressed as in (\ref{WP_v0_r1}). Regarding the states in ${\rm r}_2$, we have 
\begin{align}\label{theorem_MGF_PS_proof_1}
q_{2k+1}, 1 \leq k \leq B:\;\; 
\nonumber &\left(\mu - s\right)\bar{v}^s_{2k+1,0} = \lambda \bar{v}^s_{2k,0},\\
&\left(\lambda + \mu - s\right) \bar{v}^s_{2k+1,1} = \lambda_2 (\bar{v}^s_{2k,0} + \bar{v}^s_{2k+1,0}) + \lambda_1 (\bar{\pi}_{2k} + \bar{\pi}_{2k+1}).
\end{align}

We observe from (\ref{theorem_MGF_PS_proof_1}) that $\sum_{q \in\;{\rm r}_2}{\bar{v}^s_{q0}}$ is given by (\ref{WP_v0_r2}) and $\sum_{q\in\;{\rm r}_2}{\bar{v}^s_{q1}}$ can be expressed as
\begin{align}\label{PS_v1_r2}
\left(\lambda + \mu - s\right)\sum_{q \in\;{\rm r}_2}{\bar{v}^s_{q1}} = \lambda_2 \sum_{q \in\;\ncalQ/\{1\}}{\bar{v}^s_{q0}} + \lambda_1 \left(1 - \bar{\pi}_1\right).
\end{align}

Hence, the MGF of AoI of source 1 at the destination can be evaluated as
\begin{align}\label{MGF_PS_v0}
  \overset{\rm PS}{M}_1(\bar{s}) &= \sum_{q \in\; {\rm r}_1 \cup \;{\rm r}_2}{\bar{v}^s_{q0}} \a \dfrac{\left(\lambda + \mu - s\right)\sum_{q \in \;{\rm r}_1}{\bar{v}^s_{q0}} - \lambda \bar{v}^s_{10}}{\mu - s} \b \dfrac{\rho_1\left(1 - \bar{\pi}_1 + \bar{v}^s_{10}\right)}{\left(1 - \bar{s}\right) \left(\rho - \bar{s}\right) - \rho_{-1}},
\end{align}
where step (a) follows from substituting (\ref{WP_v0_r2}) into (\ref{MGF_PS_v0}), and step (b) follows from obtaining $\sum_{q \in \;{\rm r}_1}{\bar{v}^s_{q0}}$ from (\ref{WP_v0_r1}), (\ref{WP_v0_r2}) and (\ref{PS_v1_r2}) as $\frac{\rho_1\left(1 - \bar{\pi}_1\right)\left(1 - \bar{s}\right) + \bar{v}^s_{10}  \left(\rho_{1} - \rho \bar{s}\right)\left(1 + \rho - \bar{s}\right)}{\left(1 + \rho - \bar{s}\right)\left[\left(1 - \bar{s}\right) \left(\rho - \bar{s}\right) - \rho_{-1}\right]}$ and substituting it into (\ref{MGF_PS_v0}). Finally, $\bar{v}^s_{10}$ in (\ref{theorem:MGF_PS_v10}) can be obtained by following similar steps as in (\ref{theorem:Avg_PS_proof_2})-(\ref{theorem:Avg_PS_proof_4}). 
\hfill 
\IEEEQED

\subsection{Proof of Theorem~\ref{theorem:MGF_SA}} \label{app:theorem:MGF_SA}
Using Table \ref{table:SA}, the set of equations in (\ref{gen_vMGF}) corresponding to $q \in {\rm r}_1$ can be expressed as
\begin{align}\label{theorem:MGF_SA_proof_1}
q_1:\;\; \left(\eta - s \right)\bar{v}^s_{10} = \mu\bar{v}^s_{31} + \mu \sum_{j=4}^{N+2}{\bar{v}^s_{j0}},
\end{align}
\begin{align}\label{theorem:MGF_SA_proof_2}
q_2:\;\; \left(\eta+\lambda - s\right)\bar{v}^s_{20} = \eta\bar{v}^s_{10} + \mu\bar{v}^s_{N+4,1} + \mu \sum_{j=N+5}^{2N+3}{\bar{v}^s_{j0}},
\end{align}
\begin{align}\label{theorem:MGF_SA_proof_3}
\nonumber q_{2+k(N+1)}, 1 \leq k \leq B-2:\;\; \left(\eta+\lambda-s\right)\bar{v}^s_{2+k(N+1),0} &= \eta\bar{v}^s_{2+(k-1)(N+1),0} + \mu\bar{v}^s_{3+(k+1)(N+1),1} \\
&+ \mu \sum_{j=4+(k+1)(N+1)}^{N+2+(k+1)(N+1)}{\bar{v}^s_{j0}},
\end{align}
\begin{align}\label{theorem:MGF_SA_proof_4}
 q_{2+(B-1)(N+1)}:\;\; \left(\lambda - s\right)\bar{v}^s_{2+(B-1)(N+1),0} = \eta\bar{v}^s_{2+(B-2)(N+1),0}.
\end{align}

Further, the set of equations in (\ref{gen_vMGF}) corresponding to $q \in {\rm r}_{i+1}$, $1 \leq i \leq N$, can be expressed as
\begin{align}\label{theorem:MGF_SA_proof_5}
q_{2+i+k(N+1)}, 0 \leq k \leq B-1:\;\; \left(\mu - s\right) \bar{v}^s_{2+i+k(N+1),0} = \lambda_i \bar{v}^s_{2+k(N+1),0}.   
\end{align}

Summing the equations in (\ref{theorem:MGF_SA_proof_1})-(\ref{theorem:MGF_SA_proof_4}) gives
\begin{align}\label{theorem:MGF_SA_proof_6}
\left(\lambda - s\right) \sum_{q \in\;{\rm r}_1}{\bar{v}^s_{q0}} = \mu \sum_{q \in\; {\rm r}_2}{\bar{v}^s_{q1}} + \mu \sum_{q \in\;\ncalQ/({\rm r}_1 \cup\; {\rm r}_2)}{\bar{v}^s_{q0}} + \lambda \bar{v}^s_{10},    
\end{align}
where $\sum_{q \in\; {\rm r}_2}{\bar{v}^s_{q1}} = \dfrac{\lambda_1\sum_{q \in\;{\rm r}_1 \cup\; {\rm r}_2/\{1\}}{\bar{\pi}_q}}{\left(\mu + \lambda_1 - s\right)}$. In addition, by summing the equations in (\ref{theorem:MGF_SA_proof_5}), we get
\begin{align}\label{theorem:MGF_SA_proof_7}
\left(\mu - s\right) \sum_{q \in\;{\rm r}_{i+1}}{\bar{v}^s_{q0}} = \lambda_i \sum_{q\in\;{\rm r}_1 / \{1\}}{\bar{v}^s_{q0}},
\end{align}
where $1 \leq i \leq N$. From (\ref{theorem:MGF_SA_proof_6}) and (\ref{theorem:MGF_SA_proof_7}), $\sum_{q \in\;{\rm r}_{1}}{\bar{v}^s_{q0}}$ can be obtained as
\begin{align}\label{theorem:MGF_SA_proof_8}
\big[\left(1 - \bar{s}\right) \left(\rho - \bar{s}\right) - \rho_{-1}\big] \sum_{q \in\;{\rm r}_1}{\bar{v}^s_{q0}} = \dfrac{\rho_1\left(1 - \bar{s}\right)\sum_{q \in\;{\rm r}_1 \cup\; {\rm r}_2 /\{1\}}{\bar{\pi}_q}}{1+\rho_1 - \bar{s}} + \left(\rho_1 - \rho \bar{s}\right) \bar{v}^s_{10}.
\end{align}

Hence, the MGF of AoI of source 1 can be evaluated as 
\begin{align}\label{theorem:MGF_SA_proof_9}
\nonumber \overset{\rm SA}{M}_1(\bar{s}) = \sum_{q \in\; \ncalQ}{\bar{v}^s_{q0}} &\a \dfrac{\left(\lambda + \mu - s\right)\sum_{q \in \;{\rm r}_1}{\bar{v}^s_{q0}} - \lambda \bar{v}^s_{10}}{\mu - s} \\
 &\b \dfrac{\rho_1\Big[\left(1+\rho-\bar{s}\right)\sum_{q \in {\rm r}_1 \cup \;{\rm r}_2 /\{1\}}{\bar{\pi}_q} + \left(1 + \rho_1 - \bar{s}\right)\bar{v}^s_{10}\Big]}{\left(1 + \rho_1 - \bar{s}\right)\Big[\left(1 - \bar{s}\right) \left(\rho - \bar{s}\right) - \rho_{-1}\Big]},
\end{align}
where step (a) [step (b)] follows from substituting (\ref{theorem:MGF_SA_proof_7}) [(\ref{theorem:MGF_SA_proof_8})] into (\ref{theorem:MGF_SA_proof_9}). Finally, $\bar{v}^s_{10}$ in (\ref{theorem:MGF_SA_v10}) can be obtained by following similar steps as in (\ref{theorem:Avg_SA_proof_6})-(\ref{theorem:Avg_SA_proof_8}). 
\hfill 
\IEEEQED

\bibliographystyle{IEEEtran}
\bibliography{AoI_QT_Multi_source}

\begin{thebibliography}{10}
\providecommand{\url}[1]{#1}
\csname url@samestyle\endcsname
\providecommand{\newblock}{\relax}
\providecommand{\bibinfo}[2]{#2}
\providecommand{\BIBentrySTDinterwordspacing}{\spaceskip=0pt\relax}
\providecommand{\BIBentryALTinterwordstretchfactor}{4}
\providecommand{\BIBentryALTinterwordspacing}{\spaceskip=\fontdimen2\font plus
\BIBentryALTinterwordstretchfactor\fontdimen3\font minus
  \fontdimen4\font\relax}
\providecommand{\BIBforeignlanguage}[2]{{%
\expandafter\ifx\csname l@#1\endcsname\relax
\typeout{** WARNING: IEEEtran.bst: No hyphenation pattern has been}%
\typeout{** loaded for the language `#1'. Using the pattern for}%
\typeout{** the default language instead.}%
\else
\language=\csname l@#1\endcsname
\fi
#2}}
\providecommand{\BIBdecl}{\relax}
\BIBdecl

\bibitem{abd2018role}
M.~A. Abd-Elmagid, N.~Pappas, and H.~S. Dhillon, ``On the role of age of
  information in the {Internet} of things,'' \emph{IEEE Commun. Magazine},
  vol.~57, no.~12, pp. 72--77, Dec. 2019.

\bibitem{Roy20}
A.~Roy, F.~H. Kumbhar, H.~S. Dhillon, N.~Saxena, S.~Y. Shin, and S.~Singh,
  ``Efficient monitoring and contact tracing for covid-19: A smart {IoT}-based
  framework,'' \emph{IEEE Internet of Things Magazine}, vol.~3, no.~3, pp.
  17--23, Sept. 2020.

\bibitem{kaul2012real}
S.~Kaul, R.~Yates, and M.~Gruteser, ``Real-time status: How often should one
  update?'' in \emph{Proc., IEEE INFOCOM}, 2012.

\bibitem{abd2018coverage}
M.~A. {Abd-Elmagid}, M.~A. {Kishk}, and H.~S. {Dhillon}, ``Joint energy and
  {SINR} coverage in spatially clustered {RF}-powered {IoT} network,''
  \emph{IEEE Trans. on Green Commun. and Networking}, vol.~3, no.~1, pp.
  132--146, March 2019.

\bibitem{costa2016age}
M.~Costa, M.~Codreanu, and A.~Ephremides, ``On the age of information in status
  update systems with packet management,'' \emph{IEEE Trans. on Info. Theory},
  vol.~62, no.~4, pp. 1897--1910, Apr. 2016.

\bibitem{kaul2012status}
S.~K. Kaul, R.~D. Yates, and M.~Gruteser, ``Status updates through queues,'' in
  \emph{Proc., IEEE Conf. on Info. Sciences and Systems (CISS)}, 2012.

\bibitem{soysal}
A.~Soysal and S.~Ulukus, ``Age of information in {G/G/1/1} systems: Age
  expressions, bounds, special cases, and optimization,'' {\em IEEE Trans. on
  Info. Theory}, to appear.

\bibitem{chen2016age}
K.~Chen and L.~Huang, ``Age-of-information in the presence of error,'' in
  \emph{Proc., IEEE Intl. Symposium on Information Theory}, 2016.

\bibitem{kam2018age}
C.~Kam, S.~Kompella, G.~D. Nguyen, J.~E. Wieselthier, and A.~Ephremides, ``On
  the age of information with packet deadlines,'' \emph{IEEE Trans. on Info.
  Theory}, vol.~64, no.~9, pp. 6419--6428, Sept. 2018.

\bibitem{kavitha2018controlling}
V.~Kavitha, E.~Altman, and I.~Saha, ``Controlling packet drops to improve
  freshness of information,'' 2018, available online: arxiv.org/abs/1807.09325.

\bibitem{zou2019waiting}
P.~Zou, O.~Ozel, and S.~Subramaniam, ``Waiting before serving: A companion to
  packet management in status update systems,'' \emph{IEEE Trans. on Info.
  Theory}, vol.~66, no.~6, pp. 3864--3877, June 2019.

\bibitem{Inoue19}
Y.~{Inoue}, H.~{Masuyama}, T.~{Takine}, and T.~{Tanaka}, ``A general formula
  for the stationary distribution of the age of information and its application
  to single-server queues,'' \emph{IEEE Trans. Info. Theory}, vol.~65, no.~12,
  pp. 8305--8324, Dec. 2019.

\bibitem{kosta2020non}
A.~Kosta, N.~Pappas, A.~Ephremides, and V.~Angelakis, ``The age of information
  in a discrete time queue: Stationary distribution and non-linear age mean
  analysis,'' \emph{IEEE Journal on Sel. Areas in Commun.}, vol.~39, no.~5, pp.
  1352--1364, May 2021.

\bibitem{Champati19}
J.~P. {Champati}, H.~{Al-Zubaidy}, and J.~{Gross}, ``On the distribution of
  {AoI} for the {GI/GI/1/1 and GI/GI/1/2*} systems: Exact expressions and
  bounds,'' in \emph{Proc., IEEE INFOCOM}, 2019.

\bibitem{Chiariotti_dist}
F.~Chiariotti, O.~Vikhrova, B.~Soret, and P.~Popovski, ``Peak age of
  information distribution for edge computing with wireless links,'' \emph{IEEE
  Trans. on Commun.}, vol.~69, no.~5, pp. 3176--3191, May 2021.

\bibitem{ayan2020probability}
O.~Ayan, H.~M. G{\"u}rsu, A.~Papa, and W.~Kellerer, ``Probability analysis of
  age of information in multi-hop networks,'' \emph{IEEE Networking Letters},
  vol.~2, no.~2, pp. 76--80, June 2020.

\bibitem{olga20}
O.~{Vikhrova}, F.~{Chiariotti}, B.~{Soret}, G.~{Araniti}, A.~{Molinaro}, and
  P.~{Popovski}, ``Age of information in multi-hop networks with priorities,''
  in \emph{Proc., IEEE Globecom}, 2020.

\bibitem{yates2012real}
R.~D. Yates and S.~Kaul, ``Real-time status updating: Multiple sources,'' in
  \emph{Proc., IEEE Intl. Symposium on Information Theory}, 2012.

\bibitem{Moltafet_multisource}
M.~{Moltafet}, M.~{Leinonen}, and M.~{Codreanu}, ``On the age of information in
  multi-source queueing models,'' \emph{IEEE Trans. on Commun.}, vol.~68,
  no.~8, pp. 5003--5017, Aug. 2020.

\bibitem{pappas2015age}
N.~Pappas, J.~Gunnarsson, L.~Kratz, M.~Kountouris, and V.~Angelakis, ``Age of
  information of multiple sources with queue management,'' in \emph{Proc., IEEE
  Intl. Conf. on Commun. (ICC)}, 2015.

\bibitem{yates2017status}
R.~D. Yates and S.~K. Kaul, ``Status updates over unreliable multiaccess
  channels,'' in \emph{Proc., IEEE Intl. Symposium on Information Theory},
  2017.

\bibitem{kosta2019age}
A.~Kosta, N.~Pappas, A.~Ephremides, and V.~Angelakis, ``Age of information
  performance of multiaccess strategies with packet management,'' \emph{Journal
  of Commun. and Networks}, vol.~21, no.~3, pp. 244--255, June 2019.

\bibitem{najm2018status}
E.~Najm and E.~Telatar, ``Status updates in a multi-stream m/g/1/1 preemptive
  queue,'' in \emph{Proc., IEEE INFOCOM Workshops}, 2018.

\bibitem{huang2015optimizing}
L.~Huang and E.~Modiano, ``Optimizing age-of-information in a multi-class
  queueing system,'' in \emph{Proc., IEEE Intl. Symposium on Information
  Theory}, 2015.

\bibitem{Xu_21}
J.~{Xu} and N.~{Gautam}, ``Peak age of information in priority queuing
  systems,'' \emph{IEEE Trans. on Info. Theory}, vol.~67, no.~1, pp. 373--390,
  Jan. 2021.

\bibitem{Akar21}
N.~{Akar}, ``Discrete-time queueing model of age of information with multiple
  information sources,'' {\em IEEE Internet of Things Journal}, to appear.

\bibitem{Ozancan_2021}
O.~Dogan and N.~Akar, ``The multi-source probabilistically preemptive m/ph/1/1
  queue with packet errors,'' {\em IEEE Trans. on Commun.}, to appear.

\bibitem{yates2018age}
R.~D. Yates and S.~K. Kaul, ``The age of information: Real-time status updating
  by multiple sources,'' \emph{IEEE Trans. on Info. Theory}, vol.~65, no.~3,
  pp. 1807--1827, Mar. 2019.

\bibitem{yates2020age}
R.~D. Yates, ``The age of information in networks: Moments, distributions, and
  sampling,'' \emph{IEEE Trans. on Info. Theory}, vol.~66, no.~9, pp.
  5712--5728, Sept. 2020.

\bibitem{hespanha2006modelling}
J.~P. Hespanha, ``Modelling and analysis of stochastic hybrid systems,''
  \emph{IEE Proceedings-Control Theory and Applications}, vol. 153, no.~5, pp.
  520--535, Sept. 2006.

\bibitem{SHS_6}
A.~{Maatouk}, M.~{Assaad}, and A.~{Ephremides}, ``On the age of information in
  a csma environment,'' \emph{IEEE/ACM Trans. on Networking}, vol.~28, no.~2,
  pp. 818--831, Apr. 2020.

\bibitem{SHS_7}
M.~{Moltafet}, M.~{Leinonen}, and M.~{Codreanu}, ``Average {AoI} in
  multi-source systems with source-aware packet management,'' \emph{IEEE Trans.
  on Commun.}, vol.~69, no.~2, pp. 1121--1133, Feb. 2021.

\bibitem{SHS_5}
A.~Javani, M.~Zorgui, and Z.~Wang, ``Age of information for multiple-source
  multiple-server networks,'' 2021, available online: arxiv.org/abs/2106.07247.

\bibitem{SHS_8}
M.~Moltafet, M.~Leinonen, and M.~Codreanu, ``Moment generating function of the
  {AoI} in a two-source system with packet management,'' \emph{IEEE Wireless
  Commun. Letters}, vol.~10, no.~4, pp. 882--886, Apr. 2021.

\bibitem{Yates_EH}
R.~D. {Yates}, ``Lazy is timely: Status updates by an energy harvesting
  source,'' in \emph{Proc., IEEE Intl. Symposium on Information Theory}, 2015.

\bibitem{zheng2019closed}
X.~Zheng, S.~Zhou, Z.~Jiang, and Z.~Niu, ``Closed-form analysis of non-linear
  age of information in status updates with an energy harvesting transmitter,''
  \emph{IEEE Trans. on Wireless Commun.}, vol.~18, no.~8, pp. 4129--4142, Aug.
  2019.

\bibitem{farazi2018average}
S.~Farazi, A.~G. Klein, and D.~R. Brown, ``Average age of information for
  status update systems with an energy harvesting server,'' in \emph{Proc.,
  IEEE INFOCOM Workshops}, 2018.

\bibitem{farazi2018bverage}
S.~{Farazi}, A.~G. {Klein}, and D.~R. {Brown}, ``Age of information in energy
  harvesting status update systems: When to preempt in service?'' in
  \emph{Proc., IEEE Intl. Symposium on Information Theory}, 2018.

\bibitem{abdelmagid_2021a}
M.~A. Abd-Elmagid and H.~S. Dhillon, ``Closed-form characterization of the
  {MGF} of {AoI} in energy harvesting status update systems,'' 2021, available
  online: arxiv.org/abs/2105.07074.

\bibitem{roy_survey}
R.~D. Yates, Y.~Sun, D.~R. Brown, S.~K. Kaul, E.~Modiano, and S.~Ulukus, ``Age
  of information: An introduction and survey,'' \emph{EEE Journal on Selected
  Areas in Commun.}, vol.~39, no.~5, pp. 1183--1210, May 2021.

\bibitem{Baran_EH}
B.~T. Bacinoglu, E.~T. Ceran, and E.~Uysal-Biyikoglu, ``Age of information
  under energy replenishment constraints,'' in \emph{Proc., Information Theory
  and its Applications (ITA)}, 2015.

\bibitem{Jing_EH}
X.~Wu, J.~Yang, and J.~Wu, ``Optimal status update for age of information
  minimization with an energy harvesting source,'' \emph{IEEE Trans. on Green
  Commun. and Networking}, vol.~2, no.~1, pp. 193--204, Mar. 2018.

\bibitem{Leng19}
S.~Leng and A.~Yener, ``Age of information minimization for an energy
  harvesting cognitive radio,'' \emph{IEEE Trans. on Cognitive Commun. and
  Networking}, vol.~5, no.~2, pp. 427--439, June 2019.

\bibitem{arafa2019age}
A.~Arafa, J.~Yang, S.~Ulukus, and H.~V. Poor, ``Age-minimal transmission for
  energy harvesting sensors with finite batteries: Online policies,''
  \emph{IEEE Trans. on Info. Theory}, vol.~66, no.~1, pp. 534--556, Jan. 2020.

\bibitem{AbdElmagid2019Globecom_a}
M.~A. Abd-Elmagid, H.~S. Dhillon, and N.~Pappas, ``Online age-minimal sampling
  policy for {RF}-powered {IoT} networks,'' in \emph{Proc., IEEE Globecom},
  2019.

\bibitem{hatami2020age}
M.~Hatami, M.~Jahandideh, M.~Leinonen, and M.~Codreanu, ``Age-aware status
  update control for energy harvesting iot sensors via reinforcement
  learning,'' in \emph{Proc., IEEE PIMRC}, 2020.

\bibitem{abd2019tcom}
M.~A. Abd-Elmagid, H.~S. Dhillon, and N.~Pappas, ``A reinforcement learning
  framework for optimizing age of information in {RF}-powered communication
  systems,'' \emph{IEEE Trans. on Commun.}, vol.~68, no.~8, pp. 4747 -- 4760,
  Aug. 2020.

\bibitem{AbdElmagid_joint}
M.~A. {Abd-Elmagid}, H.~S. {Dhillon}, and N.~{Pappas}, ``Ao{I}-optimal joint
  sampling and updating for wireless powered communication systems,''
  \emph{IEEE Trans. on Veh. Technology}, vol.~69, no.~11, pp. 14\,110--14\,115,
  Nov. 2020.

\bibitem{Elvina21}
E.~Gindullina, L.~Badia, and D.~G\"{u}nd\"{u}z, ``Age-of-information with
  information source diversity in an energy harvesting system,'' \emph{IEEE
  Trans. on Green Commun. and Networking}, vol.~5, no.~3, pp. 1529--1540, Sept.
  2021.

\bibitem{khorsandmanesh2020average}
Y.~Khorsandmanesh, M.~J. Emadi, and I.~Krikidis, ``Average peak age of
  information analysis for wireless powered cooperative networks,'' {\em IEEE
  Trans. on Cognitive Commun. and Networking}, to appear.

\bibitem{nouri2020age}
N.~Nouri, D.~Ardan, and M.~M. Feghhi, ``Age of information-reliability
  trade-offs in energy harvesting sensor networks,'' 2020, available online:
  arxiv.org/abs/2008.00987.

\bibitem{sun2017update}
Y.~Sun, E.~Uysal-Biyikoglu, R.~D. Yates, C.~E. Koksal, and N.~B. Shroff,
  ``Update or wait: How to keep your data fresh,'' \emph{IEEE Trans. on Info.
  Theory}, vol.~63, no.~11, pp. 7492--7508, Nov. 2017.

\bibitem{Qing_he}
Q.~{He}, D.~{Yuan}, and A.~{Ephremides}, ``Optimal link scheduling for age
  minimization in wireless systems,'' \emph{IEEE Trans. on Info. Theory},
  vol.~64, no.~7, pp. 5381--5394, July 2018.

\bibitem{han2020fairness}
B.~Han, Y.~Zhu, Z.~Jiang, M.~Sun, and H.~D. Schotten, ``Fairness for freshness:
  Optimal age of information based {OFDMA} scheduling with minimal knowledge,''
  {\em IEEE Trans. on Wiress Commun.}, to appear.

\bibitem{ornee2019sampling}
T.~Z. Ornee and Y.~Sun, ``Sampling for remote estimation through queues: Age of
  information and beyond,'' in \emph{Proc., Modeling and Optimization in
  Mobile, Ad Hoc and Wireless Networks}, 2019.

\bibitem{abdel2018ultra}
M.~K. Abdel-Aziz, C.-F. Liu, S.~Samarakoon, M.~Bennis, and W.~Saad,
  ``Ultra-reliable low-latency vehicular networks: Taming the age of
  information tail,'' in \emph{Proc., IEEE Globecom}, 2018.

\bibitem{abd2018average}
M.~A. {Abd-Elmagid} and H.~S. {Dhillon}, ``Average peak age-of-information
  minimization in {UAV}-assisted {IoT} networks,'' \emph{IEEE Trans. on Veh.
  Technology}, vol.~68, no.~2, pp. 2003--2008, Feb. 2019.

\bibitem{AbdElmagid2019Globecom_b}
M.~A. Abd-Elmagid, A.~Ferdowsi, H.~S. Dhillon, and W.~Saad, ``Deep
  reinforcement learning for minimizing age-of-information in {UAV}-assisted
  networks,'' in \emph{Proc., IEEE Globecom}, 2019.

\bibitem{ferdowsi2021neural}
A.~Ferdowsi, M.~A. Abd-Elmagid, W.~Saad, and H.~S. Dhillon, ``Neural
  combinatorial deep reinforcement learning for age-optimal joint trajectory
  and scheduling design in uav-assisted networks,'' \emph{IEEE Journal on
  Selected Areas in Commun.}, vol.~39, no.~5, pp. 1250--1265, May 2021.

\bibitem{emara2019spatiotemporal}
M.~{Emara}, H.~{Elsawy}, and G.~{Bauch}, ``A spatiotemporal model for peak
  {AoI} in uplink {IoT} networks: Time versus event-triggered traffic,''
  \emph{IEEE Internet Things Journal}, vol.~7, no.~8, pp. 6762--6777, Aug.
  2020.

\bibitem{mankar2020stochastic_GC2}
P.~D. Mankar, M.~A. Abd-Elmagid, and H.~S. Dhillon, ``Spatial distribution of
  the mean peak age of information in wireless networks,'' \emph{IEEE Trans. on
  Wireless Commun.}, vol.~20, no.~7, pp. 4465--4479, July 2021.

\bibitem{Praful_GC1}
P.~D. Mankar, Z.~Chen, M.~A. Abd-Elmagid, N.~Pappas, and H.~S. Dhillon,
  ``Throughput and age of information in a cellular-based {IoT} network,''
  \emph{IEEE Trans. on Wireless Commun., to appear}.

\bibitem{tang2020age}
H.~Tang, P.~Ciblat, J.~Wang, M.~Wigger, and R.~Yates, ``Age of information
  aware cache updating with file-and age-dependent update durations,'' in
  \emph{Proc., Modeling and Optimization in Mobile, Ad Hoc and Wireless
  Networks}, 2020.

\bibitem{ma2020age}
M.~Ma and V.~W. Wong, ``Age of information driven cache content update
  scheduling for dynamic contents in heterogeneous networks,'' \emph{IEEE
  Trans. on Wireless Commun.}, vol.~19, no.~12, pp. 8427--8441, Dec. 2020.

\bibitem{bastopcu2020information}
M.~{Bastopcu} and S.~{Ulukus}, ``Information freshness in cache updating
  systems,'' \emph{IEEE Trans. on Wireless Commun.}, vol.~20, no.~3, pp.
  1861--1874, Mar. 2021.

\bibitem{yang2020age}
H.~H. Yang, A.~Arafa, T.~Q. Quek, and H.~V. Poor, ``Age-based scheduling policy
  for federated learning in mobile edge networks,'' in \emph{Proc., IEEE Intl.
  Conf. on Acoustics, Speech, and Sig. Proc. (ICASSP)}, 2020.

\bibitem{buyukates2020timely}
B.~Buyukates and S.~Ulukus, ``Timely communication in federated learning,'' in
  \emph{Proc., IEEE INFOCOM Workshops}, 2021.

\bibitem{Jain}
R.~Jain, D.-M. Chiu, and W.~Hawe, ``A quantitative measure of fairness and
  discrimination for resource allocation in shared computer systems,''
  \emph{DEC Research Repor, Technical Report TR-301}, Sept. 1984.

\end{thebibliography}
\end{document}